\documentclass{aa}  

\usepackage{graphicx}
\usepackage{txfonts}
\usepackage{lipsum}
\usepackage{subcaption}         
\usepackage{lscape}             
\usepackage{placeins}           

\usepackage{xcolor}
                                
\usepackage[colorlinks=true]{hyperref}
\hypersetup{
    colorlinks = true,
    citecolor = {blue},
    linkcolor = {blue}
}

\begin{document}

   \title{MUSE-ALMA Haloes}
   \subtitle{XV: Cold molecular, ionised, and stellar kinematics of HI-rich galaxies at $z\sim 0.5$}


   \author{C. Barfety \inst{1}
        \and N. M. F\"orster Schreiber\inst{1}
        \and C. P\'eroux\inst{2,3}
        \and L. J. Tacconi\inst{1}
        \and J. Chen \inst{1}
        \and S. Weng \inst{3}
        \and V. Bollo \inst{4}
        \and T. O'Beirne \inst{2, 5, 6}
        \and M. Zwaan \inst{2}
        \and R. Augustin \inst{7}
        \and B. Casavecchia \inst{8}
        \and L. Chemin \inst{9}
        \and A. L . Danhaive \inst{10,11}
        \and N. Guerra-Varas \inst{2}
        \and V. Kulkarni \inst{12}
        \and R. Potlia \inst{2}
        }

   \institute{Max-Planck-Institut für extraterrestrische Physik, Giessenbachstrasse, D-85748 Garching, Germany 
             \email{cbarfety@mpe.mpg.de}
             \and
             European Southern Observatory, Karl-Schwarzschild-Str. 2, 85748 Garching near Munich, Germany 
             \and
             Aix Marseille Univ., CNRS, LAM, (Laboratoire d’Astrophysique de Marseille), UMR 7326, F-13388 Marseille, France 
             \and
             Department of Astronomy, Stockholm University, Oscar Klein Centre, AlbaNova University Centre, 106 91 Stockholm, Sweden 
             \and
             Centre for Astrophysics and Supercomputing, Swinburne University of Technology, Hawthorn, Victoria 3122, Australia 
             \and 
             CSIRO Space \& Astronomy, PO Box 1130, Bentley, Western Australia 6102, Australia 
             \and
             Leibniz-Institut für Astrophysik Potsdam (AIP), An der Sternwarte 16, 14482 Potsdam, Germany 
             \and
             Max-Planck-Institut für Astrophysik, Karl-Schwarzschild-Strasse 1, 85748 Garching near Munich, Germany 
             \and 
             Observatoire Astronomique de Strasbourg - UMR 7550, 11 rue de l’Université, 67000 Strasbourg, France 
             \and
             Kavli Institute for Cosmology, University of Cambridge, Madingley Road, Cambridge CB3 0HA, UK 
             \and
             Cavendish Laboratory, University of Cambridge, 19 JJ Thomson Avenue, Cambridge CB3 0HE, UK 
             \and
             Department of Physics and Astronomy, University of South Carolina, Columbia, SC 29208, USA 
            }

  \abstract 
  {Spatially resolved galaxy kinematics have been most extensively studied at $z\sim0$ and $z>1$. The emerging consensus is that disks are common out to $z\sim3-6$, albeit with larger disk velocity dispersions. In this picture, $z\sim0.5$ appears as a key transition in several fundamental aspects, such as the main driver of disk gas turbulence and baryonic-dark matter (DM) mass budget. While at $z\sim0$ multi-tracer spatially resolved characterisation is most readily accessible to study galaxy dynamics, studies with more than one tracer are still sparse for most galaxies outside the local Universe.}
  {We exploited a unique dataset covering spatially resolved cold molecular gas observations with ALMA and spatially resolved ionised gas line emission and stellar continuum with MUSE to study the resolved kinematics of six galaxies at $0.43<z<0.83$. Obtained as part of the MUSE-ALMA Haloes program and selected by their association with neutral atomic hydrogen (H\,{\sc i}) absorption systems, the galaxies span stellar masses from $\mathrm{log}(M_\star/M_\odot) = 10.1 - 11.8$. }
  {Using the new ALMA/CO observations and combining them with the existing MUSE ionised gas data and \textit{Hubble Space Telescope} stellar morphologies, we determined the kinematic nature of the sources by applying morpho-kinematic criteria. For those identified as rotating disks, we employed the forward-modelling parametric kinematic fitting code {\scshape DysmalPy} to jointly model the H$\beta$ and CO(2-1) observations of the galaxies, and we compared the results to theoretical predictions for pressure support in galaxies. For the stellar kinematics, we used {\scshape pPXF} to fit rest-frame optical stellar absorption features.}
  {We find that four of the six galaxies can be classified as disks, and two of these galaxies have previously identified galaxy group members. The four disk galaxies exhibit elevated gas velocity dispersions compared to $z\sim0$ disks, in agreement with the interpolation of observed trends between $z\sim0$ and $z>1$. The results are consistent with marginally stable gaseous disks, where gravity-driving still plays a role in addition to stellar feedback, although we find tentative evidence of a differentiation between a dynamically hotter warm phase and the cold gas ($\sigma_\mathrm{0, H\beta} \sim1.7 \times \sigma_\mathrm{0, CO}$). We derived the average DM fraction $\langle f_{DM} \rangle = 0.42\pm0.14$, which is consistent with observations at comparable epochs and stellar masses and surface densities. This study highlights the use and completeness of analysing a sample of galaxies in a self-consistent manner and using multi-wavelength observations to explore the physics of disk settling in a relatively less explored redshift range.}
  {}


   \maketitle \nolinenumbers

\section{Introduction}
Understanding the kinematics of star-forming galaxies (SFGs) and how they evolve with redshift is a key ingredient of galaxy evolution \citep{sofue_rotation_2001}. In the simplest picture, the observed rotation curve reflects the enclosed mass profile within the gravitational potential. The rise and turnover of the circular velocity constrain the relative contributions of a central baryonic component (bulge plus inner disk) and an extended dark-matter halo, while departures from purely circular motion can flag additional mass components (e.g. bars, spiral arms, or recent perturbations; \citealt{genzel_evidence_2023, pastras_noema3d_2025, jolly_noema3d_2026}). Over the past two decades, and with the advent of integral field units (IFUs), a significant body of work has brought to light the prevalence of rotation-dominated disk galaxies out to $z \sim 3$ using mostly rest-frame optical lines such as H$\alpha$ or [O\,{\sc iii}] to trace the ionised gas component (see \citealt{forster_schreiber_star-forming_2020} for a recent review). This is in stark contrast with the initial picture of a chaotic merger-driven evolution of galaxies powering star-forming (SF) activity \citep{sanders_luminous_1996}. It put forward a view in which internal processes in the general SFG population play an important role in the regulated growth and evolution underlying galaxy-scale scaling relations, such as the main sequence (MS) between stellar mass ($M_\star$) and star formation rate (SFR; e.g. see review by \citealt{tacconi_evolution_2020}). More recently, the capabilities of the \textit{James Webb Space Telescope} (\textit{JWST}) have pushed the barrier of ionised gas kinematic studies to much higher redshifts (out to $z=8$), revealing the presence of ordered rotation also in the early Universe \citep{rowland_rebels-25_2024, de_graaff_ionised_2024, danhaive_dawn_2025}.  

A key result emerging from resolved kinematic surveys is that the majority of massive near–MS galaxies at \(z\sim1\)–3 are rotating disks rather than major mergers, albeit characterised by elevated gas velocity dispersion compared to $z \sim 0$ disks. In \(z\sim2-3\) disk galaxies, velocity dispersions in the ionised interstellar medium (ISM) are a factor of two higher than in present-day spirals ($\sim 40 - 70$ km/s versus $\sim 15 - 30$ km/s), implying geometrically thicker and more turbulent disks \citep{forster_schreiber_sinfoni_2006, forster_schreiber_sins_2009, kassin_epoch_2012, wisnioski_kmos3d_2015, wisnioski_kmos3d_2019, ubler_evolution_2019, genzel_evidence_2023, tsukui_emergence_2025}. This translates to lower rotational-to-random motion support, typically quantified via $V_\mathrm{rot}/\sigma_0$ (where $V_\mathrm{rot}/\sigma_0 >3$ implies a rotation-dominated disk), and some compact systems even appearing dispersion dominated \citep{newman_sinszc-sinf_2013}. This empirical evolution underpins the `disk settling' scenario, in which SFGs become progressively more kinematically ordered with time as disordered motions decline and rotational support strengthens, with the most massive systems settling earlier on average (e.g. \citealt{kassin_epoch_2012, simons_z_2017}). Physically, the elevated dispersions and their redshift dependence are often linked to the higher gas fractions and stronger feedback and accretion processes at early times \citep{peroux_cosmic_2020, peroux_multi-scale_2024}. Processes such as inflows, SF, and outflows, together with gravitational instability and continued accretion and interactions, can repeatedly disturb and then rapidly re-establish disks \citep{bouche_impact_2010, dave_analytic_2012, lilly_gas_2013, krumholz_unified_2018, wisnioski_evolution_2025}. A complementary theoretical perspective argues that the emergence and long-term survival of extended gas disks may be primarily governed by a mass threshold that depends only weakly on redshift. \citealt{dekel_mass_2020} predict that stable, long-lived rotating gas disks preferentially persist in haloes above \( M_{\rm h}\sim2\times10^{11}\,M_\odot\) (corresponding to \(M_\star\sim10^{9}\,M_\odot\)), whereas in lower mass haloes several mechanisms (e.g. feedback) prevent gas from settling into extended disks.

At the same time, substantial scatter in estimates of $V_\mathrm{rot}/\sigma_0$ persists at a fixed redshift, in part due to sample selection effects or differences in analysis \citep{lee_cristal_2025, lee_disk_2025}. The past decade has seen rapid progress in tracing galaxy dynamics with cold(er) ISM lines such as CO and [C\,{\sc ii}]158 $\mu$m (at \(z\gtrsim4\)) using the Atacama Large Millimeter/submillimeter Array (ALMA) and the Northern Extended Millimeter Array (NOEMA; e.g. \citealt{rizzo_alma-alpaka_2023, parlanti_alma_2023, tsukui_spatially_2023, lee_cristal_2025, jolly_noema3d_2026}). However, cross-phase comparisons have not yet converged on a cohesive picture. In the nearby Universe, direct multi-tracer measurements in the same galaxies often indicate that molecular gas is dynamically `colder' than ionised gas, with modestly higher rotation velocities and systematically lower dispersions in CO than in ionised gas \citep{cortese_alma_2017,levy_edge-califa_2018, klitsch_almacal_2018, peroux_multiphase_2019, szakacs_muse-alma_2021, girard_systematic_2021}. At $z > 1$, the limited set of multiphase datasets finds both comparable dispersions in CO and ionised gas within current uncertainties \citep{ubler_ionized_2018, genzel_evidence_2023} and higher ionised gas velocity dispersion \citep{lelli_cold_2023, liu_600_2023, liu_detailed_2024}. Nonetheless, compilations suggest that at a fixed gas mass, ionised-gas dispersions tend to be higher on average than molecular-gas dispersions, often by factors of around two. This is consistent with scenarios in which turbulence in the cold gas is tied more directly to how disks regulate gravitational instability, whereas the warm ionised medium is additionally broadened and made more stochastic by feedback \citep{klitsch_almacal_2018, peroux_multiphase_2019, szakacs_muse-alma_2021, girard_systematic_2021, rathjen_silcc_2023, wisnioski_evolution_2025}. At \(z>4\), [C\,{\sc ii}]158$\mu$m measurements span a wide range and show substantial scatter (e.g. \citealt{rizzo_dynamically_2020, rowland_rebels-25_2024, lee_cristal_2025}), to which is added the uncertainty of the gas phase traced by the [C\,{\sc ii}] line \citep{casavecchia_atomic_2025}. Taken together, the large intrinsic scatter, the susceptibility of ionised gas to non-gravitational broadening, and the still-uncertain phase-to-phase relationship motivate resolved and consistent kinematic analyses of multiple tracers within the same galaxies with careful control of resolution and modelling systematics.

In this framework, the intermediate-redshift Universe at \(z\sim0.5\) occupies a key but comparatively less explored regime for disk evolution. While large programs at \(z\sim0\) (e.g. SAMI, \citealt{bryant_sami_2015}; CALIFA, \citealt{sanchez_califa_2012}; MaNGA, \citealt{bundy_overview_2015}) and \(z\sim 1-3\) (e.g. KMOS3D, \citealt{wisnioski_kmos3d_2019}; KROSS, \citealt{stott_kmos_2016}; KGES, \citealt{tiley_kmos_2021}) have studied the statistical trends in galaxy kinematics, both theory and observations indicate that the largest change in characteristic gas velocity dispersions occurs between \(0<z<1\) \citep{krumholz_unified_2018, ubler_evolution_2019, mai_magpi_2024}. In particular, \(z\sim0.5\) is the epoch where disks are expected to transition from the thicker, more turbulent systems common at high redshift to the thin, dynamically cold disks seen today. Consequently, \(z\sim0.5\) provides a unique test bed for connecting observed kinematic settling (declining \(\sigma_0\), rising \(V/\sigma\)) to the interplay between accretion, disk stability, and feedback at the time when the dominant turbulence-regulation mechanism may be changing. However, self-consistent, multi-wavelength, and spatially resolved measurements across this interval are sparse. Some studies probe the ionised gas kinematics (e.g. \citealt{puech_3d_2007, contini_deep_2016, bouche_muse_2022}). In particular, the MAGPI and MAGIC surveys investigate the stellar and ionised gas kinematics of galaxies at $z \sim 0.3$ and $z\sim0.7$, respectively \citep{mai_magpi_2024, derkenne_magpi_2024, mercier_scaling_2022, mercier_stellar_2023, munoz_lopez_stellar_2024}, but multi-wavelength observations usually target individual objects (e.g. \citealt{klitsch_almacal_2018, peroux_multiphase_2019, szakacs_muse-alma_2021}).

In this context, the MUSE-ALMA Haloes project provides a unique opportunity. The program targets known circumgalactic medium (CGM) H\,{\sc i} absorption features in the spectra of background quasars and follows up with the Very Large Telescope (VLT) Multi Unit Spectroscopic Explorer (MUSE) and ALMA observations to identify the associated galaxies and thus provide a sample not selected through stellar mass or SFR. The program provides spatially resolved observations of the ionised gas, the cold molecular gas, and the stars, providing a near-complete census of the distribution and kinematics of the baryons in H\,{\sc i}-rich galaxies. The overall kinematics of the ionised gas and connections to the absorbers for the statistical sample of MUSE-detected galaxies were analysed in \citealt{weng-gasflows-2023}, where it was found that $\sim 60\%$ of the absorbers trace the extended gas in the galaxy disk. For two of the sources presented here, previous kinematic analysis has found that the H\,{\sc i}-absorbing gas can be related to a mixture of accretion, outflows, and extended disk halo \citep{peroux_nature_2017, norris_discovery_2021}.

In this paper, we exploit the new ALMA CO observations and combine them with MUSE data to characterise the kinematic nature of the sources and compare properties of the cold and warm gas phases (and stellar component in targets where possible) in a sub-sample of the MUSE-ALMA Haloes galaxies at $z \sim 0.5$. The paper is organized as follows: In Sec. \ref{sec:mah}, we present the survey and sample. In Sec. \ref{sec:kin} and \ref{sec:dysmalpy} we present the kinematics and kinematic modelling of the cold molecular and ionised gas component of the galaxies using {\scshape DysmalPy}, which we complement with a systematic search for stellar feedback in Sec. \ref{sec:outflows}. In Sec. \ref{sec:ppxf} we present the {\scshape pPXF} fitting and stellar kinematic retrieval. Finally, we discuss the implications of the results in Sec. \ref{sec:disc} and conclude in Sec. \ref{sec:conc}. Throughout the paper, we adopt the Planck 2018 cosmology \citep{planck-collaboration_2020}. 

\section{The MUSE-ALMA Haloes survey}\label{sec:mah}
\subsection{Survey design}
\begin{figure}
        \centering
        \includegraphics[width=\linewidth]{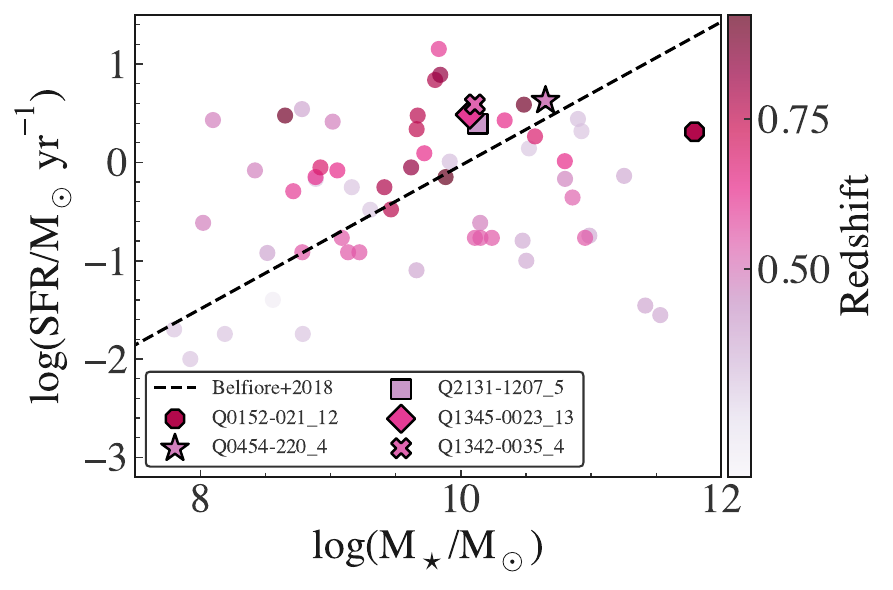}
        \caption{Distribution of the MUSE-ALMA Haloes galaxies identified in the MUSE observations \citep{weng_muse-alma_2023} according to their SFR as a function of stellar mass, colour-coded by redshift. The black dashed line shows the SF MS from \cite{belfiore_sdss_2018}, as shown in \cite{augustin_muse-alma_2024}. The galaxies outlined in black are five of the six galaxies analysed in this paper (the sixth one, at $z = 0.83$, Q1229-021\_6 has no reported stellar mass estimate).}
        \label{fig:sample}
    \end{figure}

The MUSE-ALMA Haloes project is a multi-wavelength, multi-facility project whose aim is to study the CGM and its associated galaxies \citep{peroux_muse-alma_2022, peroux_muse-alma_2026}. By identifying absorption features (in this case H\,{\sc i} absorption) in the spectrum of a background quasar, one can obtain a direct detection of CGM gas in the line of sight. Using this method, the MUSE-ALMA Haloes sample was built by targeting 32 known H\,{\sc i} absorbers in the spectra of 19 quasars with emission redshifts between $ 0.48 < z_\mathrm{quasar} < 2.19$ with the \textit{Hubble Space Telescope} (\textit{HST}) Space Telescope Imaging Spectrograph (STIS) or Faint Object Spectrograph (FOS), where the only selection criteria for the absorbers were 1) $z_{abs} < 0.85$, such that lines up to [O\,{\sc iii}]5007$\AA$ can be observed with VLT/MUSE, and 2) log[N(HI)/cm$^2$] > 18. Based on MUSE observations of these fields, \cite{weng_muse-alma_2023} identified a total of 79 galaxies with redshift within $\pm 500 \: \mathrm{km/s}$ of 27 of the absorbers. The galaxies were also targeted with \textit{HST}/Wide Field Camera 3 (WFC3)-UVIS and IR, providing valuable photometry used to estimate galaxy properties such as stellar mass and effective radius \citep{karki_muse-alma_2023,augustin_muse-alma_2024}. This builds a unique sample of galaxies and their CGM at $0.2 < z < 1.4$ ($\langle z \rangle = 0.6$). In Fig. \ref{fig:sample}, we show the galaxies' distribution along the SF MS, colour-coded by redshift. In addition to 21 sources with archival observations, 39 of the 79 galaxies were observed as part of an ALMA large program, of which 9 were detected above $3 \sigma$ \citep{bollo_muse-alma_2026, peroux_muse-alma_2026}.

    \begin{table*}[h!]
    \caption{Summary of the relevant physical properties and observational characteristics for the galaxies analysed in this paper.}                
    \label{table:sample1}    
    \centering                       
    \begin{tabular}{c | c c c c c c c c c c }      
    \hline\hline              
    Galaxy & $z$ & MUSE & ALMA & $\mathrm{log}(M_\star)$\tablefootmark{a} & $\mathrm{log}(M_\mathrm{H2})$\tablefootmark{b}  & SFR\tablefootmark{c}                    & $R_\mathrm{e}$\tablefootmark{d} & $b/a$\tablefootmark{d} & $n_\mathrm{disk}$\tablefootmark{d}   & Class \\         
           &     & PSF  & beam &                         &                                &                        &                &       &                     &       \\
           &     & $''$ & $''$ &  $M_\odot$              & $M_\odot$                      & $M_\odot$/yr           & kpc            &       &                     &       \\
    \hline                      
       Q0454-220\_4   & 0.484 & 0.57 & $0.59 \times 0.73$ & $10.7 ^{+0.05} _{-0.07}$ & $9.7 \pm 0.1$  & $4.3 \pm 1.3$ & $7.55 \pm 0.17$ & $0.86 \pm 0.01$ & 0.83 & Disk\\    
       Q0152-2001\_12 & 0.780 & 0.72 & $0.81 \times 0.71$ & $11.8 ^{+0.2} _{-0.4}$   & $9.7 \pm 0.1$  & $2.1 \pm 0.7$ & $5.85 \pm 0.17$ & $0.80 \pm 0.01$ & 1.56 & Disk\\    
       Q2131-1207\_5  & 0.430 & 0.72 & $0.65 \times 0.82$ & $10.1 ^{+0.2} _{-0.1}$   & $9.5 \pm 0.1$  & $2.5 \pm 0.8$ & $5.86 \pm 0.07$ & $0.75 \pm 0.01$ & 0.87 & Disk\\
       Q1345-0023\_13 & 0.607 & 0.64 & $0.69 \times 0.88$ & $10.1 ^{+0.6} _{-0.6}$   & $9.8 \pm 0.1$  & $3.1 \pm 1.2$ & $5.67 \pm 0.42$ & $0.58 \pm 0.02$ & 1.67 & Disk\\
       Q1342-0035\_4  & 0.538 & 1.30 & $0.60 \times 0.78$ & $10.1 ^{+0.1} _{-0.1}$   & $9.9 \pm 0.1$  & $3.9 \pm 1.0$ & $6.29 \pm 0.19$ & $0.50 \pm 0.01$ & 0.84 & -\\        
       Q1229-021\_6   & 0.832 & 0.80 & $0.70 \times 0.90$ & N/A                      & $10.3 \pm 0.1$ & $6.2 \pm 2.1$ & $2.31 \pm 0.06$ & $0.58 \pm 0.01$ & 1.07 & -\\   
    \hline                                 
    \end{tabular}
    \tablefoot{\\\tablefoottext{a}{From \cite{augustin_muse-alma_2024} and based on an SED fitting of the \textit{HST} photometry.}\tablefoottext{b}{From \cite{bollo_muse-alma_2026}.}\tablefoottext{c}{From \citealt{weng_muse-alma_2023} using H$\alpha$ or [O\,{\sc ii}].}\tablefoottext{d}{From \cite{karki_muse-alma_2023} using {\scshape GALFIT}.}}
    \end{table*}

\subsection{Observations}
The galaxies were observed in broadband imaging from \textit{HST}/WFC3-UVIS and -IR, as well as the WFPC2 (Programs: 15939, 5098, 5143, 5351, 6557, 7329, 7451, 9173, and 14 594; PIs: P\'eroux, Burbidge, Macchetto, Bergeron, Steidel, Malkan, Smette, Bechtold, Bielby). The detailed description of the HST observations and the subsequent data reduction and analysis can be found in \cite{peroux_muse-alma_2022} and \cite{karki_muse-alma_2023}. 

Using the reduced, point-spread function (PSF)-subtracted \textit{HST} images, \citealt{karki_muse-alma_2023} extracted the morphological information of the galaxies using {\scshape GALFIT} \citep{peng_detailed_2010}, and \citealt{augustin_muse-alma_2024} obtained estimates of the stellar masses through SED fitting using {\scshape LePhare} \citep{arnouts_measuring_1999, ilbert_accurate_2006}. The stellar masses of the galaxies in this analysis range from $ 9.8  < \mathrm{log}(M_*/M_\odot)<11.8$ \citep{augustin_muse-alma_2024}, and the SFRs from $2-14 \: M_\odot/\mathrm{yr}$ (\citealt{weng_muse-alma_2023}; see Fig. \ref{fig:sample} and Table \ref{table:sample1}).

The VLT/MUSE data of the galaxies studied in this paper were observed under the programs ESO 95.A-0303, 100.A-0753, 101.A-0660, 102.A-0370 (PI. P\'eroux), as well as ESO 298.A-0517 (PI. Klitsch). The observations have integration times between 47 mins and 1h20mins, with an angular resolution full-width-at-half-maximum (FWHM) of $0.54'' - 1.3''$. The MUSE observations cover the lines from [O\,{\sc ii}]3726,3729$\AA$ to [O\,{\sc iii}]5007$\AA$. All cubes have $0.2''$ pixelscale, with a spectral sampling of 1.25$\AA$, and spectral resolution FWHM $92 - 111$ km/s. A detailed description of the observations is presented in \citealt{peroux_muse-alma_2022}. 

Finally, 39 of the 79 MUSE-identified galaxies were followed up through an ALMA Large Program (LP; PI: P\'eroux, ID: 2023.1.00127.L), targeting CO transitions lines (CO(2-1), CO(3-2) and CO(4-3)) using Band 4 and Band 6. The data reduction was performed using the Common Astronomy Software Applications (CASA, version 6.5.6), creating both a `naturally' weighted cube with velocity resolution $\sim 60$ km/s and a `Briggs' weighted cube (robust = 0.5) with velocity resolution $\sim 15$ km/s, and the observations have angular resolution from $0.57'' - 1.0''$. Self-calibration was used when possible to optimize the noise characteristics of the cubes. A detailed summary of the observations and data reduction can be found in \citealt{peroux_muse-alma_2026}, and the analysis of those observations and their gas mass estimates are presented in \citealt{bollo_muse-alma_2026}. For the analysis presented here, we use the $15$ km/s resolution cubes to better sample the line profile and spatially resolved kinematics, which have average root-mean-square sensitivity $\sim 230\:\mu$Jy/beam. The beam sizes and gas masses are reported in Table \ref{table:sample1}. All the galaxies analysed here were detected in the CO(2-1) emission line. Prior to the analysis, all observations were carefully aligned using the position of stars from the \textit{Gaia} Data Release 3 catalogues \citep{collaboration-gaia-2016, collaboration-gaia-2023} to correct the astrometry of the \textit{HST} and MUSE observations.

\section{Kinematic sample}\label{sec:kin}
Based on the available information of the galaxies, we determined which are disks by adapting the disk criteria developed in previous disk kinematic studies \citep{forster_schreiber_sinszc-sinf_2018} to the multi-tracer analysis presented here. The criteria used are the following: 
\begin{enumerate}
    \item The 2D velocity fields of the ionised and cold molecular gas both have a smooth monotonic velocity gradient.
    \item Both tracers have a value of $V_\mathrm{obs}/2\sigma_\mathrm{int} > 0.4$ (see definition below, \citealt{lee_cristal_2025}).
    \item The photometric position angle (PA) obtained from the \textit{HST} imaging \citep{karki_muse-alma_2023}, ionised gas kinematic PA, and cold molecular gas PA are all aligned within $30^\circ$.
    \item The photometric centre aligned with the kinematic centres of both tracers are within spatial uncertainties.
    \item The kinematic centres of both tracers are aligned with the peak of their respective velocity dispersions within spatial uncertainties.
\end{enumerate}
The kinematic PA of the galaxies was determined as the counterclockwise angle from the north from which a line passing through the centre connects both the maximum and minimum rotation velocity. The kinematic centre was defined to be the midpoint of the velocity extrema. We determined $V_\mathrm{obs}/2\sigma_\mathrm{int}$ for the purpose of this criterion as the ratio of the maximum velocity difference along the kinematic major-axis to twice the standard deviation obtained by fitting a single Gaussian profile to the integrated spectra of the galaxy (see Sec. \ref{sec:outflows} and Appendix \ref{ap:outflow} for the integrated spectra). We note that this $\sigma_\mathrm{int}$ is a galaxy integrated estimate of the velocity dispersion and not the true intrinsic local turbulence in the disk. To be classified as a disk, we required that galaxies fulfil at least three of the above criteria.

\begin{figure*}
        \centering
        \includegraphics[width=\textwidth]{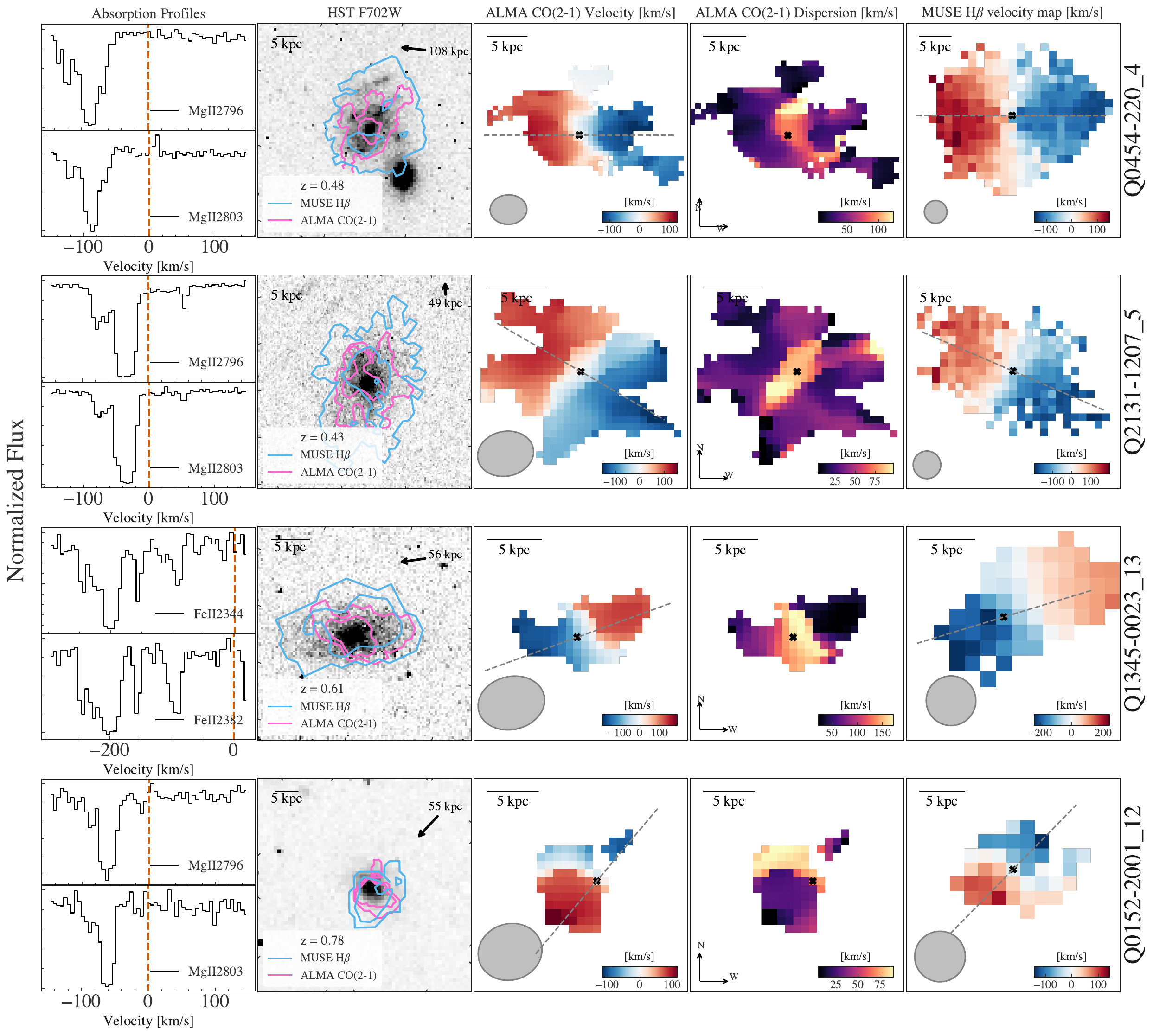}
        \caption{Presentation of the four galaxies identified as rotating disks. From left to right: Absorption in the normalised quasar spectra associated with the galaxy, with the systemic velocity of the galaxy marked by the red dashed line; HST/WFC3 F702W filter image, with the  ALMA CO(2-1) and MUSE H$\beta$ 3, 5, and 10 $\sigma$ contours; and ALMA/CO(2-1) velocity map, ALMA/CO(2-1) dispersion map, and MUSE/H$\beta$ velocity map. The black arrows indicate the direction of the background quasar used for the absorption, and they are accompanied by a number showing the associated impact parameter. The grey shaded ellipses indicate the sizes of the synthetic beam and PSF for the ALMA and MUSE observations, respectively. On the kinematic maps, the black cross and grey dashed lines indicate the kinematic centres and PA for each galaxy. For Q0454-220\_4, the companion galaxy, visible in the \textit{HST} image, is a foreground galaxy at $z = 0.38$.}
        \label{fig:disks}
    \end{figure*}

\begin{figure*}
        \centering
        \includegraphics[width=\textwidth]{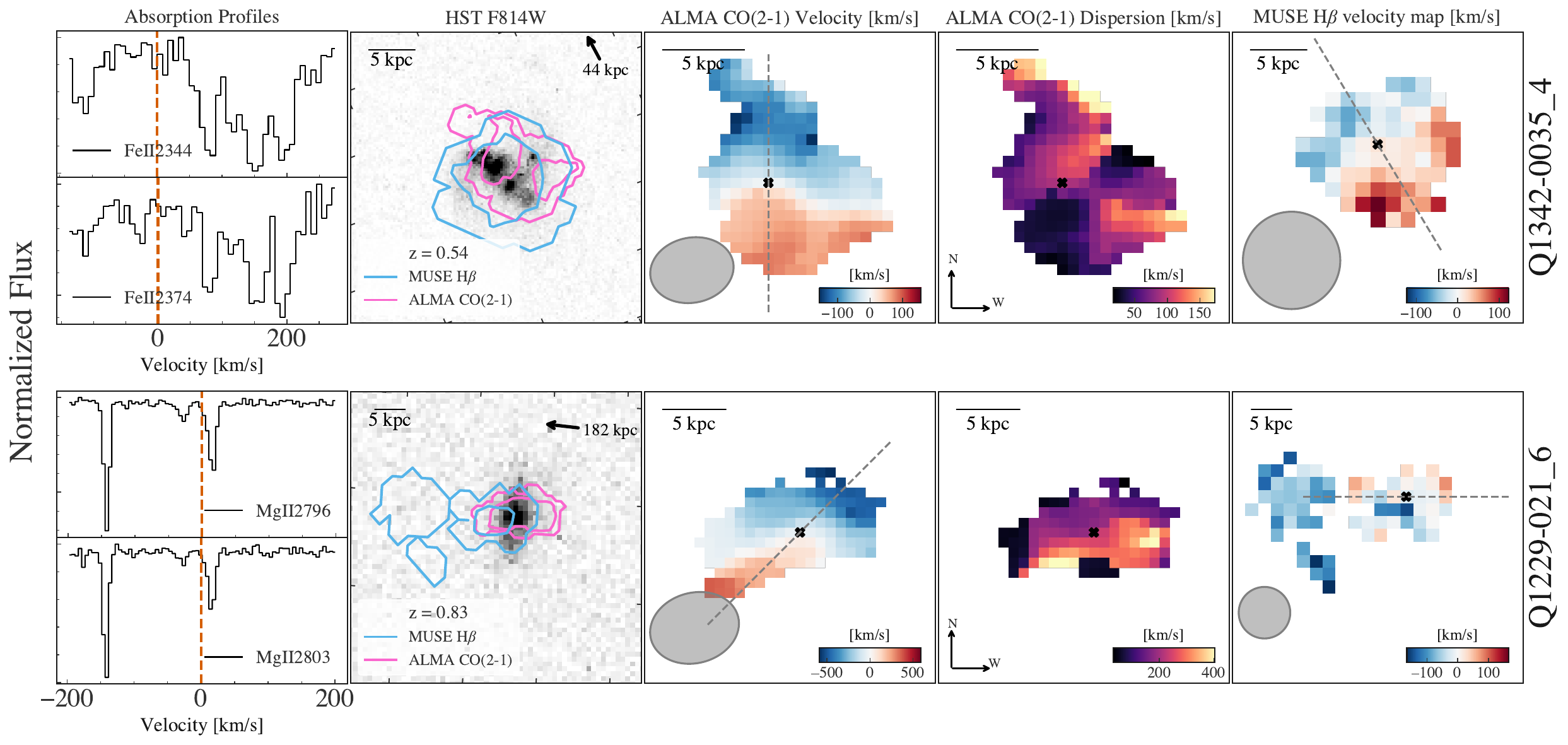}
        \caption{Same as Fig. \ref{fig:disks} but this time showcasing the two galaxies identified as non-disks.}
        \label{fig:mergers}
    \end{figure*}

Of the nine MUSE-ALMA Haloes galaxies detected in CO in the LP, three are not spatially resolved, which does not allow for kinematic analysis of the cold molecular gas. We extracted the galaxy kinematics from the remaining six, as the systems are detected with signal-to-noise ratio (S/N) per beam $> 3$ over at least two beams. We show in the second panel of Figures \ref{fig:disks} and \ref{fig:mergers} the \textit{HST}/F702W or F814W imaging of all six galaxies, with the contours from the cold molecular gas and ionised gas emission overlaid on top. The leftmost panel shows some of the main absorption features of the absorber at the galaxy's redshift. 

As a first investigation of the galaxies' kinematics, we used the software {\scshape linefit}, first developed for SINFONI data analysis (\citealt{davies_how_2011}; see Appendix \ref{ap:linefit} for a description of the code). We applied {\scshape linefit} on the CO observations and show the resulting velocity and velocity dispersion maps in Fig. \ref{fig:disks} and \ref{fig:mergers} (third and fourth panels from the left). Then, we exploited the existing observations of rest-frame optical emission lines (H$\beta$, [O\,{\sc ii}]3726,3729$\AA$ and [O\,{\sc iii}]5007$\AA$) from MUSE to extract the kinematic maps for the ionised gas component of the galaxies. We used H$\beta$ for five of the galaxies and [O\,{\sc iii}]5007$\AA$ for Q1229-021\_6, based on which is best detected. {\scshape linefit} also de-convolves the observed line and the line spread function (LSF) of the instrument using an input template constructed using the {\scshape MPDAF} package \citep{bacon_mpdaf_2016, piqueras_mpdaf_2019}. The resulting velocity maps are shown in the rightmost panels of Figs. \ref{fig:disks} \& \ref{fig:mergers}. We performed a S/N cut on the extracted velocity and velocity dispersion maps (where the threshold value varies from galaxy to galaxy based on visual inspection), as well as a manual removal of poorly fitted spaxels (e.g. pixels with largely deviant velocity and/or velocity dispersion compared to the neighbouring pixels), in order to have clean kinematic maps of the data.

Four of the targets (Q0454-220\_4, Q2131-1207\_5, Q1345-0023\_13 and Q0152-2001\_12) fulfil criteria 1 to 5 in at least one gas tracer, and generally in both CO and ionised gas. For Q0152-2001\_12, the low S/N combined with asymmetric CO flux distribution likely affects the determination of the centre; we fixed it at the better determined centre from the MUSE ionised gas data. We classified these four systems as rotating disks, suitable for kinematic modelling (see Fig. \ref{fig:disks}). Qualitatively, the kinematics of both the ionised and cold molecular gas in our disk galaxies are in good agreement, and a precise comparison of their kinematics is reported in Sec. \ref{sec:dysmalpy}.

For Q1229-021\_6 and Q1342-0035\_4, the velocity dispersion maps in CO(2-1) both show extremely high values of velocity dispersion ($\sigma_\mathrm{CO} > 300$ km/s and $\sigma_\mathrm{CO} > 100$ km/s, respectively). For both galaxies, the $\sigma_\mathrm{H\beta}$ and $\sigma_\mathrm{CO}$ are both not peaked in their respective kinematic centres, and the PA of the ionised gas kinematics is misaligned with the CO, and both are misaligned with the photometric PA. In addition, the \textit{HST} imaging of Q1342-0035\_4 shows a disturbed morphology. Similarly, the ionised gas component of Q1229-021\_6 shows no clear monotonic velocity gradient, and some extended emission away from the galaxy. For both of these systems, we determined that they are not disks and thus cannot be modelled as such. The galaxies are shown in Fig. \ref{fig:mergers}.

\section{Kinematic modelling with {\scshape DysmalPy}}\label{sec:dysmalpy}
To model the kinematics of the MUSE-ALMA Haloes galaxies, we used the {\scshape{DysmalPy}}\footnote{https://www.mpe.mpg.de/resources/IR/DYSMALPY} software, developed first as an IDL package, and then adapted to {\scshape Python} \citep{davies_nuclear_2004, davies_how_2011, cresci_sins_2009, wuyts_kmos3d_2016, lang_falling_2017, price_rotation_2021, lee_disk_2025}. The software is a parametric forward modelling code that can receive as input a rotation and dispersion profile (1D), a velocity and velocity dispersion map (2D), or a full data cube (3D). In addition, {\scshape DysmalPy} takes the beam or PSF information from the observations (in this case ALMA and MUSE, listed in Table \ref{table:sample1}) as well as the spectral resolution of the instruments to properly account for instrumental effects in the modelling. The software returns a best-fit model cube of the galaxy, built from the best-fit mass profile and geometry, convolved with the same instrumental properties as the input data. A more complete description of the code and its most recent implementation can be found in \cite{price_rotation_2021} and \cite{lee_disk_2025}.

To fit the identified rotating disks, we modelled the galaxies based on their 1D kinematic profiles (velocity, velocity dispersion) along the kinematic major axis. We are here interested in the global kinematic properties of the galaxies: $V_\mathrm{rot}$ and $\sigma_0$. For this purpose, we performed the fitting using the 1D profile rather than the 2D maps, as the kinematic information stored along the major axis of the galaxy has the most weight and is less prone to the influence of possible radial motions \citep[][see also Sec. 5 in \citealt{price_rotation_2021}]{van_der_kruit_kinematics_1978, lelli_wisdom_2022, lee_disk_2025}. Furthermore, the 2D maps presented in Fig \ref{fig:disks} are used to determine the kinematic centre and PA to extract the profile, leveraging the spatial resolution of the data. Having kinematics data from both CO and H$\beta$, we exploited the {\scshape{DysmalPy}} capability to simultaneously fit the profiles of both tracers from independent datasets, and the model assumptions used in this analysis are described in Sec. \ref{sec:model}. The ionised gas kinematics of the full MUSE-detected sample of the MUSE-ALMA Haloes galaxies were also modelled in \cite{weng-gasflows-2023} using {\scshape GALPAK$^\mathrm{3D}$} \citep{bouche_galpak_2015, bouche_galpak3d_2015}. We investigated in Sec. \ref{sec:cosmo} how the results compare using the two software.

\subsection{Slit extraction}
We first extracted the 1D velocity and dispersion profiles by drawing a pseudo-slit on the galaxies oriented in the direction of the kinematic major axis, following the observed PA of the MUSE observations. The centre of the slit was defined based on the midpoint between the steepest velocity gradient and the peak of the dispersion. The width of the synthetic slit was equal to one beam FWHM, and we extracted the profiles every one-fourth beam FWHM within apertures of half the beam size along the slight length. We fitted a Gaussian model to the resulting lines to obtain the central velocity and velocity dispersion of the gas in each slitlet, thus constructing the velocity and dispersion profile (see left panels of Fig. \ref{fig:0454_dysmal} but also Fig. \ref{fig:2131_dysmal}, \ref{fig:0152_dysmal}, and \ref{fig:1345_dysmal} in Appendix \ref{ap:dysmal}). 
In the MUSE cubes, we extracted the 1D profile using the H$\beta$ line as it is the brightest single line available. While the [O\,{\sc ii}] doublet is also covered, the lines are not fully de-blended due to the spectral resolution of the cubes. This would add another layer of uncertainty to the kinematic profile, especially with regards to the velocity dispersion.

\subsection{Model}\label{sec:model}
The galaxies are modelled as a thick turbulent disk plus bulge, embedded in a dark matter profile. The bulge is spheroidal, with $R_\mathrm{e,bulge} = 1 \ \mathrm{kpc}$ (following \citealt{price_rotation_2021} and \citealt{nestor_shachar_rc100_2023}) and $n_\mathrm{bulge} = 4$; whereas the disk is oblate with effective radius $R_\mathrm{e,disk}$, Sersic index $n_\mathrm{disk}$ and fixed intrinsic disk scale height to scale length ratio $q_0 = 0.15$, which is motivated by the typical values of $R_\mathrm{disk}/h_\mathrm{disk}$ from kinematics in samples at $z\sim0.6-1$ (e.g. \citealt{wisnioski_kmos3d_2015, price_rotation_2021}). We also try fitting with $q_0 = 0.25$, following the average for galaxies with this mass at this redshift \citep{2014ApJ...792L...6V}, and find that the best-fit results agree within uncertainties. The baryonic components of the disk and bulge are modelled with de-projected Sersic indices using the \cite{noordermeer_disc_2008} prescriptions to account for the geometry of a disk with finite thickness, and are characterised by the total baryonic mass $M_\mathrm{tot}$ and the bulge-to-total B/T. Following \cite{burkert_turbulence_2010}, we do not include adiabatic contraction and add a pressure support correction from a constant isotropic velocity dispersion to $V_\mathrm{circ}$. The dark matter profile is modelled according to a Navarro-Frenk-White (NFW) profile \citep{navarro_structure_1996}. While the software takes into account different dark matter profiles, the resolution of the data does not allow one model to be constrained over the other. Changes in the profile would result in changes in the dark matter fraction within the effective radius, for example lower DM fractions for a profile with a flat core \citep[e.g.][]{burkert_structure_1995, genzel_rotation_2020}. However, we do not expect it to change the best-fit results beyond the uncertainties \citep{genzel_strongly_2017}. The halo concentration was fixed to $c = 10.9 \times (1+z)^{-0.83}$, following the approach in \cite{genzel_rotation_2020}. 

In addition, to perform the full kinematic fitting, we put Gaussian priors on some of the physical parameters based on previously published measurements: the total baryonic mass (which we compute using the stellar masses of the galaxies and the molecular gas mass; \citealt{augustin_muse-alma_2024, bollo_muse-alma_2026}), the effective radius $R_\mathrm{e}$, ellipticity and Sersic index \citep{karki_muse-alma_2023}. We derived the inclination from the ellipticity using $\mathrm{cos} \ i = \sqrt{[(b/a)^2 - q_0^2] / [1-q_0^2] }$ \citep{law_kiloparsec-scale_2009, wisnioski_kmos3d_2019, ubler_evolution_2019, lee_cristal_2025}. All the parameters and their associated uncertainties are reported in Table \ref{table:sample1}. As a prior, we adopted the derived values, and as the standard deviation of the prior, we took the uncertainty on the measurement, except for the Sersic index of the disk, which we fixed. By estimating the total baryonic mass from only the stellar and molecular gas mass, we are under-estimating the true total baryonic mass by not including the ionised and atomic gas components. Using the H$\beta$ flux in our galaxies and the theoretical flux ratio $H\alpha/H\beta \approx2.86$, we made an order of magnitude estimate for the ionised gas mass in our galaxies of $\sim 10^5-10^6 \: \rm M_\odot$ \citep{genzel_sins_2011}. This represents $\lesssim 1\%$ of the total mass and can therefore be neglected in the mass estimate. For the atomic gas mass, it is important to note that the code considers the mass enclosed as a function of radius, which our data probes only on scale $\sim 1-2 R_{\rm e}$. In SFGs, while H\,{\sc i} dominates at large radii, within $\sim R_{\rm e}$ where the column densities are higher, the cold molecular gas tends to dominate the cold gas budget \citep{bigiel_star_2008, lagos_cosmic_2011, eibensteiner_phangs-meerkat_2024}. Following \cite{obeirne_MAH_2026}, we estimated the atomic mass within 2$R_{\rm e}$ while assuming a constant H\,{\sc i} surface density of $\sim 10 \: \rm M_\odot/pc^2$. We find that the H\,{\sc i} mass within 2$R_{\rm e}$ is always inferior or similar to the H$_2$ mass and always within the uncertainties of the total baryonic masses given to {\scshape DysmalPy} \citep[see also][]{genzel_combined_2015, ubler_evolution_2017, wuyts_kmos3d_2016}.

We fitted the ionised and cold molecular gas simultaneously. Assuming that all the mass in the galaxy traces the same underlying gravitational potential, we created a model galaxy with a single geometry, mass distribution and profile, but with two distinct dispersion profiles for both tracers. The fitted parameters were therefore the total baryonic mass $M_\mathrm{tot}$, the disk effective radius $R_\mathrm{e,disk}$, the bulge-to-total ratio $B/T$, the dark matter fraction $f_{DM}$, the inclination and the velocity dispersions $\sigma_\mathrm{0,CO}$ and $\sigma_\mathrm{0,H\beta}$. The virial mass and scale heights (for both tracers) were tied to the dark matter fraction and velocity dispersions respectively.

\subsection{Fitting}
\begin{figure*}
    \centering
    \includegraphics[width=\textwidth]{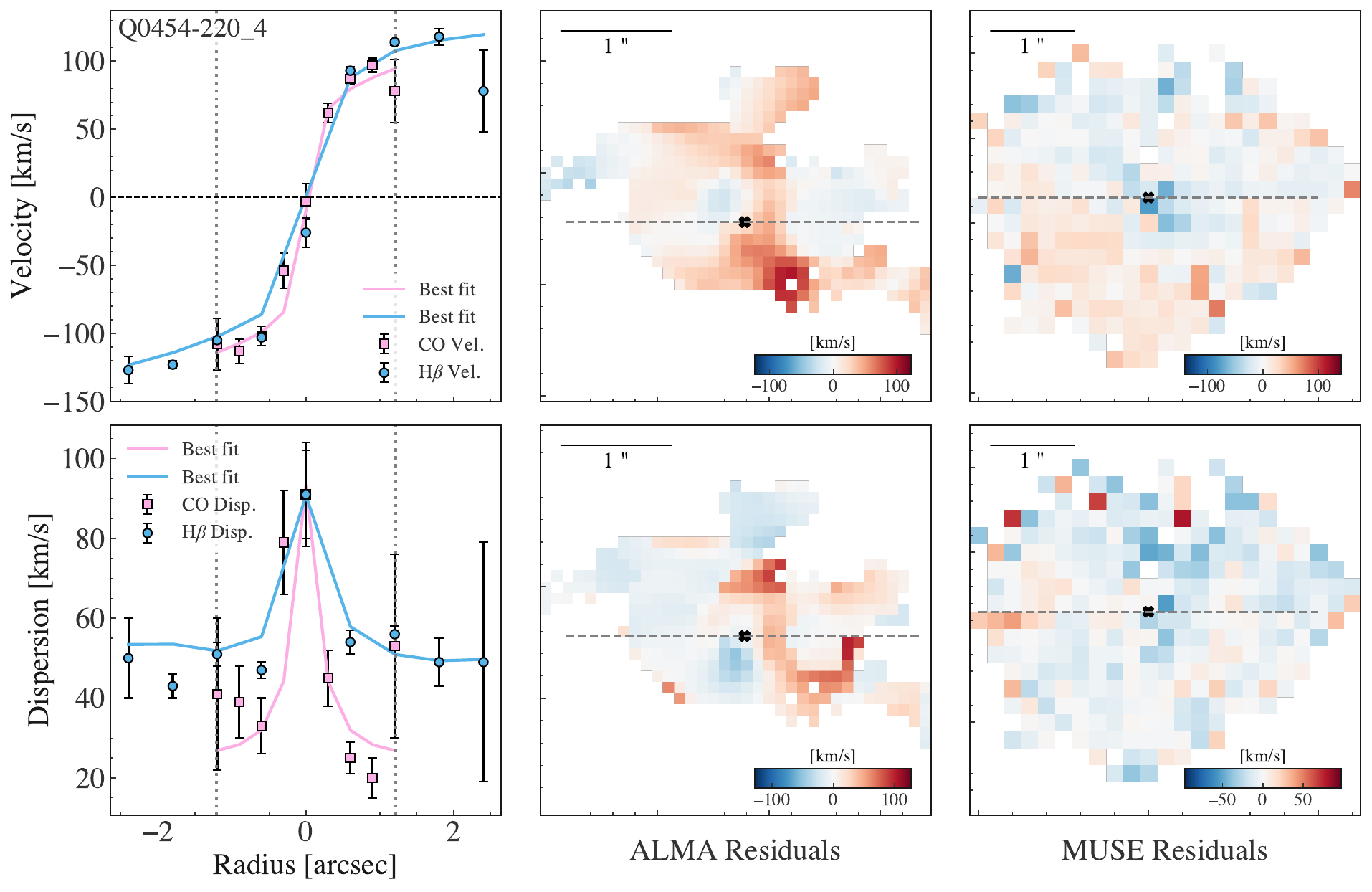}
    \caption{Fit results from the kinematic modelling with {\scshape DysmalPy} for Q0454-220\_4. The top row shows the fit and residuals of the velocity profile, whereas the bottom row shows the fit and residuals for the velocity dispersion profile. The left column shows the 1D profiles extracted from the ALMA (pink data points) and MUSE (blue data points) data as well as the corresponding best-fit profiles from {\scshape DysmalPy}. The grey dotted lines delimit the effective radius of the galaxy from \cite{karki_muse-alma_2023}. The middle and right columns show the reconstructed 2D velocity and velocity dispersion residual maps for the ALMA and MUSE data, respectively. The black cross indicates the kinematic centre of the maps, and the grey dashed line shows the kinematic PA.}
    \label{fig:0454_dysmal} 
\end{figure*}

The data were fit using a Markov chain Monte Carlo (MCMC) approach, where {\scshape DysmalPy} uses the package {\scshape emcee} \citep{foreman-mackey_emcee_2013} to perform the simulation. The data are run on chains with 300 walkers (independent agents simultaneously exploring the parameter space), a burn-in phase of 60 steps (discarded to ensure convergence to the target distribution), and 550 steps (the number of iterations each walker takes to sample the posterior distribution). The resulting fitted parameters are reported in Table \ref{table:dysmalpy}, the corner plots can be found in Appendix \ref{ap:dysmal}, and the fitted 1D profiles are displayed in Figures \ref{fig:0454_dysmal}, \ref{fig:2131_dysmal}, \ref{fig:0152_dysmal} and \ref{fig:1345_dysmal}. The uncertainty on the best-fit parameters is defined as the smallest interval including 68\% of the marginalized posteriors \citep{price_rotation_2021}.

    \begin{table*}[h!]
    \caption{Galaxy kinematic properties fitted with {\scshape DysmalPy} from simultaneously fitting CO(2-1) and H$\beta$.}              
    \label{table:dysmalpy}   
    \centering                       
    \begin{tabular}{c | c c c c c c c c c}      
    \hline\hline               
    Galaxy & $\mathrm{log}(M_\mathrm{tot}/M_\odot)$ & $R_\mathrm{e,disk}$ & $B/T$ & $f_{DM}$ & $i$ & $V_\mathrm{circ}(R_\mathrm{e})$ & $\sigma_\mathrm{0,CO}$ & $\sigma_\mathrm{0, H\beta}$ & $\chi^2_\mathrm{red}$ \\         
           &  dex  & kpc & & & deg & km/s & km/s & km/s  & \\
    \hline                      
       Q0454-220\_4    & $10.60 ^{+0.14} _{-0.17}$ & $7.57 ^{+0.15} _{-0.16}$  & $0.51 ^{+0.30} _{-0.15}$  & $0.64 ^{+0.06} _{-0.13}$ & $32 ^{+5} _{-2}$  & $245 ^{+17} _{-15}$   & $25.2 ^{+4.3} _{-3.2}$   & $23.6 ^{+1.3} _{-1.9}$  & 16.8 \\
       Q0152-2001\_12  & $10.91 ^{+0.05} _{-0.1}$ & $5.92 ^{+0.15} _{-0.19}$   & $[0.2]$                   & $0.27 ^{+0.20} _{-0.16}$ & $[30]$            & $226 ^{+47} _{-30}$   & $17.4 ^{+7.1} _{-6.0}$   & $32.2 ^{+11} _{-13}$    & 13.1 \\
       Q2131-1207\_5   & $10.73 ^{+0.16} _{-0.18}$ & $5.86 ^{+0.06} _{-0.07}$  & $0.20 ^{+0.16} _{-0.07}$  & $0.34 ^{+0.22} _{-0.13}$ & $46 ^{+5} _{-5}$  & $198 ^{+25} _{-17}$   & $13.3 ^{+3.8} _{-3.8}$   & $17.5 ^{+4.2} _{-2.7}$  & 4.2  \\
       Q1345-0023\_13  & $10.74 ^{+0.17} _{-0.22}$ & $5.87 ^{+0.34} _{-0.42}$  & $0.41 ^{+0.26} _{-0.30}$  & $0.44 ^{+0.13} _{-0.20}$ & $60 ^{+8} _{-7}$  & $234 ^{+32} _{-26}$   & $11.0 ^{+11} _{-7.9}$    & $28.1 ^{+8.7} _{-11}$   & 4.5  \\
    \hline                                 
    \end{tabular}
    \tablefoot{The reported uncertainties constitute the smallest interval including 68\% of the marginalized posterior. The values in brackets were fixed to aid the fitting.}
    \end{table*}

From the output best-fit cube, we recreated the 2D velocity and velocity dispersion map, following the same procedure as applied to the data, and we subtracted the model to the maps extracted from the data. The resulting velocity and velocity dispersion residuals for both the cold molecular and ionised gas are shown for all galaxies the middle and right panels of Figures \ref{fig:0454_dysmal}, \ref{fig:2131_dysmal}, \ref{fig:0152_dysmal} and \ref{fig:1345_dysmal}. In detail, the results of the fitting for each galaxy are the following:

\begin{itemize}
    \item Q0454-220\_4: The galaxy is well detected in both the MUSE and ALMA observations (Nres = 9 and 44 beams for ALMA and MUSE respectively). Our best-fit results for the galaxy's inclination and rotation velocity (after accounting for inclination) are in line with the estimates from \citealt{kacprzak_halo_2010}. The 2D velocity residuals show prominent residuals (up to $\sim 100$ km/s) along the minor axis of the galaxy (top middle panel, Fig. \ref{fig:0454_dysmal}), which reflect the twist in isovelocity contours already apparent in the velocity field (middle panel, top row; Fig. \ref{fig:disks}). 
    \item Q2131-1207\_5: The galaxy is well detected in both the MUSE and ALMA observations (Nres = 7 and 23 beams for ALMA and MUSE respectively). The 2D residuals show no note-worthy features (Fig. \ref{fig:2131_dysmal}). We note that the MUSE [O\,{\sc iii}]5007$\AA$ and CO(3-2) had already been fitted in \citealt{peroux_nature_2017} and \citealt{szakacs_muse-alma_2021} respectively, using {\scshape GALPAK$^\mathrm{3D}$} \citep{bouche_galpak_2015, bouche_galpak3d_2015}. Our resulting estimates for the mass and rotation velocity are in agreement, whereas there are some discrepancies with their best-fit inclination for the ionised gas ($60^\circ$) and effective radius ($7.9 \pm 0.1$ and $3.7 \pm 0.1$ kpc, respectively). However, $i$ is notoriously difficult to constrain in modest S/N and resolution data, and our inclination fit using both CO(2-1) and H$\beta$ agrees with that of \citealt{szakacs_muse-alma_2021} for the CO(3-2). The difference in $R_\mathrm{e}$ is most likely driven by the prior we impose on {\scshape DysmalPy} from \citealt{karki_muse-alma_2023}, which was derived after the aforementioned analysis.
    \item Q1345-0023\_13: This galaxy is best resolved in H$\beta$ (Nres = 8 beams) and more marginally resolved in CO (Nres = 2 beams). As a result, the CO best-fit parameters, and in particular the velocity dispersion, have the largest uncertainties. Fig. \ref{fig:1345_dysmal} shows the resulting best-fit and associated 2D residuals. 
    \item Q0152-2001\_12: The galaxy is marginally resolved above the beam size in both the ALMA and MUSE cubes -- Nres $\approx$ 2 beams for CO(2-1) and $\approx 4$ for H$\beta$ -- and the kinematic fitting is more poorly constrained. As a result, to help constrain the fit we fixed $B/T$ and the inclination and varied their values (in increments of 0.1 for $B/T$ and 10 degrees for $i$) to pinpoint which reduces the $\chi^2$ best. These values are reported in brackets in Table \ref{table:dysmalpy}. Fig. \ref{fig:0152_dysmal} shows the resulting best-fit and associated 2D residuals. We find values of $\mathrm{log}(M_\mathrm{tot}/M_\odot)$ 1 dex smaller than reported in \citealt{augustin_muse-alma_2024} for the stellar mass.
\end{itemize}

\section{Stellar feedback in the MUSE-ALMA Haloes galaxies}\label{sec:outflows}

\begin{figure*}
    \centering
    \begin{subfigure}[b]{0.5\textwidth}
        \centering
        \includegraphics[width=\textwidth]{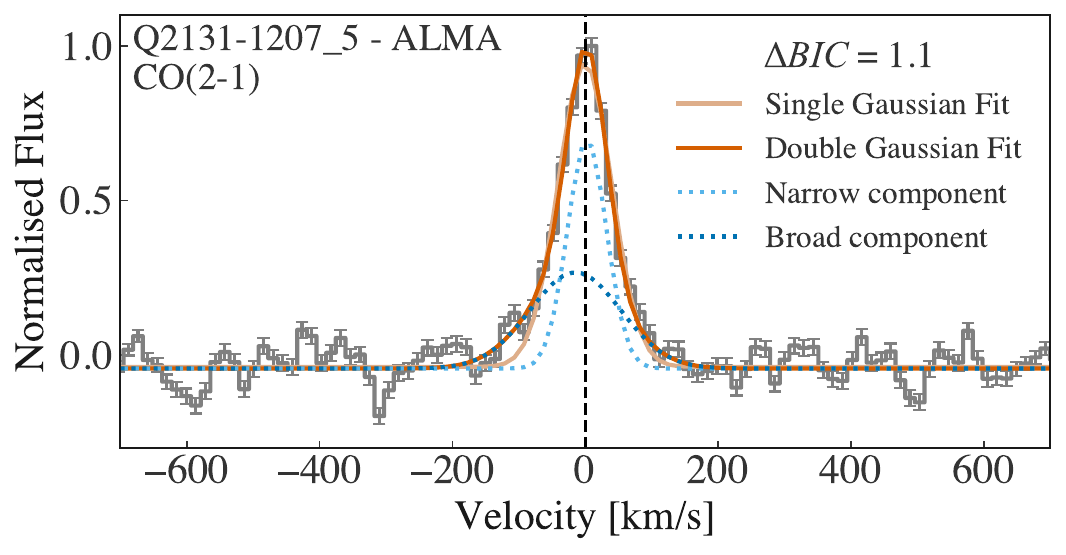}
    \end{subfigure}%
    ~ 
    \begin{subfigure}[b]{0.5\textwidth}
        \centering
        \includegraphics[width=\textwidth]{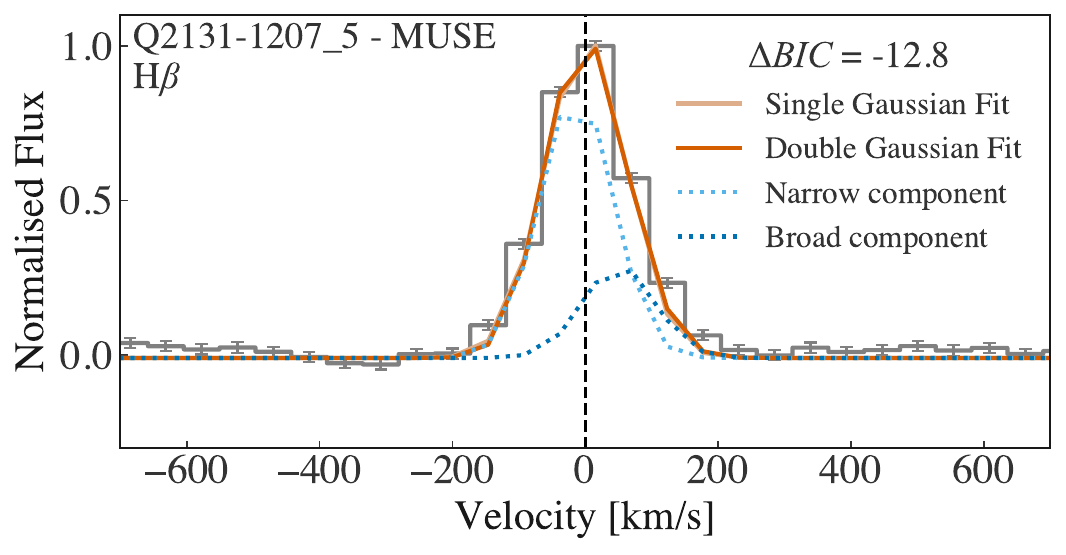}
    \end{subfigure}
    \caption{Integrated spectra of the CO(2-1) (left) and H$\beta$ (right) emission of Q2131-1207\_5. In both plots, the yellow line shows the best-fit single Gaussian, the red line shows the double Gaussian fit, and the light and dark blue dotted lines show the narrow and broad components of the double Gaussian component, respectively. The $\Delta\mathrm{BIC} = \mathrm{BIC_{sing}} - \mathrm{BIC_{doub}}$ indicates a mild favouring of the double Gaussian fit for the ALMA data (though this can be attributed to beam smearing; see Appendix \ref{ap:outflow}) and a strong favouring of the single Gaussian fit for the MUSE data. No convincing outflow signatures were found in the galaxy in either phase of the gas.}
    \label{fig:2131_outflows}
\end{figure*}

To fully characterise the kinematic properties of the four galaxies, we also performed a systematic search for a galaxy-wide signature of outflows in the form of a high-velocity component in the emission line profile of the ionised and cold molecular gas. To do so, we extracted the pixel-by-pixel spectra of the galaxies within the areas of the velocity maps displayed in Fig \ref{fig:disks} and integrated over the entire area. To account for the effects of the galaxies' rotation, we shifted all spectra by the velocity estimated at a given pixel in the velocity maps (e.g. \citealt{forster_schreiber_sinszc-sinf_2014, forster_schreiber_sinszc-sinf_2018, genzel_evidence_2014}). The resulting integrated spectra are shown in Fig. \ref{fig:2131_outflows} for Q2131-1207\_5, in both tracers (see also Appendix \ref{ap:outflow} for the other galaxies).

Once we had an integrated spectrum, we performed a single Gaussian fit and a double Gaussian fit to the data. If there is an outflow signature in the galaxy, the double Gaussian fit should reproduce the data better, with the second component tracing the broad fast moving gas, and the primary component tracing the narrow systemic emission. To quantify which model fits the data best, we computed the reduced $\chi^2$ of both fits, as well as the $\Delta\mathrm{BIC} = \mathrm{BIC_{sing}} - \mathrm{BIC_{doub}}$ of the two fits, where the BIC is the Bayesian information criterion of the single and double Gaussian fits respectively \citep{schwarz_estimating_1978}.\footnote{The BIC is defined as $\mathrm{BIC} = \mathrm{kln(n)} + \chi^2$, where k is the number of free parameters, and n is the number of fitted data points \citep{hogg_data_2010}.} In general, $\Delta\mathrm{BIC} > 0$ mildly favours the double Gaussian fit, and $\Delta\mathrm{BIC} > 10$ strongly favours the double Gaussian fit \citep{liddle_information_2007}. For all galaxies and tracers but two, the obtained $\Delta\mathrm{BIC} < 0$ indicate that there is no need for an additional broad component to fit the data, i.e. there are no galaxy-wide detected outflow signature above the noise level. For the integrated spectrum of the CO emission in Q0454-220\_4 and Q2131-1207\_5, we get $\Delta\mathrm{BIC} =7.6$ and $\Delta\mathrm{BIC} =1.1$ respectively (see left panels of Figs. \ref{fig:2131_outflows} and \ref{fig:0454_outflows}), indicating that the spectra are marginally better fit with a double Gaussian model. Indeed, there is excess flux on the side of the line with respect to a single Gaussian fit. However, beam smearing effects can introduce artificial wings in integrated spectra of galaxies, mimicking a secondary broad Gaussian component \citep{concas_being_2022}. Using our {\scshape DysmalPy} best-fit model to extract a purely rotating integrated spectrum in the same area which includes beam smearing, we see similar low-amplitude broad wings in the resulting spectrum (see Fig. \ref{fig:dysmalpy_outflows}). Performing a $\Delta\mathrm{BIC}$ analysis again, we see that the excess flux can be attributed to instrumental effects, rather than outflows (see Appendix \ref{ap:outflow} for a detailed description of the method). Therefore, we conclude that for the four MUSE-ALMA Haloes galaxies, there are no convincing outflow signatures, which is not surprising given the low SFR surface density of the galaxies ($\Sigma_\mathrm{SFR} < 0.02 \: M_\odot/\mathrm{kpc^2}$; \citealt{karki_muse-alma_2023}).

\section{Stellar kinematics from pPXF}\label{sec:ppxf}

\subsection{{\scshape pPXF} fitting}
\begin{figure*}
        \centering
        \includegraphics[width=\textwidth]{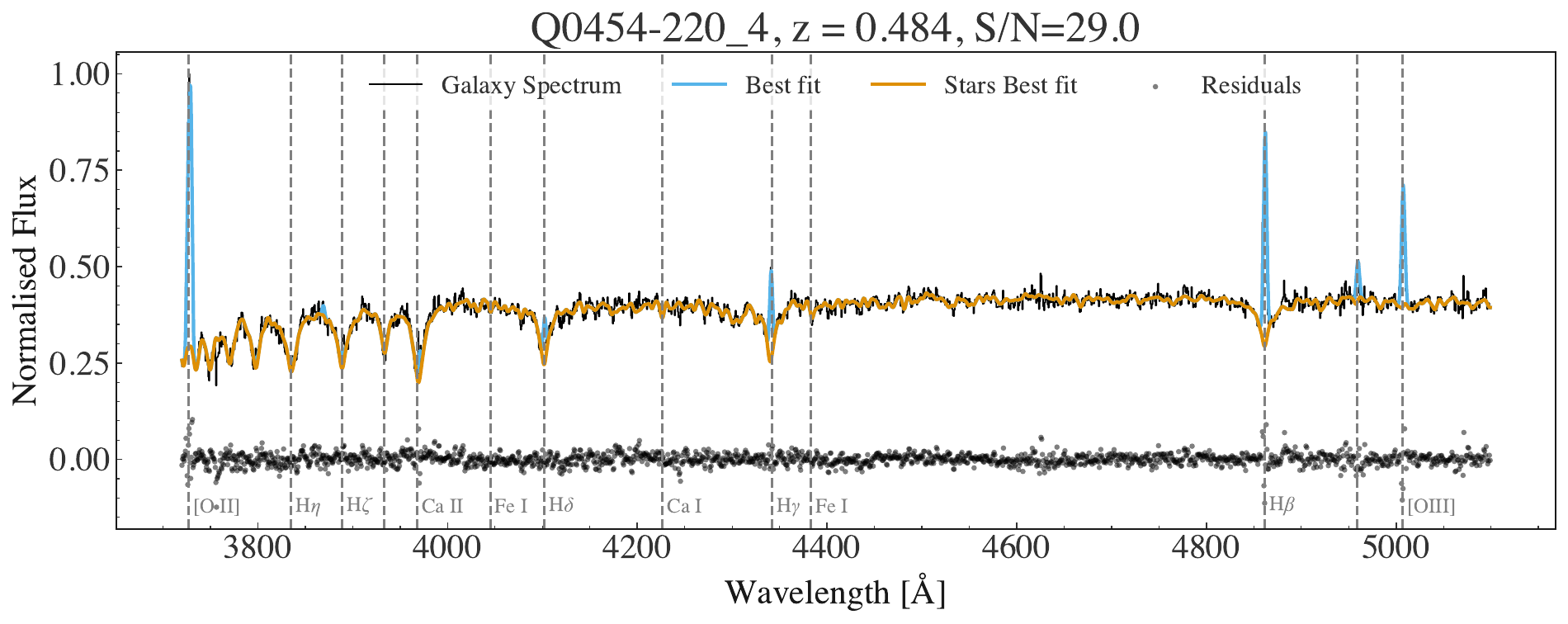}
        \caption{Best fit to the integrated spectrum of Q0454-220\_4 from {\scshape pPXF}. The black solid line represents the data, the orange solid line represents the fit to the stellar continuum, and the blue solid line represents the fit to the stellar continuum. The grey dots are the fit residuals. The dashed grey lines show the location of detected absorption and emission lines.}
        \label{fig:0454_ppxf}
    \end{figure*}

For the four disk galaxies presented in this analysis, the MUSE observations cover key stellar absorption features in the rest-frame optical, which can be used to determine stellar kinematics. The absorption lines are detected in all four integrated spectra of the galaxies (See Fig. \ref{fig:0454_ppxf}, but also Fig. \ref{fig:2131_ppxf}, \ref{fig:1345_ppxf} and \ref{fig:0152_ppxf}). We fitted the stellar continuum using the software {\scshape pPXF} \citep{cappellari_full_2023}, which fits the input galaxy spectrum with a linear combination of template spectra convolved with line-of-sight velocity distribution and additive and multiplicative polynomials, using $\chi^2$ minimization to find the best-fit. 

We performed two separate fits. We describe our assumptions and the spectral stellar library used in Appendix \ref{ap:ppxf}. At first, we fitted the integrated spectra of the four galaxies, extracted within the effective radii of the galaxies reported in \cite{karki_muse-alma_2023}, which agree within uncertainties with the values from {\scshape DysmalPy} (see Table \ref{table:dysmalpy}). The best example is shown in Fig. \ref{fig:0454_ppxf} for Q0454-220\_4, which reaches S/N $=29$ per spectral channel. The other three integrated spectral fits are shown in Appendix \ref{ap:ppxf}. For the four galaxies, we retrieved estimates for the global integrated velocity dispersion of the stars $\sigma_\star$, which are reported in Table \ref{table:ppxf}. 

\begin{table*}[h!]
    \caption{Global line-of-sight velocity dispersion of the stars in the four MUSE-ALMA Haloes galaxies from {\scshape pPXF}. }          
    \label{table:ppxf}    
    \centering                       
    \begin{tabular}{c | c | c c c c}      
    \hline\hline               
                       &                           & Q0454-220\_4         & Q0152-2001\_12    & Q2131-1207\_5          & Q1345-0023\_13 \\       
    \hline                     
    $\sigma_\star$     & km/s                      & $102 \pm 21$          &  $145 \pm 30$     & $98 \pm 14$            & $208 \pm 103$   \\
    $M_\mathrm{dyn}$   & $10^{10}M_\odot$          & $7.9(7) \pm 3.2$      &  $13 \pm 5$       & $5.6(3.7) \pm 1.7$     & $25\pm 24$           \\
    $f_{DM,\star}$     &                           & $0.65(0.59) \pm 0.20$ &  $0.63 \pm 0.30$  & $0.45(0.18) \pm 0.25$  & $0.85 \pm 0.40$          \\
    \hline                                  
    \end{tabular}
    \tablefoot{For Q0454-220\_4 and Q2131-1207\_5, the values in parenthesis are derived from spatially resolved stellar kinematic fitting.}
    \end{table*}

\begin{figure*}
        \centering
        \includegraphics[width=\textwidth]{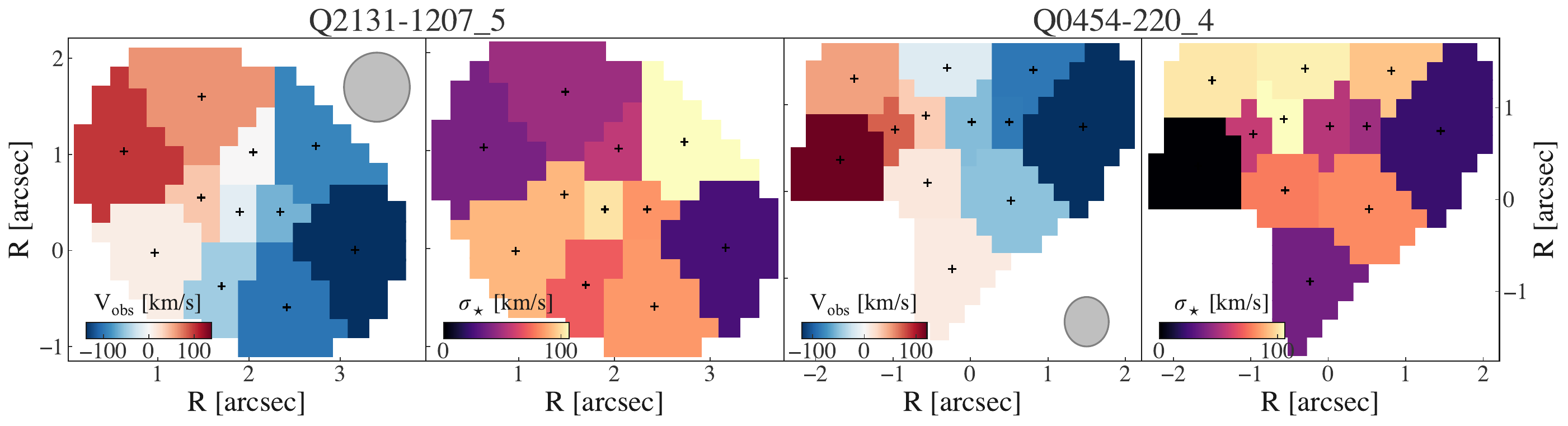}
        \caption{Spatially resolved stellar kinematics extracted in Q2131-1207\_5 (left) and Q0454-220\_4 (right). The PSF of the observations are shown as grey ellipses. The shape of the Q0454-220\_4 maps is due to the removal of a foreground galaxy (bottom-left corner) and a diffraction spike from a nearby star (north) that contaminates the stellar continuum.}
        \label{fig:2d_ppxf}
\end{figure*}

The second fit was done in a spatially resolved manner, only on Q2131-1207\_5 and Q0454-220\_5, as the extent and S/N of the MUSE detections in the other two do not allow for reliable 2D fitting. Even for the two best galaxies, individual spaxels do not have sufficient S/N to allow for pixel-by-pixel fitting of the stellar absorption, and thus we used the {\scshape PowerBin} method (which bins the data by maximizing the signal-to-noise ratio by adaptively determining bin sizes based on the data distribution; \citealt{cappellari_powerbin_2025}) to group spaxels in bins with S/N $>12$ following \cite{van_de_sande_fundamental_2014} and \cite{deugenio_fast-rotator_2024}. The resulting maps are shown in Fig. \ref{fig:2d_ppxf}.

\subsection{Dynamical masses and dark matter fractions from stellar kinematics}
Using the integrated line-of-sight velocity dispersions reported in Table \ref{table:ppxf}, we made a separate independent estimate for the dynamical mass within $R_\mathrm{e}$ of the galaxies. For consistency, we used $R_\mathrm{e}$ listed in Table \ref{table:sample1} from \cite{karki_muse-alma_2023}. We used the following relation:

\begin{equation}
   M_\mathrm{dyn} = K \frac{\sigma_\star^2 R_\mathrm{e}}{G},
\end{equation}

where the geometry- and mass-dependent scaling factor $K\sim 4.3$, using the virial coefficients from \cite{price_kinematics_2022} for an oblate de-projected Sersic model with axis ratio $q_0 < 0.2$, when computing the mass with the velocity dispersion. This is similar to the value of 3.4 derived by \cite{erb_stellar_2006} for a disk geometry. We used our estimates of $\sigma_\star$ estimated within $R_\mathrm{e}$, and the resulting dynamical mass are reported in Table \ref{table:ppxf}. 

For the two galaxies for which we were able to retrieve the spatially resolved stellar kinematics, we also estimated a lower limit on dynamical mass estimate using the measured $V_\mathrm{\star,obs}$, which we correct for inclination using the {\scshape GALFIT} estimates from \cite{karki_muse-alma_2023}, as a proxy for $V_\mathrm{circ}$ for a rotation-dominated system. The resulting estimates are reported in Table \ref{table:ppxf}, in parenthesis, for Q2131-1207\_5 and Q0454-220\_4. 

The values of $M_\mathrm{dyn}$ agree within uncertainties with the results from {\scshape DysmalPy}. They are also within 1$\sigma$ (for Q0454-220\_4 and Q1345-0023\_13) and 2$\sigma$ (for Q0152-2001\_12 and Q2131-1207\_5) with the estimates made by \cite{obeirne_MAH_2026} in a census of the condensed baryons (molecular gas, atomic gas and stars) of the MUSE-ALMA Haloes galaxies.

From $M_\mathrm{dyn}$, we also retrieved the dark matter fraction within the effective radii of the galaxies, where $f_{DM, \star} = 1 - \frac{M_\star + M_\mathrm{gas}}{M_\mathrm{dyn}}$ \citep{tacconi_evolution_2020, nestor_shachar_rc100_2023, danhaive_dark_2026}. For $M_\star + M_\mathrm{gas}$, we used the values of enclosed baryonic mass within $R_\mathrm{e}$ returned by {\scshape DysmalPy}, rather than the values of stellar mass obtained in \cite{augustin_muse-alma_2024} through SED fitting and of $H_2$ mass derived in \cite{bollo_muse-alma_2026} from CO observations. This was done for the following two reasons: 1) the values reported in the studies mentioned above are computed based on the total galaxy emission, and not within the effective radius, in contrast with our estimates for $M_\mathrm{dyn}$; 2) the stellar mass of Q0152-2001\_12 reported in \cite{augustin_muse-alma_2024} is extremely high (see Table \ref{table:sample1}) and would return a negative value of $f_{DM}$ which is unphysical. This might be the result of having only two \textit{HST} filters to constrain the SED fit around the 4000 $\AA$ break. We therefore used the best-fit results from {\scshape DysmalPy} for this galaxy, and for consistency also for the other three galaxies. The resulting fractions are reported in Table \ref{table:ppxf} and discussed in Sec. \ref{sec:fdm}.

\section{Discussion}\label{sec:disc}
\subsection{The scatter in kinematic measurements}\label{sec:scatter}
At any given redshift, the scatter in kinematic measurements of disk galaxies is quite substantial (\citealt{ubler_evolution_2019, lee_cristal_2025, wisnioski_evolution_2025}; see also Sec. \ref{sec:cosmo}) and can have multiple origins. 
Firstly, the inhomogeneity of methods used to estimate $\sigma_0$ and $V_\mathrm{rot}$. \cite{lee_disk_2025} and \cite{yttergren_kinematics_2025} both investigated the retrieval of the same physical parameters on sets of simulated galaxies using {\scshape DysmalPy}, {\scshape qubefit} \citep{neeleman_qubefit_2020}, {\scshape GALPAK$^\mathrm{3D}$} \citep{bouche_galpak_2015, bouche_galpak3d_2015} and {\scshape 3DBarolo} \citep{di_teodoro_3d_2015}. Both studies find that there are systematics inherent to the difference in methodology in the parameters retrieved. At higher redshifts, where large IFU surveys are not necessarily available, some codes reconstruct the kinematics based on slit spectroscopy or grism (wide-field slitless spectroscopy) observations, using photometry to inform the geometry of galaxies \citep{de_graaff_ionised_2024, danhaive_dawn_2025}. In addition, sample selection effects might play an important role in the average kinematic properties: \cite{wisnioski_evolution_2025} investigated the effect of various physical properties on kinematics and found, among others, that there is a strong correlation between the inclination corrected velocity and stellar mass of the galaxies, which translates into higher $V_\mathrm{rot}/\sigma_0$ for higher mass systems. Alternatively, the scatter could also be a consequence of variation in the velocity dispersion on short timescale, for example due to stellar feedback \citep{hung_what_2019, ubler_evolution_2019}. Another instrumental effect is the spectral resolution, which limits the minimum recoverable intrinsic velocity dispersion \citep{boselli_virgo_2021, lee_disk_2025}. Finally, the scatter can also be explained by the different tracers used. While there is evidence in the Local Universe of a systematic offset between the velocity dispersion of the ionised and cold molecular gas (e.g. \citealt{levy_edge-califa_2018, girard_systematic_2021}), at higher redshifts the picture is less clear, with some studies showing an offset (e.g. \citealt{ubler_evolution_2019, lelli_cold_2023}), and others not (e.g. \citealt{parlanti_alma_2023, girard_towards_2019}). In addition, at $z>4$, observing kinematics with CO transition lines becomes virtually impossible, and as a result studies have used the luminous [C\,{\sc ii}]158 $\mu$m line instead. However, this line traces more than one phase of the ISM, and the proportion of the dominant phase is still under study (for example, \citealt{casavecchia_atomic_2025} find that [C\,{\sc ii}] mostly traces neutral atomic gas). \cite{wisnioski_evolution_2025} investigated the effect of the different gas tracers on kinematic measurements, and found a systematic offset between the ionised gas and cold molecular gas kinematics at fixed mass, with [C\,{\sc ii}] measurements following either of the tracers.

\begin{figure}
    \centering
    \includegraphics[width=\linewidth]{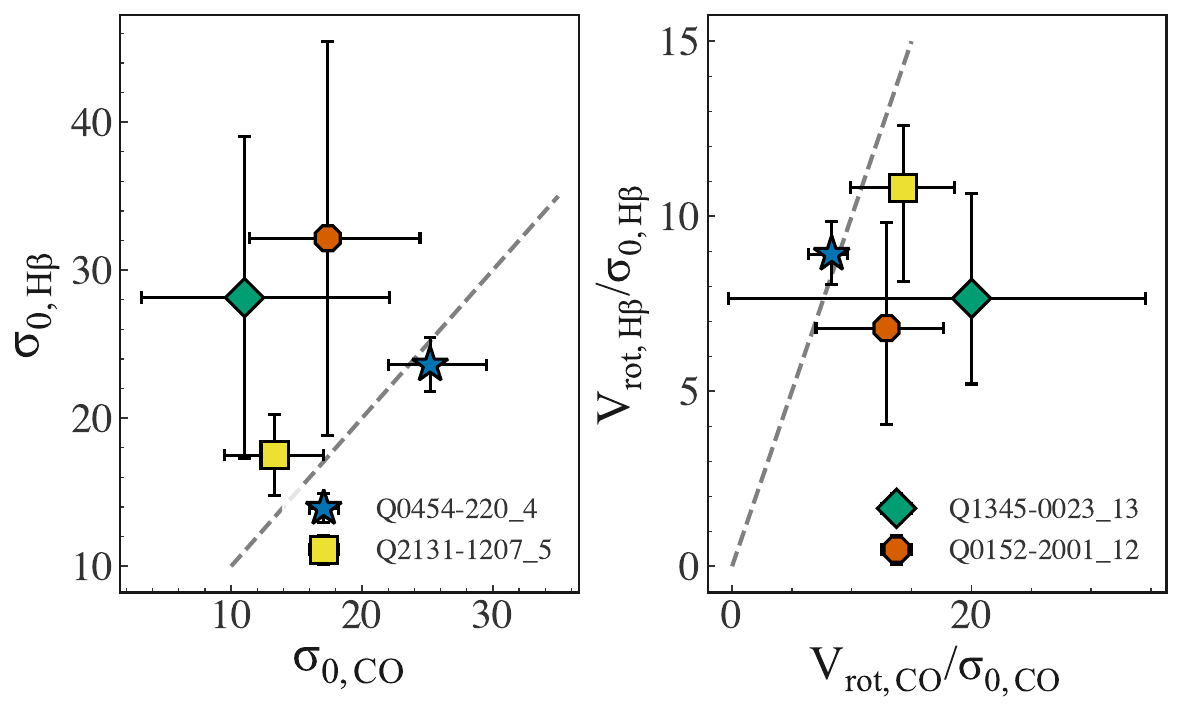}
    \caption{Comparison of the multi-wavelength kinematics obtained for each galaxy. On the x-axis are the $\mathrm{\sigma_0}$ and $V_\mathrm{rot}/\sigma_0$ obtained from CO, whereas on the y-axis are those extracted from H$\beta$. The dashed grey line represents the 1:1 ratio. The kinematics of Q0454-220\_4 show no difference linked to gas phase, whereas the ionised gas is marginally less rotation dominated than the cold molecular gas in Q2131-1207\_5, Q0152-2001\_12, and Q1345-0023\_13.}
    \label{fig:co_vs_hb}
\end{figure}

Through the MUSE-ALMA Haloes program, we have the opportunity to study a well characterised sample of galaxies at $z \sim 0.5$, and to perform a consistent analysis of both the cold molecular and ionised gas kinematics. In Fig. \ref{fig:co_vs_hb}, we explore how the retrieved $\mathrm{\sigma_0}$ and $V_\mathrm{rot}/\sigma_0$ compare with each other for the same objects. For Q0454-220\_4, the velocity dispersion and $V_\mathrm{rot}/\sigma_0$ of the cold molecular gas agrees remarkably well with that of the ionised gas. On the other hand, Q0152-2001\_12, Q2131-1207\_5 and Q1345-0023\_13 have ionised gas velocity dispersions $1.8 \times$ higher than the cold molecular gas, although the discrepancy remains within $2\sigma$. As a consequence, the ionised gas kinematics appear less rotation dominated than the cold molecular gas. 

\subsection{The MUSE-ALMA Haloes sample gas kinematics in a cosmological context}\label{sec:cosmo}
   \begin{figure*}
        \centering
        \includegraphics[width=\textwidth]{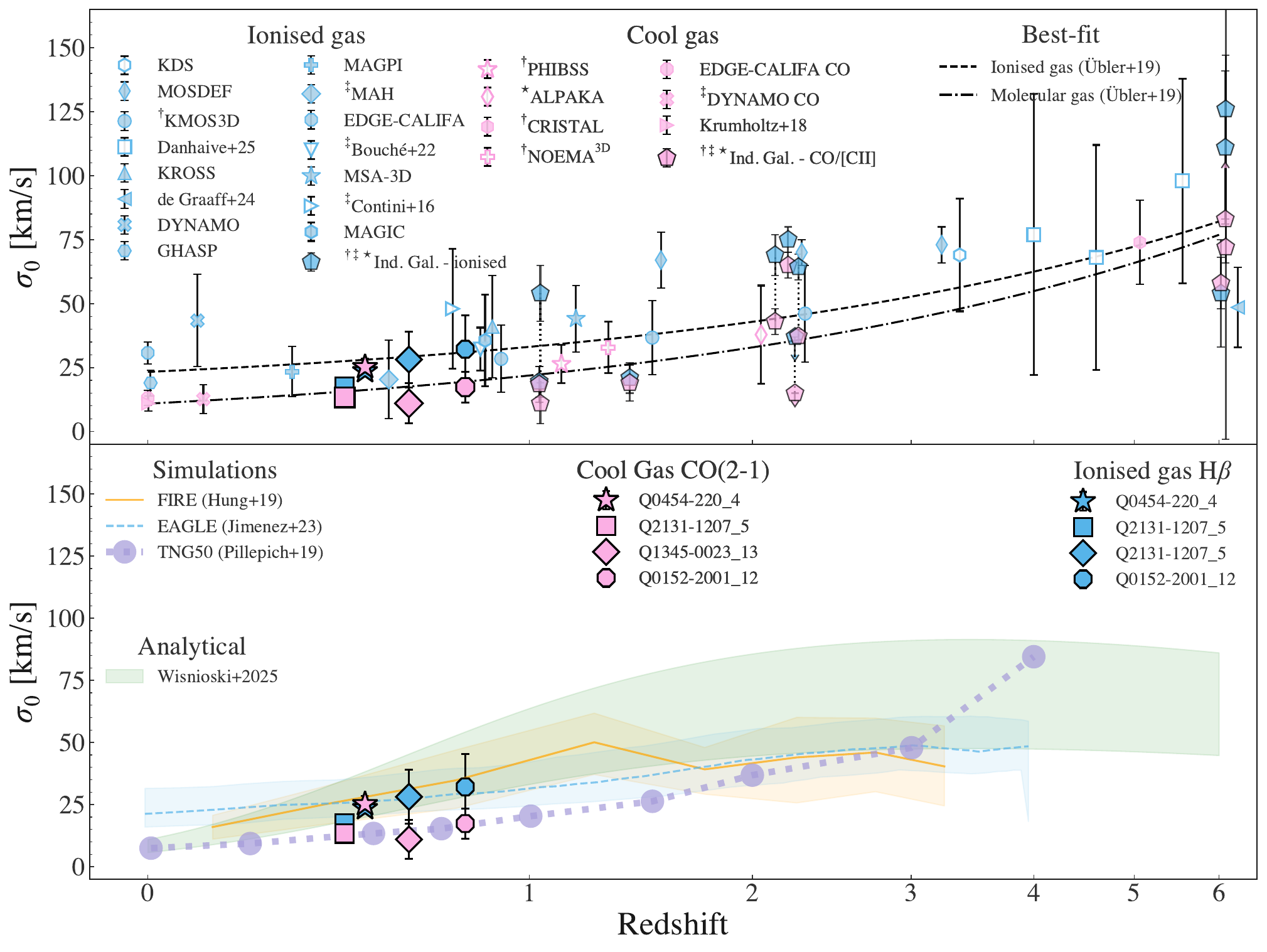}
        \caption{Evolution of $\sigma_0$ as a function of redshift. For all markers, the blue and pink colours denote the ionised (from H$\alpha$, H$\beta$ and [O\,{\sc iii}]) and cool gas phases (CO or [C\,{\sc ii}]), respectively. The values from {\scshape DysmalPy} for the four disk galaxies are shown with different shapes outlined in black. In the top panel, the transparent pentagons outlined in black are individual galaxies for which both the ionised gas and cool gas phases have been modelled. Values from the same galaxy are connected by a dotted line \citep{ubler_ionized_2018, girard_towards_2019, lelli_cold_2023, liu_600_2023, genzel_evidence_2023, parlanti_alma_2023, liu_detailed_2024}. The transparent markers not outlined in black represent the average $\sigma_0$ from various galaxy samples in the literature. The dashed and dashed-dotted lines show the empirical best-fit redshift evolution of $\sigma_0$ for ionised and cold molecular gas, respectively, extrapolated to $z \sim 6$ \citep{ubler_evolution_2019}. The dagger ($\dagger$) symbol denotes galaxies modelled using {\scshape DysmalPy}, the double dagger ($\ddagger$) denotes galaxies modelled using {\scshape GALPAK$^\mathrm{3D}$}, and the star ($\star$) denotes galaxies modelled using {\scshape 3DBarolo}. In the bottom panel, the redshift evolution of $\sigma_0$ in the FIRE \citep{hung_what_2019}, EAGLE \citep{jimenez_physical_2023}, and TNG50 \citep{pillepich_first_2019} simulations are denoted with the orange solid line and shaded area, the light blue dashed line and shaded area, and the purple markers and dotted line, respectively. The green shaded area shows the expectation on the evolution of $\sigma_0$ with redshift from the \citealt{wisnioski_evolution_2025} formalism. The $\sigma_0$ values in both tracers for the individual MUSE-ALMA Haloes galaxies are in line with observations and predictions from theory, and they show marginally higher ionised gas dispersions compared to CO.}
        \label{fig:sigma0}%
    \end{figure*}
    
Taking advantage of the panchromatic dataset on all four disk galaxies, we investigated how the velocity dispersion of the disk galaxies in both gas tracers compare to that of spatially resolved observations between $z = 0 - 6$ for both ionised gas (through the H$\alpha$, H$\beta$, [O\,{\sc ii}]3726,3729$\AA$ and [O\,{\sc iii}]5007$\AA$ lines) and cool gas (which includes both CO transitions and [C\,{\sc ii}] observations; \citealt{epinat_ghasp_2008, green_dynamo_2014, contini_deep_2016, turner_kmos_2017, ubler_ionized_2018, levy_edge-califa_2018, ubler_evolution_2019, girard_towards_2019, price_mosdef_2020, girard_systematic_2021, bouche_muse_2022, mercier_scaling_2022,weng-gasflows-2023, genzel_evidence_2023, lelli_cold_2023, rizzo_alma-alpaka_2023, parlanti_alma_2023, liu_600_2023, liu_detailed_2024, mai_magpi_2024, de_graaff_ionised_2024, lee_cristal_2025, danhaive_dawn_2025, ju_msa-3d_2026}). The results are presented in Fig. \ref{fig:sigma0}. We also compared to theoretical predictions \citep{hung_what_2019, pillepich_first_2019, jimenez_physical_2023, wisnioski_evolution_2025}, and the overall cosmic time evolution of this parameter (bottom panel of Fig. \ref{fig:sigma0}. Compared to the Local Universe and cosmic noon, the $\sim 0.5$ redshift range remains poorly explored. The MAGPI survey has investigated the ionised gas kinematics of 87 SFGs at $z\sim0.3$ \citep{mai_magpi_2024}, \cite{mercier_scaling_2022} fitted the [O\,{\sc ii}] doublet of 593 galaxies at $z\sim0.7$, and a few individual objects have observations of both the cold molecular and ionised gas phase (e.g. \citealt{klitsch_almacal_2018, peroux_multiphase_2019, szakacs_muse-alma_2021}), but the multi-wavelength sampling of this redshift range remains sparse, despite the epoch representing a key transition phase for the settling of galaxies into disks.

    \begin{figure*}
        \centering
        \includegraphics[width=\textwidth]{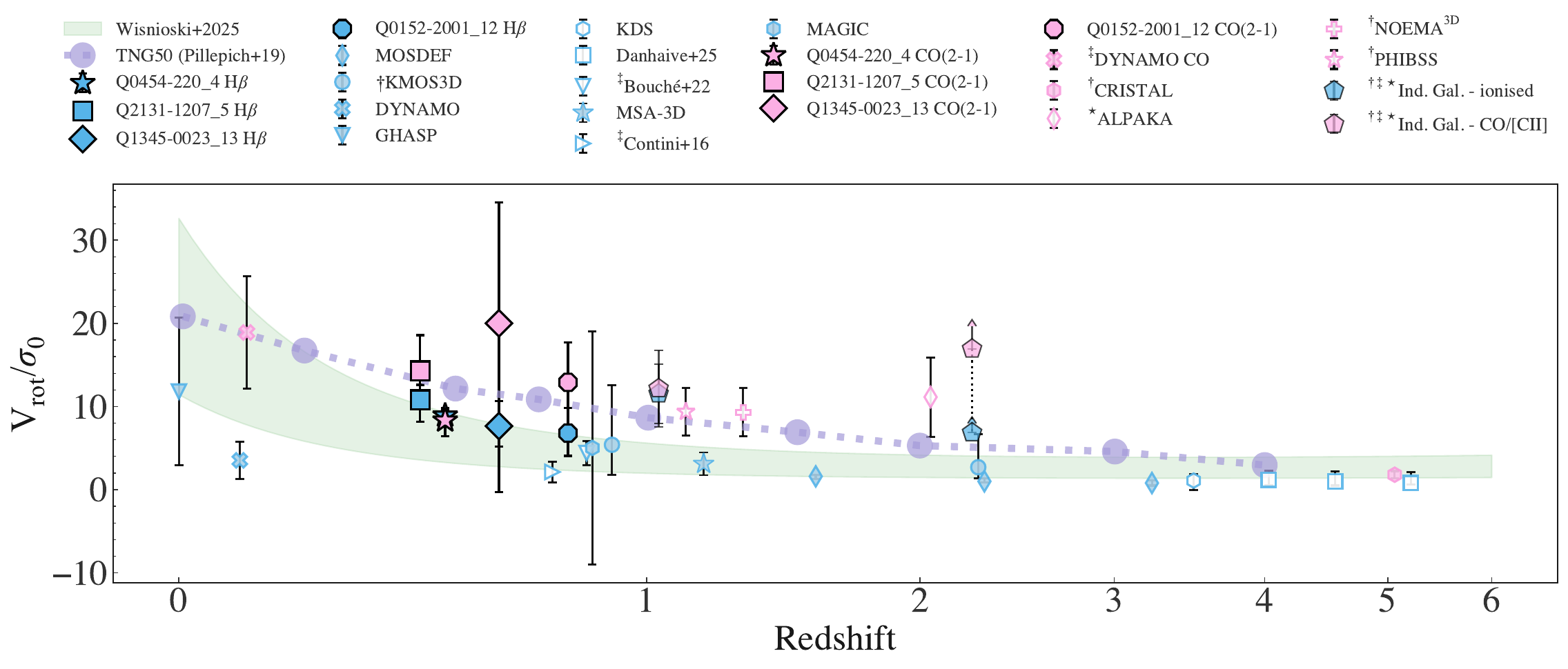}
        \caption{Evolution of $V/\sigma_0$ as a function of redshift. The legend follows the same logic and format as in Fig. \ref{fig:sigma0}. For all markers, the blue and pink colours denote the ionised (from H$\alpha$, H$\beta$ and [O\,{\sc iii}]) and cold gas phases (CO or [C\,{\sc ii}]), respectively. The values from {\scshape DysmalPy} for the four disk galaxies are indicated by the shapes outlined in black. The outlined pentagons  are individual galaxies for which both the ionised gas and cool gas phase have been modelled. Values from the same galaxy are connected by a dotted line \citep{girard_towards_2019, lelli_cold_2023}. The empty markers denote galaxy samples probing different stellar mass or mass density ranges compared to the MUSE-ALMA Haloes galaxies. The MUSE-ALMA Haloes galaxies show a high $V/\sigma_0$, in line with observations and expectations in this redshift range.}
        \label{fig:vsigma0}%
    \end{figure*}

The measurements from the MUSE-ALMA Haloes galaxies are consistent with trends of a progressive decline in velocity dispersion between $z\sim2$ and $z=0$. Comparing our results to the other estimates of velocity dispersion from observations, the values of $\sigma_{0}$ for both CO and H$\beta$ extracted for the MUSE-ALMA Haloes galaxies are in agreement within uncertainties with the best-fit from \cite{ubler_evolution_2019}. In addition, the galaxies' scatter is consistent with other studies, with the average ionised gas velocity dispersion in the MUSE-ALMA Haloes galaxies fitted using {\scshape GALPAK$^\mathrm{3D}$} \citep{weng-gasflows-2023}. Our values for the ionised gas velocity dispersion are marginally larger than that of the cold molecular gas, by a factor of 1.7 on average, which matches observed trends in the local Universe between ionised and molecular gas dispersions ($\sim 15$ km/s offset between CO and ionised gas tracers; \citealt{levy_edge-califa_2018, ubler_evolution_2019}). 

In addition, we compared with cosmological hydro-dynamical simulations: results from the FIRE simulation tracing all the gas within $10\%$ of the Virial radius \citep{hung_what_2019}, from TNG50 tracing the H$\alpha$ light, which for their model traces SF gas \citep{pillepich_first_2019}, and the EAGLE simulation which studies the kinematics of the `cold' gas, defined as any gas particle with SFR $>0$ (or $T\leq10^4$K for the few particles which fulfil this criterion; \citealt{jimenez_physical_2023}). Our values of the ionised gas velocity dispersion are in agreement with the various simulations, within the scatter and overlap; however, none of the aforementioned works resolve gas $T <10^4$K, limiting the possible comparison to the measurements of $\sigma_\mathrm{0,CO}$ for the MUSE-ALMA Haloes galaxies. Recent works have shown that galaxy morphology and kinematic classifications can strongly depend on the tracer used: the SPICE simulation finds that [C\,{\sc ii}] is highly sensitive to feedback, affecting both the disk fraction and the timing of disk settling (e.g. \citealt{bhagwat_stellar_2026}), whereas ionised gas tracers  may still exhibit ordered rotational signatures \citep{casavecchia_new_2026}. Similarly, \cite{kohandel_dynamically_2024} finds that at $z > 5$, $\sigma_\mathrm{0,H\alpha} \gtrsim 2 \sigma_\mathrm{0,[C\,{\sc ii}]}$, highlighting how inferred galaxy dynamical state vary across tracers (see also \citealt{phillips_lessons_2025}). This suggests that future numerical studies should increasingly focus on multi-phase and multi-wavelength kinematic modelling to robustly assess the impact of feedback on galaxy morphology and disk formation. Observational multi-wavelength studies such as this one therefore provide essential constraints for interpreting galaxy kinematics and for testing feedback models across cosmic time, particularly if such simulations are extended to lower redshifts.

We also show in Fig. \ref{fig:sigma0} the predicted kinematics in the framework of marginally stable gas-rich disks (light green shading), following the analytical prescriptions described in \cite{wisnioski_evolution_2025} (see also e.g. \citealt{genzel_sins_2011, wisnioski_kmos3d_2015, lee_cristal_2025}). We used the \cite{tacconi_evolution_2020} parametrization of the depletion time $t_{dep}(t)$ and the specific SFR $sSFR(t)$ to compute the time-dependent gas fraction $f_\mathrm{gas}(t)$, and we used the \citealt{genzel_sins_2011} derivation of the Toomre relation:
\begin{equation}
    Q_\mathrm{gas} = \frac{\sigma}{V_\mathrm{circ}} \frac{a}{f_\mathrm{gas}},
\end{equation}
where $Q_\mathrm{gas} < 1$ is the necessary criterion for a thin, rotating gas disk to be unstable to gravitational fragmentation. In this formalism, $a = \sqrt{2}$ for constant rotation velocity, and $V_\mathrm{circ}$ is the circular velocity. We adopted $Q_{gas} = 1$ as the canonical marginal stability criterion for direct comparison with previous studies. The shaded region in Fig. \ref{fig:sigma0} corresponds to the expectations for $V_\mathrm{circ} \in [130; 250]$ km/s, which corresponds to the range of circular velocities in our sample. While our values of ionised gas kinematics are in agreement with the expectations from the model, the cold molecular gas kinematics is almost systematically underneath the predictions. This can be explained by the choice of $Q_{gas} = 1$ as value for disk stability. Indeed, this value is the theoretical expectation for a single-phase infinitesimally thin disk \citep{binney_galactic_2008}, which is not necessarily physical. Studies have found $Q_{gas} \sim 0.7$ is more appropriate for quasi-stable thick disks \citep{behrendt_structure_2015, bacchini_3d_2024}, with studies at high-redshift finding even $Q_{gas} \sim 0.3$ for the cold molecular gas (\citealt{fujimoto_primordial_2025}; see also \citealt{wisnioski_evolution_2025} for a discussion on the value of $Q_{gas}$).

To investigate the overall disk stability of our sample, and place it in a cosmological context, we plotted $V_\mathrm{rot}/\sigma_0$ as a function of redshift on Fig. \ref{fig:vsigma0}, along with reported values from the literature. In general, there exist from observations a decrease in $V_\mathrm{rot}/\sigma_0$ with increasing redshift. The MUSE-ALMA Haloes galaxies sit in line with predictions and observations, and the cool gas is marginally more rotationally dominated than the ionised gas.

\subsection{The origin of turbulence}
    \begin{figure*}
        \centering
        \includegraphics[width=\textwidth]{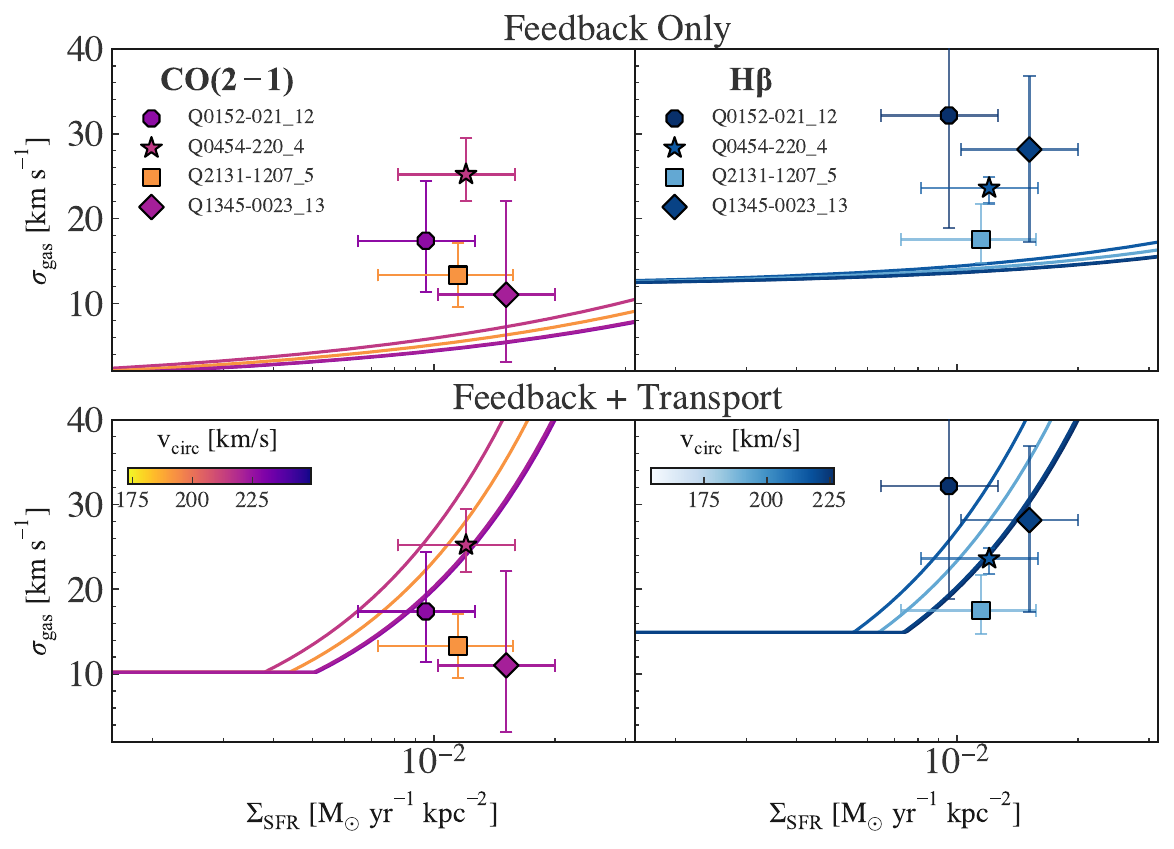}
        \caption{Comparison of the $\sigma_\mathrm{gas}$ versus $\Sigma_{\mathrm{SFR}}$ of our observations with the theoretical predictions, using the `feedback-only' model at a fixed Q=1 (top) and the `feedback + transport' model (bottom) from \cite{krumholz_unified_2018} for both the cold molecular (left) and ionised (right) gas. The points and curves are colour-coded according to the circular velocity of the galaxy. The measurements of $\sigma_0$ for the MUSE-ALMA Haloes galaxies are more in line with `feedback + transport' model, implying that gravity-induced pressure support is required in addition to feedback.}
        \label{fig:feedback}%
    \end{figure*}
    
To investigate the origin of the gas turbulence in the galaxies, we compared to the theoretical predictions from \cite{krumholz_unified_2018}, which proposes a unified view of the evolution of the gas in disks with cosmic time. The model takes into account both turbulence from SF feedback and radial flows to explain the pressure support in disks, and it shows that the former dominates the driving of turbulence in low redshift systems, while the latter dominates at earlier times, with a shift between dominant mechanism at $z \sim 0.5$, the redshift probed by our sample. 

Using the `feedback-only' model at fixed $Q=1$, we used Eq. 61 of \citealt{krumholz_unified_2018} to plot the expected values of $\sigma_{gas}$ as a function of the SFR surface density:

\begin{equation} \label{eq:sig_sfr}
    \Sigma_{\mathrm{SFR}} = \frac{8 (1+\beta) \pi \eta \sqrt{\phi_{mp} \phi_{nt}^3} \phi_Q \ \sigma_\mathrm{gas}^2}{G Q^2 \langle p_*/m_* \rangle f_{g,P} \ t_{orb}^2}.
\end{equation}

We defined the orbital time $t_{orb} = 2 \pi\ R_\mathrm{e} / V_\mathrm{circ}$ in megayears and the fraction of non-thermal velocity dispersion as $\phi_{nt} = 1 - \sigma_{th}^2 / \sigma_g^2$, where $\sigma_{th}$ is the thermal velocity dispersion (i.e. the broadening of the line due to particle movement within the gas) and is $\approx 0.2$ km/s for the cold molecular gas and $\approx 12$ km/s for the ionised gas \citep{krumholz_unified_2018, ubler_evolution_2019, bland-hawthorn_turbulent_2024}. All other parameters and their associated values are described in Appendix \ref{ap:krumholz}. The resulting theoretical predictions of $\sigma_{gas}$ with $\Sigma_{\mathrm{SFR}}$ are shown in the top row of Fig. \ref{fig:feedback} as the solid lines, colour-coded by the circular velocity of the galaxies. The difference in zeropoints is explained by the difference in $\sigma_{th}$, which varies depending on gas phase. We also display the four disk galaxies on Fig. \ref{fig:feedback}, using the best-fit values of $\sigma_{gas}$ from {\scshape DysmalPy} for both tracers, and the values of $\Sigma_{\mathrm{SFR}}$ estimated in \cite{karki_muse-alma_2023} using the effective radius determined through fitting with {\scshape GALFIT}, which is used as prior in the kinematic fitting. The remaining assumptions in Eq. \ref{eq:sig_sfr} follow that described in \cite{krumholz_unified_2018} for local galaxies and are presented in Table \ref{table:params_theory}. Even though these assumptions are made for galaxies at $z=0$, which is lower than our sample, they are deemed the closest approximation as other values reported in \cite{krumholz_unified_2018} pertain to galaxies at $z > 1$. 

Next, we investigated the `feedback + transport' model of \cite{krumholz_unified_2018}. This model has two modes, one for which $Q \geq 1$ and $\sigma_{gas} = \sigma_{sf}$, where $\sigma_{sf}$ is the velocity dispersion expected purely from stellar feedback; and one mode for which $Q = 1$ and $\sigma_{gas} > \sigma_{sf}$. To investigate this model in more detail, we first compute the expected value of $\sigma_{sf}$ for our sample using the average effective radius and circular velocities at $R_\mathrm{e}$ of CO and H$\beta$, and find $\sigma_{sf,\mathrm{CO}} \sim 10$ km/s and $\sigma_{sf, \mathrm{H\beta}} \sim 15$ km/s. Values of velocity dispersion above this cannot be explained solely through stellar feedback. Finally, we used Eq. 59 of \cite{krumholz_unified_2018} to obtain the expected values of $\sigma_{gas}$ as a function of the SFR surface density for the `feedback + transport' model, using the parameters reported in Table \ref{table:params_theory}:

\begin{equation}
    \begin{split}
    \Sigma_{\mathrm{SFR}} = \frac{f_{sf} \sqrt{8(1+\beta)} f_{g,Q} \sigma_{gas}}{GQt_{orb}^2} \\
    \times \ \mathrm{max} \Bigg [\frac{8 \epsilon_{ff} f_{g,Q}}{Q}  \sqrt{\frac{2(1+\beta)}{3f_{g,P} \phi_{mp}}} , \frac{t_{orb}}{t_{SF,max}} \Bigg].
    \end{split}
\end{equation}

From Fig. \ref{fig:feedback}, we see that the velocity dispersions of the MUSE-ALMA Haloes galaxies are more in agreement with the predictions of the `feedback + transport' model, in particular for the ionised gas. This implies that at $z\sim0.5$, gravity-induced pressure support is required in addition to feedback. Given the uncertainties on the measurements and model assumptions, a case-by-case analysis is difficult, but we note that the one galaxy for which the velocity dispersion in both tracers is most in line with the `feedback + transport' model at equivalent circular velocity and effective radius is Q0454-220\_4. This is also the target which has strong velocity residuals 2D CO(2-1) maps (Fig. \ref{fig:0454_dysmal}), indicative of non-axisymmetric motions in the galaxy. This outlines how the processes in play in a galaxy at a given time can impact the observed velocity dispersion.

In addition, part of the discrepancy between our measurements and the theoretical values (in particular for the cold molecular gas) could stem from the model assumptions. As the MUSE-ALMA Haloes galaxies are located at $z\sim0.5$, we have chosen the values associated to local spiral galaxies reported in \cite{krumholz_unified_2018}. When using the $z>1$ assumptions where the disks have higher gas and SF fractions, the $\sigma_{sf}$ increases and the knee of the relation is pushed to higher values of $\mathrm{\Sigma_{SFR}}$. In this case, the galaxies sit in the region of the model where $\sigma_{gas} = \sigma_{sf}$ (and $Q > 1$). This underlines further the transitional aspect of $z \sim 0.5$ in the mode of pressure support between SF feedback and gas transport.

\subsection{Dark matter content of $z\sim0.5$ galaxies}\label{sec:fdm}
    \begin{figure*}
        \centering
        \includegraphics[width=\linewidth]{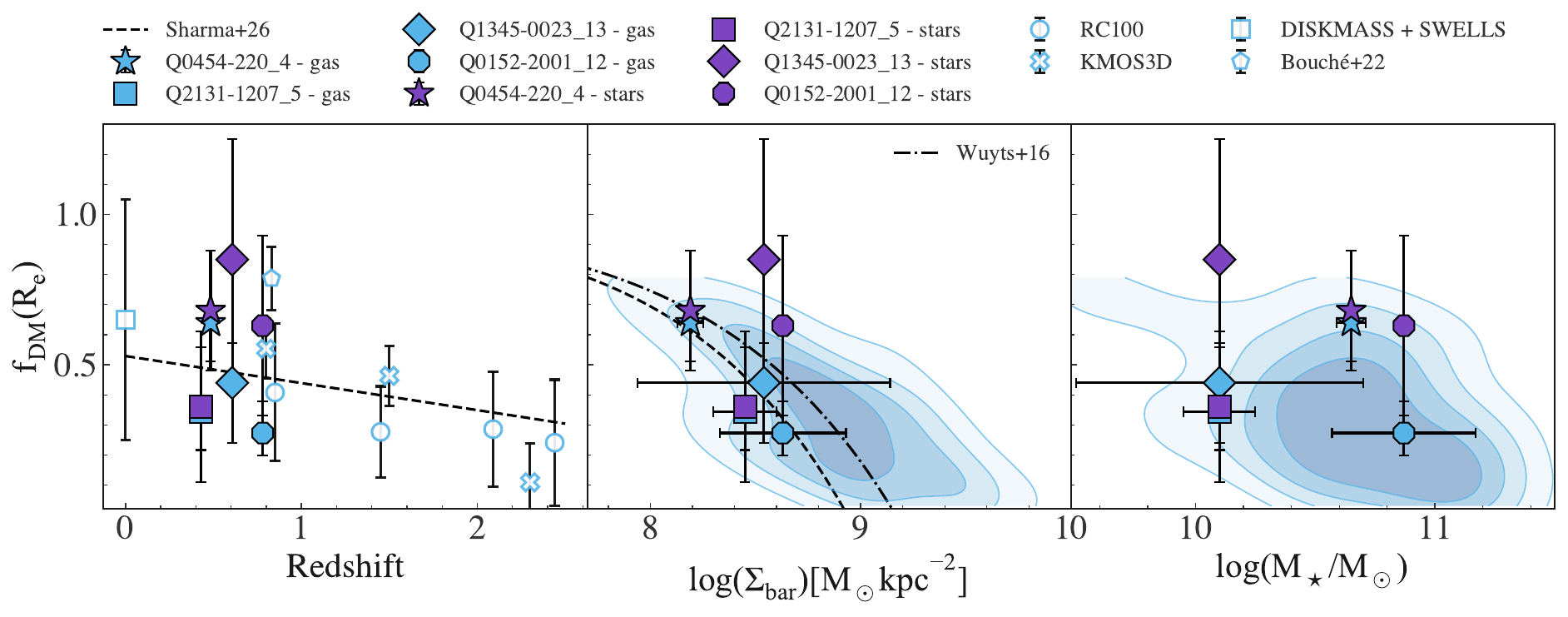}
        \caption{Dark matter fraction for the four MUSE-ALMA Haloes galaxies modelled here. In blue are the values from the fit to the gas, and in purple are the values from the fit to the stars. For Q0454-220\_4 and Q2131-1207\_5, the points come from the spatially resolved fitting; for the other two, they come from the integrated fitting. Left: Evolution of $f_{DM}$ with redshift along with literature values \citep{barnabe_swells_2012, martinsson_diskmass_2013, wuyts_kmos3d_2016, bouche_muse_2022, nestor_shachar_rc100_2023}. The black dashed line shows the best-fit correlation extracted from the MAGPI survey for galaxies with a similar stellar mass as our sample ($10.0 < \mathrm{log}(M_\star) < 10.5$; \citealt{sharma_magpi_2026}). Middle: Evolution of $f_{DM}$ with baryonic surface density along with the values from RC100 (blue contours; \citep{nestor_shachar_rc100_2023}) and the parametrization from \cite{wuyts_kmos3d_2016} and \cite{sharma_magpi_2026} (dot-and-dashed and dashed line, respectively). Right: Evolution of $f_{DM}$ with stellar mass along with the values from RC100 (blue contours; \citep{nestor_shachar_rc100_2023}). In all three plots, our extracted values fall in line with the expectations from the literature, with the exception of $f_{\rm DM, \star}$ for Q1345-0023\_13.}
        \label{fig:fdm}%
    \end{figure*}

From the kinematic modelling of the MUSE-ALMA Haloes galaxies, we derived the dark matter fraction of the systems. Those values are reported in Table \ref{table:dysmalpy}, and plotted as a function of redshift, baryonic surface density $\Sigma_\mathrm{bar}$ and stellar mass in Fig. \ref{fig:fdm}. By using $M_\mathrm{tot} = M_\star + M_\mathrm{gas}$ as prior on our model, we did not include the dust and neutral atomic gas mass, which might lead to lower $f_{DM}(R_\mathrm{e})$. However, as {\scshape DysmalPy} fits for total baryonic mass even with the input prior, we expect that the difference would fall within the uncertainties. We estimate $\Sigma_\mathrm{bar}$ at $R_\mathrm{e}$ using the total baryonic mass enclosed at the effective radius returned by {\scshape DysmalPy}, and the effective radii. 

In general, studies based on ionised gas kinematic modelling find a decreasing dark matter fraction with redshift out to $z \sim 3$ \citep{barnabe_swells_2012, martinsson_diskmass_2013, wuyts_kmos3d_2016, genzel_strongly_2017, genzel_rotation_2020, nestor_shachar_rc100_2023, sharma_magpi_2026} and for all four galaxies, the best-fit values for $f_{DM}$ from the gas are in line with expectations from these surveys, and systematically lower than the average dark matter fraction recovered from SFGs in the MUSE Extremely Deep Field \citep{bouche_muse_2022}. The latter, however, probes stellar masses almost one order of magnitude lower than the MUSE-ALMA Haloes galaxies, and studies report an anti-correlation of $f_{DM}$ with galaxy mass and central mass surface density \citep{wuyts_kmos3d_2016, genzel_strongly_2017, genzel_rotation_2020, nestor_shachar_rc100_2023, bouche_muse_2022, sharma_magpi_2026}, out to $z = 4-6$ \citep{danhaive_dark_2026}. 
On Fig. \ref{fig:fdm}, we explored the correlation between stellar mass and baryonic surface density, which is known to hold the tightest correlation \citep{wuyts_kmos3d_2016}. Aside from the $f_{\rm DM, \star}$ of Q1345-0023\_13, our data show excellent agreement with the results from the RC100 sample (100 massive, SFGs at cosmic noon; \citealt{nestor_shachar_rc100_2023}), as well as the parameterisations from \cite{sharma_magpi_2026}, using MAGPI in a similar redshift and mass range, and of \cite{wuyts_kmos3d_2016}, using KMOS$\mathrm{^{3D}}$; following a distinct trend of decreasing $f_{DM}$ with increasing $\Sigma_\mathrm{bar}$. This also agrees with the results from \cite{obeirne_MAH_2026} for the whole MUSE-ALMA Haloes survey, finding increasing baryon fractions with increasing stellar surface densities. This could be attributed to more extended systems probing further into the dark matter halo, which translates to higher $f_{DM}$ \citep{wuyts_kmos3d_2016}, and agrees with a scenario where massive galaxies undergo efficient gas infall and central baryonic condensation, increasing the central mass surface density and thereby lowering the inferred dark matter fraction within $R_\mathrm{e}$ \citep{genzel_rotation_2020, nestor_shachar_rc100_2023, sharma_magpi_2026}.

In addition, we reported in Fig. \ref{fig:fdm} the dark matter fractions retrieved from fitting the stellar kinematics (see Sec. \ref{sec:ppxf}), as the purple markers (see values in Table \ref{table:dysmalpy} and \ref{table:ppxf}). For our two galaxies with the highest S/N, Q2131-1207\_5 and Q0454-220\_5, we used the values from spatially resolved stellar kinematics and observe an excellent agreement between the two $f_{DM}$ estimates. For the remaining two galaxies, $f_{DM,\star}$ are $\sim2 \times$ higher than those estimated from the gas kinematics. However, the low S/N of the integrated spectra (S/N$< 8$; see Appendix \ref{ap:ppxf}; Figs. \ref{fig:1345_ppxf} \& \ref{fig:0152_ppxf}) is likely responsible for the offset between the two $f_{DM}$ estimates, although the measurements agree within $< 2\sigma$. 

The four galaxies are in line with the redshift and mass trends, with consistent $f_{DM}(R_\mathrm{e})$ estimates from gas and from stars in the two cases with spatially resolved, higher S/N constraints on stellar kinematics. In a scenario where gas-rich galaxies at cosmic noon are both more turbulent (lower $V/\sigma_0$, sustained by gravitational instabilities and high accretion rates) and more centrally baryon-dominated; and as the gas fraction decreases through SF and feedback, the disks settle and $f_{DM}(R_\mathrm{e})$ increases \citep{genzel_strongly_2017, pillepich_first_2019}; the agreement of the $f_{DM}(R_\mathrm{e})$ and $V/\sigma_0$ of the MUSE-ALMA Haloes galaxies with these well-characterized samples suggests that the same physical conditions are likely at play.

\subsection{The MUSE-ALMA Haloes galaxy population}
Studies of the MUSE–ALMA Haloes sample show that strong H\,{\sc i} absorbers trace a heterogeneous set of galaxies and CGM structures (see also \citealt{obeirne_MAH_2026}). The MUSE analysis by \cite{weng_muse-alma_2023} finds that 18/27 (67\%) of the absorbers are linked to two or more galaxies, implying overdense environments. In addition, kinematic analysis of the ionised gas of the galaxies suggests only $\sim 15\%$ of absorbers are consistent with co-rotation with the gas disk, while $\sim10\%$ are consistent with gas accretion and up to $\sim 30 \%$ fall near minor axes (outflow-like, though only a few are clear cases). For the molecular phase of the gas, deep ALMA observations yield a CO detection rate of $\sim 20 \%$ for a non-metallicity and -stellar mass selected sample and push $H_2$ detections to $> 1$dex below previous H{\sc i}-selected limits, revealing systems ranging from metal-rich, active SFGs with substantial molecular reservoirs to gas-rich but low-efficiency star formers, which fall off the SF scaling relations \citep{bollo_muse-alma_2026}. A comparison of the ionised gas and the cool, neutral CGM of MaNGA galaxies at $z<0.1$ by \cite{klimenko_baryonic_2023} shows consistency with the neutral gas co-rotating with the ionised gas in the inner disk out to $\sim 10 R_\mathrm{e}$.

For the galaxies analysed in this paper, four of the six ($67\%$) targets are identified as rotation-dominated disks which are kinematically modelled. From \cite{peroux_nature_2017} and \cite{weng-gasflows-2023}, three of the six are part of groups with at least three identified members (Q0454-220\_4, Q1229-021\_6, Q2131-1207\_5), and one is in an identified pair (Q1342-0035\_4). For the two disks which are in groups, there is no evidence of the other group members having an influence on their kinematics. However, the number of galaxies detected for each absorber is also a function of the depth of the MUSE observations (which is not uniform across the survey), and as such should be considered a lower limit on the group numbers. 

For Q0152-2001\_12 and Q1342-0035\_4, no conclusions can be made on the origin of the absorber with respect to the detected galaxies. For the absorber of Q1229-021\_6, it is identified to be outflowing gas from another galaxy in the group (which was also detected in CO as part of the ALMA observations; see \citealt{bollo_muse-alma_2026}). For both Q2131-1207\_5 and Q1345-0023\_13, the absorber is co-rotating with the disk of the galaxy and based on geometrical arguments is identified as gas accreting on the galaxies \citep{peroux_nature_2017, weng-gasflows-2023}. Interestingly, both galaxies show no sign of being more disturbed than the rest, indicating that the accretion of gas onto the galaxies is not raising the turbulence in the disks, or at least not yet. Finally, the absorber of Q0454-220\_4 reveal a complex combination of relic outflow, gas accretion and intragroup material \citep{norris_discovery_2021}. High-metallicity low-(high-) ionisation clouds in absorption reveal the presence of inflowing(outflowing) gas along the major(minor) axis of the galaxy, at impact parameter $108$kpc from the galaxy. The metallicity of the inflowing gas in absorption implies that the gas is being re-accreted after being previously ejected from the galaxy. In addition, evidence from a low metallicty absorption component tentatively suggests ongoing pristine gas accretion onto the galaxy moving along a halo filament. This complex absorption profile in the CGM could be reflected in the excellent agreement of the galaxy with the `feedback + transport' model (see Fig. \ref{fig:feedback}; \citealt{krumholz_unified_2018}).

\section{Conclusions}\label{sec:conc}
In this work, we have presented a near-complete kinematic analysis of the baryons in four galaxies at z$\sim 0.5$. The galaxies are part of the MUSE-ALMA Haloes program, which targets known H\,{\sc i} absorbers and their associated galaxies in this redshift range. Our analysis offers a detailed view of the ionised and molecular gas kinematics as well as stellar kinematics when possible, in a so-far less well explored redshift range. The MUSE-ALMA Haloes project presents us with a rare opportunity to study multiple phases of the gas in a consistent manner and make comparisons with theoretical predictions. The main results of this paper are summarised here:

\begin{itemize}
    \item As previous studies observed, the kinematic analysis of the targets shows a diverse sample, with four of the six galaxies identified as rotation-dominated systems. Four of the galaxies are in a group environment with two or more members. For the four disk galaxies, we spatially resolved the ionised and cold molecular gas emission as well as galaxy integrated stellar absorption. For two of the objects, we retrieved spatially resolved stellar kinematics.

    \item Fitting the 1D gas kinematics with {\scshape DysmalPy}, we derived the total mass, dark matter fraction, circular velocity, and velocity dispersion of the ionised and cold molecular gas of the four disk galaxies. Our sample has mean values of $\langle \mathrm{log}(M_\mathrm{tot}/M_\odot) \rangle = 10.8 \pm 0.1$, $\langle f_{DM} \rangle = 0.42 \pm 0.14$, $\langle \mathrm{\sigma_{0,CO}} \rangle = 17 \pm 5$ km/s, and $\langle \mathrm{\sigma_{0,H\beta}} \rangle = 25 \pm 5$ km/s. Our ionised gas velocity dispersions are on average a factor of 1.7 larger than the cold molecular gas, tentatively pointing to a differentiation between the phases. Overall, our estimates of $\sigma_{0}$ ($V_\mathrm{rot}/\sigma_0$) are in line with the increasing (decreasing) evolution with redshift reported in observations and simulations and theoretical predictions (see also Figs. \ref{fig:sigma0} \& \ref{fig:vsigma0}).

    \item Comparing our velocity dispersion measurements with the models of turbulence in disks from \cite{krumholz_unified_2018}, we found that our values agree better with the `feedback + transport' model, in which disk dispersion is driven by a combination of stellar feedback and radial gas transport in the plane of the disk, rather than the `feedback-only' model (see Fig. \ref{fig:feedback}). 
    
    \item A systematic search for galaxy-wide outflows revealed no significant feedback contribution to the observations, in line with a scenario in which stellar feedback contributes little to the turbulence in the disk (see Sec. \ref{sec:outflows}). 

    \item The dark matter fractions derived from the kinematic modelling are in line with previous observations of galaxies of similar mass across a redshift range of $z = 0 - 2.5$. When comparing with the dark matter fractions obtained through stellar kinematic fitting, we find that the $f_{DM}$ agree well for the two galaxies with the highest S/N ($>25$) stellar absorption features and even more so for the values coming from the resolved kinematic fitting (see Fig. \ref{fig:fdm}). 
\end{itemize}

This study highlights the value of performing multi-wavelength analysis of the same samples in order to obtain consistent analysis and measurements of the kinematic properties of multiple phases. The MUSE-ALMA Haloes program provides a unique opportunity to study the resolved multi-tracer kinematic properties of SFGs at $z\sim0.5$. In the future, additional resolved CO observations of the sample combined with the existing MUSE data will allow for a more statistically robust investigation of this relatively unexplored redshift range.

\begin{acknowledgements}
We thank the referee for their comments and feedback on the paper. We thank J.-B. Jolly, J. M. Espejo Salcedo, S. Pastras, C. Lagos and F. Belfiore for useful discussions on the analysis and interpretation of the data. C.B., J.C., and N.M.F.S. acknowledge funding by the European Union
(ERC Advanced Grant GALPHYS, 101055023).
Views and opinions expressed are, however, those of the author(s)
only and do not necessarily reflect those of the European Union or
the European Research Council. Neither the European Union nor the
granting authority can be held responsible for them. VPK acknowledges support from NSF grant AST/2007538  
and NASA grant 80NSSC25K7582 (PI Kulkarni). RA acknowledges funding from the European Research Council (ERC) under the European Union's Horizon 2020 research and innovation programme (grant agreement 101020943, SPECMAP-CGM). This paper makes use of the following ALMA data: ADS/JAO.ALMA\#2023.1.00127.L ALMA is a partnership of ESO (representing its member states), NSF (USA) and NINS (Japan), together with NRC (Canada), MOST and ASIAA (Taiwan), and KASI (Republic of Korea), in cooperation with the Republic of Chile. The Joint ALMA Observatory is operated by ESO, AUI/NRAO and NAOJ. This work has made use of data from the European Space Agency (ESA) mission
{\it Gaia} (\url{https://www.cosmos.esa.int/gaia}), processed by the {\it Gaia}
Data Processing and Analysis Consortium (DPAC,
\url{https://www.cosmos.esa.int/web/gaia/dpac/consortium}). Funding for the DPAC
has been provided by national institutions, in particular the institutions
participating in the {\it Gaia} Multilateral Agreement.
This work made use of the following softwares and Python packages: \texttt{Astropy} \citep{collaboration_astropy_2022}, \texttt{DysmalPy} \citep{davies_nuclear_2004, davies_how_2011, cresci_sins_2009, wuyts_kmos3d_2016, lang_falling_2017, price_rotation_2021, lee_disk_2025}, \texttt{pPXF} \citep{cappellari_full_2023}, \texttt{linefit} \citep{davies_how_2011}, \texttt{emcee} \citep{foreman-mackey_emcee_2013}, \texttt{Lmfit} \citep{newville_lmfit_2025}.
\end{acknowledgements}

\bibliographystyle{bibtex/aa.bst}
\bibliography{references}

@article{peroux_muse-alma_2022,
    title = {{MUSE}-{ALMA} haloes {VII}: survey science goals \& design, data processing and final catalogues},
    volume = {516},
    issn = {0035-8711},
    shorttitle = {{MUSE}-{ALMA} haloes {VII}},
    url = {https://ui.adsabs.harvard.edu/abs/2022MNRAS.516.5618P},
    doi = {10.1093/mnras/stac2546},
    urldate = {2024-12-11},
    journal = {MNRAS},
    publisher = {OUP},
    author = {Péroux, C. and Weng, S. and Karki, A. and Augustin, R. and Kulkarni, V. P. and Szakacs, R. and Klitsch, A. and Hamanowicz, A. and Fresco, A. Y. and Zwaan, M. A. and Biggs, A. and Fox, A. J. and Hayes, M. and Howk, J. C. and Kacprzak, G. G. and Kassin, S. and Kuntschner, H. and Nelson, D. and Pettini, M.},
    month = nov,
    year = {2022},
    note = {ADS Bibcode: 2022MNRAS.516.5618P},
    pages = {5618--5636},
}

@article{weng_muse-alma_2023,
    title = {{MUSE}-{ALMA} {Haloes} - {VIII}. {Statistical} study of circumgalactic medium gas},
    volume = {519},
    issn = {0035-8711},
    url = {https://ui.adsabs.harvard.edu/abs/2023MNRAS.519..931W/abstract},
    doi = {10.1093/mnras/stac3497},
    language = {en},
    number = {1},
    urldate = {2025-09-29},
    journal = {MNRAS},
    author = {{Weng}, Simon and Péroux, C. and Karki, A. and Augustin, R. and Kulkarni, V. P. and Szakacs, R. and Zwaan, M. A. and Klitsch, A. and Hamanowicz, A. and Sadler, E. M. and Biggs, A. and Fresco, A. Y. and Hayes, M. and Howk, J. C. and Kacprzak, G. G. and Kuntschner, H. and Nelson, D. and Pettini, M.},
    month = feb,
    year = {2023},
    pages = {931--947},
}

@article{augustin_muse-alma_2024,
    title = {{MUSE}-{ALMA} {Haloes} {X}: the stellar masses of gas-rich absorbing galaxies},
    volume = {528},
    issn = {0035-8711},
    shorttitle = {{MUSE}-{ALMA} {Haloes} {X}},
    url = {https://ui.adsabs.harvard.edu/abs/2024MNRAS.528.6159A},
    doi = {10.1093/mnras/stae387},
    urldate = {2025-02-11},
    journal = {MNRAS},
    publisher = {OUP},
    author = {Augustin, Ramona and Péroux, Céline and Karki, Arjun and Kulkarni, Varsha and Weng, Simon and Hamanowicz, A. and Hayes, M. and Howk, J. C. and Kacprzak, G. G. and Klitsch, A. and Zwaan, M. A. and Fox, A. and Biggs, A. and Fresco, A. Y. and Kassin, S. and Kuntschner, H.},
    month = mar,
    year = {2024},
    note = {ADS Bibcode: 2024MNRAS.528.6159A},
    pages = {6159--6166},
}

@article{karki_muse-alma_2023,
    title = {{MUSE}-{ALMA} {Haloes} - {IX}. {Morphologies} and stellar properties of gas-rich galaxies},
    volume = {524},
    issn = {0035-8711},
    url = {https://ui.adsabs.harvard.edu/abs/2023MNRAS.524.5524K},
    doi = {10.1093/mnras/stad2134},
    urldate = {2024-12-11},
    journal = {MNRAS},
    publisher = {OUP},
    author = {Karki, Arjun and Kulkarni, Varsha P. and Weng, Simon and Péroux, Céline and Augustin, Ramona and Hayes, Matthew and Ayromlou, Mohammadreza and Kacprzak, Glenn G. and Howk, J. Christopher and Szakacs, Roland and Klitsch, Anne and Hamanowicz, Aleksandra and Fresco, Alejandra and Zwaan, Martin A. and Biggs, Andrew D. and Fox, Andrew J. and Kassin, Susan and Kuntschner, Harald},
    month = oct,
    year = {2023},
    note = {ADS Bibcode: 2023MNRAS.524.5524K},
    pages = {5524--5547},
}

@article{bollo_muse-alma_2026,
	title = {{MUSE}-{ALMA} {Haloes}: {XII}. {Molecular} gas in z ∼ 0.5 {H} {I} ─ selected galaxies},
	volume = {707},
	issn = {0004-6361},
	shorttitle = {{MUSE}-{ALMA} {Haloes}},
	url = {https://ui.adsabs.harvard.edu/abs/2026A&A...707A..81B/abstract},
	doi = {10.1051/0004-6361/202556531},
	language = {en},
	urldate = {2026-07-17},
	journal = {A\&A},
	author = {Bollo, Victoria and Péroux, Céline and Zwaan, Martin and Chen, Jianhang and Kulkarni, Varsha P. and Barfety, Capucine and Weng, Simon and Förster Schreiber, Natascha M. and Tacconi, Linda and Casavecchia, Benedetta and O'Beirne, Tamsyn and Chemin, Laurent and Augustin, Ramona and Halley, Mitchell},
	month = mar,
	year = {2026},
	pages = {A81},
}

@article{price_rotation_2021,
    title = {Rotation {Curves} in z 1-2 {Star}-forming {Disks}: {Comparison} of {Dark} {Matter} {Fractions} and {Disk} {Properties} for {Different} {Fitting} {Methods}},
    volume = {922},
    issn = {0004-637X},
    shorttitle = {Rotation {Curves} in z 1-2 {Star}-forming {Disks}},
    url = {https://ui.adsabs.harvard.edu/abs/2021ApJ...922..143P/abstract},
    doi = {10.3847/1538-4357/ac22ad},
    language = {en},
    number = {2},
    urldate = {2026-02-09},
    journal = {ApJ, Volume 922, Issue 2, id.143, pp.},
    author = {Price, S. H. and Shimizu, T. T. and Genzel, R. and Übler, H. and Förster Schreiber, N. M. and Tacconi, L. J. and Davies, R. I. and Coogan, R. T. and Lutz, D. and Wuyts, S. and Wisnioski, E. and Nestor, A. and Sternberg, A. and Burkert, A. and Bender, R. and Contursi, A. and Davies, R. L. and Herrera-Camus, R. and Lee, M.-J. and Naab, T. and Neri, R. and Renzini, A. and Saglia, R. and Schruba, A. and Schuster, K.},
    month = dec,
    year = {2021},
    pages = {143},
}

@article{lee_disk_2025,
    title = {Disk {Kinematics} at {High} {Redshift}: {DysmalPy}'s {Extension} to {3D} {Modeling} and {Comparison} with {Different} {Approaches}},
    volume = {978},
    issn = {0004-637X},
    shorttitle = {Disk {Kinematics} at {High} {Redshift}},
    url = {https://ui.adsabs.harvard.edu/abs/2025ApJ...978...14L/abstract},
    doi = {10.3847/1538-4357/ad90b5},
    language = {en},
    number = {1},
    urldate = {2025-06-23},
    journal = {ApJ, Volume 978, Issue 1, id.14, 31 pp.},
    author = {{Lee}, Lilian L. and Förster Schreiber, Natascha M. and Price, Sedona H. and Liu, Daizhong and Genzel, Reinhard and Davies, Ric and Tacconi, Linda J. and Shimizu, Taro T. and Nestor Shachar, Amit and Espejo Salcedo, Juan M. and Pastras, Stavros and Wuyts, Stijn and Lutz, Dieter and Renzini, Alvio and Übler, Hannah and Herrera-Camus, Rodrigo and Sternberg, Amiel},
    month = jan,
    year = {2025},
    pages = {14},
}

@ARTICLE{lee_cristal_2025,
       author = {{Lee}, Lilian L. and {F{\"o}rster Schreiber}, Natascha M. and {Herrera-Camus}, Rodrigo and {Liu}, Daizhong and {Price}, Sedona H. and {Genzel}, Reinhard and {Tacconi}, Linda J. and {Lutz}, Dieter and {Davies}, Ric and {Naab}, Thorsten and {{\"U}bler}, Hannah and {Aravena}, Manuel and {Assef}, Roberto J. and {Barcos-Mu{\~n}oz}, Loreto and {Bowler}, Rebecca A.~A. and {Burkert}, Andreas and {Chen}, Jianhang and {Davies}, Rebecca L. and {De Looze}, Ilse and {Diaz-Santos}, Tanio and {Gonz{\'a}lez-L{\'o}pez}, Jorge and {Ikeda}, Ryota and {Mitsuhashi}, Ikki and {Posses}, Ana and {Rela{\~n}o Pastor}, M{\'o}nica and {Renzini}, Alvio and {Solimano}, Manuel and {Spilker}, Justin S. and {Sternberg}, Amiel and {Tadaki}, Kenichi and {Telikova}, Kseniia and {Veilleux}, Sylvain and {Villanueva}, Vicente},
        title = "{The ALMA-CRISTAL survey: Resolved kinematic studies of main sequence star-forming galaxies at 4 < z < 6}",
      journal = {\aap},
         year = 2025,
        month = sep,
        urldate = {2025-09-23},
       volume = {701},
          eid = {A260},
        pages = {A260},
          doi = {10.1051/0004-6361/202555362},
archivePrefix = {arXiv},
       eprint = {2507.11600},
 primaryClass = {astro-ph.GA},
       adsurl = {https://ui.adsabs.harvard.edu/abs/2025A&A...701A.260L}
}

@ARTICLE{noordermeer_disc_2008,
       author = {{Noordermeer}, E. and {Merrifield}, M.~R. and {Arag{\'o}n-Salamanca}, A.},
        title = "{Exploring disc galaxy dynamics using integral field unit data}",
      journal = {\mnras},
         year = 2008,
        month = aug,
       volume = {388},
       number = {3},
        pages = {1381-1393},
          doi = {10.1111/j.1365-2966.2008.13487.x},
archivePrefix = {arXiv},
       eprint = {0805.3230},
 primaryClass = {astro-ph},
       adsurl = {https://ui.adsabs.harvard.edu/abs/2008MNRAS.388.1381N}
}

@ARTICLE{obeirne_MAH_2026,
       author = {{O'Beirne}, T.O. and {P\'eroux}, C. and {Zwaan}, M.},
        title = "{MUSE-ALMA Haloes XV: a census of the condensed baryons in z 0.5 H i absorption-selected galaxies}",
      journal = {A\&A subm.},
         year = 2026,
}

@ARTICLE{burkert_turbulence_2010,
       author = {{Burkert}, A. and {Genzel}, R. and {Bouch{\'e}}, N. and {Cresci}, G. and {Khochfar}, S. and {Sommer-Larsen}, J. and {Sternberg}, A. and {Naab}, T. and {F{\"o}rster Schreiber}, N. and {Tacconi}, L. and {Shapiro}, K. and {Hicks}, E. and {Lutz}, D. and {Davies}, R. and {Buschkamp}, P. and {Genel}, S.},
        title = "{High-redshift Star-forming Galaxies: Angular Momentum and Baryon Fraction, Turbulent Pressure Effects, and the Origin of Turbulence}",
      journal = {\apj},
         year = 2010,
        month = dec,
       volume = {725},
       number = {2},
        pages = {2324-2332},
          doi = {10.1088/0004-637X/725/2/2324},
archivePrefix = {arXiv},
       eprint = {0907.4777},
 primaryClass = {astro-ph.CO},
       adsurl = {https://ui.adsabs.harvard.edu/abs/2010ApJ...725.2324B}
}

@article{ubler_evolution_2019,
    title = {The {Evolution} and {Origin} of {Ionized} {Gas} {Velocity} {Dispersion} from z ∼ 2.6 to z ∼ 0.6 with {KMOS}{\textless}{SUP}{\textgreater}{3D}{\textless}/{SUP}{\textgreater}},
    volume = {880},
    issn = {0004-637X},
    url = {https://ui.adsabs.harvard.edu/abs/2019ApJ...880...48U/abstract},
    doi = {10.3847/1538-4357/ab27cc},
    language = {en},
    number = {1},
    urldate = {2025-06-12},
    journal = {ApJ},
    author = {Übler, H. and Genzel, R. and Wisnioski, E. and Förster Schreiber, N. M. and Shimizu, T. T. and Price, S. H. and Tacconi, L. J. and Belli, S. and Wilman, D. J. and Fossati, M. and Mendel, J. T. and Davies, R. L. and Beifiori, A. and Bender, R. and Brammer, G. B. and Burkert, A. and Chan, J. and Davies, R. I. and Fabricius, M. and Galametz, A. and Herrera-Camus, R. and Lang, P. and Lutz, D. and Momcheva, I. G. and Naab, T. and Nelson, E. J. and Saglia, R. P. and Tadaki, K. and van Dokkum, P. G. and Wuyts, S.},
    month = jul,
    year = {2019},
    pages = {48},
}

@article{wisnioski_evolution_2025,
    title = {Evolution of gas velocity dispersion in discs from z ∼ 8 to z ∼ 0.5},
    volume = {544},
    issn = {0035-8711},
    url = {https://ui.adsabs.harvard.edu/abs/2025MNRAS.544.2777W/abstract},
    doi = {10.1093/mnras/staf1606},
    language = {en},
    number = {3},
    urldate = {2026-02-09},
    journal = {MNRAS},
    author = {Wisnioski, E. and Mendel, J. T. and Leaman, R. and Tsukui, T. and Übler, H. and Förster Schreiber, N. M.},
    month = dec,
    year = {2025},
    pages = {2777--2794},
}

@article{krumholz_unified_2018,
    title = {A unified model for galactic discs: star formation, turbulence driving, and mass transport},
    volume = {477},
    issn = {0035-8711},
    shorttitle = {A unified model for galactic discs},
    url = {https://ui.adsabs.harvard.edu/abs/2018MNRAS.477.2716K/abstract},
    doi = {10.1093/mnras/sty852},
    language = {en},
    number = {2},
    urldate = {2025-12-05},
    journal = {MNRAS},
    author = {Krumholz, Mark R. and Burkhart, Blakesley and Forbes, John C. and Crocker, Roland M.},
    month = jun,
    year = {2018},
    pages = {2716--2740},
}

@ARTICLE{2014ApJ...792L...6V,
       author = {{van der Wel}, A. and {Chang}, Yu-Yen and {Bell}, E.~F. and {Holden}, B.~P. and {Ferguson}, H.~C. and {Giavalisco}, M. and {Rix}, H.-W. and {Skelton}, R. and {Whitaker}, K. and {Momcheva}, I. and {Brammer}, G. and {Kassin}, S.~A. and {Martig}, M. and {Dekel}, A. and {Ceverino}, D. and {Koo}, D.~C. and {Mozena}, M. and {van Dokkum}, P.~G. and {Franx}, M. and {Faber}, S.~M. and {Primack}, J.},
        title = "{Geometry of Star-forming Galaxies from SDSS, 3D-HST, and CANDELS}",
      journal = {\apjl},
         year = 2014,
        month = sep,
       volume = {792},
       number = {1},
          eid = {L6},
        pages = {L6},
          doi = {10.1088/2041-8205/792/1/L6},
archivePrefix = {arXiv},
       eprint = {1407.4233},
 primaryClass = {astro-ph.GA},
       adsurl = {https://ui.adsabs.harvard.edu/abs/2014ApJ...792L...6V}
}

@article{peroux_nature_2017,
    title = {Nature of the absorbing gas associated with a galaxy group at z∼0.4},
    volume = {464},
    issn = {0035-8711},
    url = {https://ui.adsabs.harvard.edu/abs/2017MNRAS.464.2053P/abstract},
    doi = {10.1093/mnras/stw2444},
    language = {en},
    number = {2},
    urldate = {2026-03-02},
    journal = {MNRAS, Volume 464, Issue 2, p.2053-2065},
    author = {Péroux, Céline and Rahmani, Hadi and Quiret, Samuel and Pettini, Max and Kulkarni, Varsha and York, Donald G. and Straka, Lorrie and Husemann, Bernd and Milliard, Bruno},
    month = jan,
    year = {2017},
    pages = {2053},
}

@article{kacprzak_halo_2010,
    title = {Halo {Gas} and {Galaxy} {Disk} {Kinematics} {Derived} from {Observations} and Λ{CDM} {Simulations} of {Mg} {II} {Absorption}-selected {Galaxies} at {Intermediate} {Redshift}},
    volume = {711},
    issn = {0004-637X},
    url = {https://ui.adsabs.harvard.edu/abs/2010ApJ...711..533K/abstract},
    doi = {10.1088/0004-637X/711/2/533},
    language = {en},
    number = {2},
    urldate = {2026-03-03},
    journal = {ApJ},
    author = {Kacprzak, Glenn G. and Churchill, Christopher W. and Ceverino, Daniel and Steidel, Charles C. and Klypin, Anatoly and Murphy, Michael T.},
    month = mar,
    year = {2010},
    pages = {533--558},
}

@article{parlanti_alma_2023,
    title = {{ALMA} hints at the presence of turbulent disk galaxies at z \&gt; 5},
    volume = {673},
    issn = {0004-6361},
    url = {https://ui.adsabs.harvard.edu/abs/2023A%26A...673A.153P/abstract},
    doi = {10.1051/0004-6361/202245603},
    language = {en},
    urldate = {2026-02-27},
    journal = {A\&A, Volume 673, id.A153, pp.},
    author = {Parlanti, E. and Carniani, S. and Pallottini, A. and Cignoni, M. and Cresci, G. and Kohandel, M. and Mannucci, F. and Marconi, A.},
    month = may,
    year = {2023},
    pages = {A153},
}

@article{girard_towards_2019,
    title = {Towards sub-kpc scale kinematics of molecular and ionized gas of star-forming galaxies at z ∼ 1},
    volume = {631},
    issn = {0004-6361},
    url = {https://ui.adsabs.harvard.edu/abs/2019A&A...631A..91G/abstract},
    doi = {10.1051/0004-6361/201935896},
    language = {en},
    urldate = {2026-02-27},
    journal = {A\&A},
    author = {Girard, M. and Dessauges-Zavadsky, M. and Combes, F. and Chisholm, J. and Patrício, V. and Richard, J. and Schaerer, D.},
    month = nov,
    year = {2019},
    pages = {A91},
}

@article{lelli_cold_2023,
    title = {Cold gas disks in main-sequence galaxies at cosmic noon: {Low} turbulence, flat rotation curves, and disk-halo degeneracy},
    volume = {672},
    issn = {0004-6361},
    shorttitle = {Cold gas disks in main-sequence galaxies at cosmic noon},
    url = {https://ui.adsabs.harvard.edu/abs/2023A&A...672A.106L/abstract},
    doi = {10.1051/0004-6361/202245105},
    language = {en},
    urldate = {2026-02-27},
    journal = {A\&A},
    author = {Lelli, Federico and Zhang, Zhi-Yu and Bisbas, Thomas G. and Lin, Lingrui and Papadopoulos, Padelis and Schombert, James M. and Di Teodoro, Enrico and Marasco, Antonino and McGaugh, Stacy S.},
    month = apr,
    year = {2023},
    pages = {A106},
}

@article{liu_600_2023,
    title = {An  600 pc {View} of the {Strongly} {Lensed}, {Massive} {Main}-sequence {Galaxy} {J0901}: {A} {Baryon}-dominated, {Thick} {Turbulent} {Rotating} {Disk} with a {Clumpy} {Cold} {Gas} {Ring} at z = 2.259},
    volume = {942},
    issn = {0004-637X},
    shorttitle = {An  600 pc {View} of the {Strongly} {Lensed}, {Massive} {Main}-sequence {Galaxy} {J0901}},
    url = {https://ui.adsabs.harvard.edu/abs/2023ApJ...942...98L},
    doi = {10.3847/1538-4357/aca46b},
    urldate = {2025-04-07},
    journal = {ApJ},
    publisher = {IOP},
    author = {Liu, Daizhong and Förster Schreiber, N. M. and Genzel, R. and Lutz, D. and Price, S. H. and Lee, L. L. and Baker, Andrew J. and Burkert, A. and Coogan, R. T. and Davies, R. I. and Davies, R. L. and Herrera-Camus, R. and Kodama, Tadayuki and Lee, Minju M. and Nestor, A. and Pulsoni, C. and Renzini, A. and Sharon, Chelsea E. and Shimizu, T. T. and Tacconi, L. J. and Tadaki, Ken-ichi and Übler, H.},
    month = jan,
    year = {2023},
    note = {ADS Bibcode: 2023ApJ...942...98L},
    pages = {98},
}

@article{genzel_evidence_2023,
    title = {Evidence for {Large}-scale, {Rapid} {Gas} {Inflows} in z   2 {Star}-forming {Disks}},
    volume = {957},
    issn = {0004-637X},
    url = {https://ui.adsabs.harvard.edu/abs/2023ApJ...957...48G},
    doi = {10.3847/1538-4357/acef1a},
    urldate = {2025-04-07},
    journal = {ApJ},
    publisher = {IOP},
    author = {Genzel, R. and Jolly, J. -B. and Liu, D. and Price, S. H. and Lee, L. L. and Förster Schreiber, N. M. and Tacconi, L. J. and Herrera-Camus, R. and Barfety, C. and Burkert, A. and Cao, Y. and Davies, R. I. and Dekel, A. and Lee, M. M. and Lutz, D. and Naab, T. and Neri, R. and Nestor Shachar, A. and Pastras, S. and Pulsoni, C. and Renzini, A. and Schuster, K. and Shimizu, T. T. and Stanley, F. and Sternberg, A. and Übler, H.},
    month = nov,
    year = {2023},
    note = {ADS Bibcode: 2023ApJ...957...48G},
    pages = {48},
}

@article{liu_detailed_2024,
    title = {Detailed study of a rare hyperluminous rotating disk in an {Einstein} ring 10 billion years ago},
    volume = {8},
    issn = {2397-3366},
    url = {https://ui.adsabs.harvard.edu/abs/2024NatAs...8.1181L/abstract},
    doi = {10.1038/s41550-024-02296-7},
    language = {en},
    urldate = {2026-03-03},
    journal = {Nature Astronomy, Volume 8, p. 1181-1194},
    author = {Liu, Daizhong and Förster Schreiber, Natascha M. and Harrington, Kevin C. and Lee, Lilian L. and Kamieneski, Patrick S. and Davies, Richard I. and Lutz, Dieter and Renzini, Alvio and Wuyts, Stijn and Tacconi, Linda J. and Genzel, Reinhard and Burkert, Andreas and Herrera-Camus, Rodrigo and Alcalde Pampliega, Belén and Vishwas, Amit and Kaasinen, Melanie and Wang, Q. Daniel and Jiménez-Andrade, Eric F. and Lowenthal, James and Foo, Nicholas and Frye, Brenda L. and Shangguan, Jinyi and Cao, Yixian and Agapito, Guido and Agudo Berbel, Alex and Barfety, Capucine and Baruffolo, Andrea and Berman, Derek and Black, Martin and Bonaglia, Marco and Briguglio, Runa and Carbonaro, Luca and Chapman, Lee and Chen, Jianhang and Cikota, Aleksandar and Concas, Alice and Cooper, Olivia and Cresci, Giovanni and Dallilar, Yigit and Deysenroth, Matthias and Di Antonio, Ivan and Di Cianno, Amico and Di Rico, Gianluca and Doelman, David and Dolci, Mauro and Eisenhauer, Frank and Espejo, Juan and Esposito, Simone and Fantinel, Daniela and Ferruzzi, Debora and Feuchtgruber, Helmut and Gao, Xiaofeng and Garcia Diaz, Carlos and Gillessen, Stefan and Grani, Paolo and Hartl, Michael and Henry, David and Huber, Heinrich and Jolly, Jean-Baptiste and Keller, Christoph U. and Kenworthy, Matthew and Kravchenko, Kateryna and Lee, Minju M. and Lightfoot, John and Lunney, David and Macintosh, Mike and Mannucci, Filippo and Ott, Thomas and Pascale, Massimo and Pastras, Stavros and Pearson, David and Puglisi, Alfio and Pulsoni, Claudia and Rabien, Sebastian and Rau, Christian and Riccardi, Armando and Salasnich, Bernardo and Shimizu, Taro and Snik, Frans and Sturm, Eckhard and Taylor, William and Valentini, Angelo and Waring, Christopher and Wiezorrek, Erich and Xompero, Marco and Yun, Min S.},
    month = sep,
    year = {2024},
    pages = {1181},
}

@article{ubler_ionized_2018,
    title = {Ionized and {Molecular} {Gas} {Kinematics} in a z = 1.4 {Star}-forming {Galaxy}},
    volume = {854},
    issn = {0004-637X},
    url = {https://ui.adsabs.harvard.edu/abs/2018ApJ...854L..24U/abstract},
    doi = {10.3847/2041-8213/aaacfa},
    language = {en},
    number = {2},
    urldate = {2026-02-27},
    journal = {ApJL},
    author = {Übler, H. and Genzel, R. and Tacconi, L. J. and Förster Schreiber, N. M. and Neri, R. and Contursi, A. and Belli, S. and Nelson, E. J. and Lang, P. and Shimizu, T. T. and Davies, R. and Herrera-Camus, R. and Lutz, D. and Plewa, P. M. and Price, S. H. and Schuster, K. and Sternberg, A. and Tadaki, K. and Wisnioski, E. and Wuyts, S.},
    month = feb,
    year = {2018},
    pages = {L24},
}

@article{hung_what_2019,
    title = {What drives the evolution of gas kinematics in star-forming galaxies?},
    volume = {482},
    issn = {0035-8711},
    url = {https://ui.adsabs.harvard.edu/abs/2019MNRAS.482.5125H/abstract},
    doi = {10.1093/mnras/sty2970},
    language = {en},
    number = {4},
    urldate = {2025-12-04},
    journal = {MNRAS},
    author = {Hung, Chao-Ling and Hayward, Christopher C. and Yuan, Tiantian and Boylan-Kolchin, Michael and Faucher-Giguère, Claude-André and Hopkins, Philip F. and Kereš, Dušan and Murray, Norman and Wetzel, Andrew},
    month = feb,
    year = {2019},
    pages = {5125--5137},
}

@article{pillepich_first_2019,
    title = {First results from the {TNG50} simulation: the evolution of stellar and gaseous discs across cosmic time},
    volume = {490},
    issn = {0035-8711},
    shorttitle = {First results from the {TNG50} simulation},
    url = {https://ui.adsabs.harvard.edu/abs/2019MNRAS.490.3196P/abstract},
    doi = {10.1093/mnras/stz2338},
    language = {en},
    number = {3},
    urldate = {2026-03-03},
    journal = {MNRAS},
    author = {Pillepich, Annalisa and Nelson, Dylan and Springel, Volker and Pakmor, Rüdiger and Torrey, Paul and Weinberger, Rainer and Vogelsberger, Mark and Marinacci, Federico and Genel, Shy and van der Wel, Arjen and Hernquist, Lars},
    month = dec,
    year = {2019},
    pages = {3196--3233},
}

@article{jimenez_physical_2023,
    title = {The physical drivers of gas turbulence in simulated disc galaxies},
    volume = {524},
    issn = {0035-8711},
    url = {https://ui.adsabs.harvard.edu/abs/2023MNRAS.524.4346J/abstract},
    doi = {10.1093/mnras/stad2119},
    language = {en},
    number = {3},
    urldate = {2026-03-03},
    journal = {MNRAS},
    author = {Jiménez, Esteban and Lagos, Claudia del P. and Ludlow, Aaron D. and Wisnioski, Emily},
    month = sep,
    year = {2023},
    pages = {4346--4366},
}

@article{green_dynamo_2014,
    title = {{DYNAMO} - {I}. {A} sample of {Hα}-luminous galaxies with resolved kinematics},
    volume = {437},
    issn = {0035-8711},
    url = {https://ui.adsabs.harvard.edu/abs/2014MNRAS.437.1070G/abstract},
    doi = {10.1093/mnras/stt1882},
    language = {en},
    number = {2},
    urldate = {2025-06-23},
    journal = {MNRAS},
    author = {Green, Andrew W. and Glazebrook, Karl and McGregor, Peter J. and Damjanov, Ivana and Wisnioski, Emily and Abraham, Roberto G. and Colless, Matthew and Sharp, Robert G. and Crain, Robert A. and Poole, Gregory B. and McCarthy, Patrick J.},
    month = jan,
    year = {2014},
    pages = {1070--1095},
}

@article{girard_systematic_2021,
    title = {Systematic {Difference} between {Ionized} and {Molecular} {Gas} {Velocity} {Dispersions} in z ∼ 1-2 {Disks} and {Local} {Analogs}},
    volume = {909},
    issn = {0004-637X},
    url = {https://ui.adsabs.harvard.edu/abs/2021ApJ...909...12G/abstract},
    doi = {10.3847/1538-4357/abd5b9},
    language = {en},
    number = {1},
    urldate = {2026-02-27},
    journal = {ApJ},
    author = {Girard, M. and Fisher, D. B. and Bolatto, A. D. and Abraham, R. and Bassett, R. and Glazebrook, K. and Herrera-Camus, R. and Jiménez, E. and Lenkić, L. and Obreschkow, D.},
    month = mar,
    year = {2021},
    pages = {12},
}

@article{epinat_ghasp_2008,
    title = {{GHASP}: an {Hα} kinematic survey of 203 spiral and irregular galaxies - {VII}. {Revisiting} the analysis of {Hα} data cubes for 97 galaxies},
    volume = {390},
    issn = {0035-8711},
    shorttitle = {{GHASP}},
    url = {https://ui.adsabs.harvard.edu/abs/2008MNRAS.390..466E/abstract},
    doi = {10.1111/j.1365-2966.2008.13796.x},
    language = {en},
    number = {2},
    urldate = {2025-06-23},
    journal = {MNRAS},
    author = {Epinat, B. and Amram, P. and Marcelin, M.},
    month = oct,
    year = {2008},
    pages = {466--504},
}

@article{mai_magpi_2024,
    title = {The {MAGPI} {Survey}: the evolution and drivers of gas turbulence in intermediate-redshift galaxies},
    volume = {533},
    issn = {0035-8711},
    shorttitle = {The {MAGPI} {Survey}},
    url = {https://ui.adsabs.harvard.edu/abs/2024MNRAS.533.3878M/abstract},
    doi = {10.1093/mnras/stae2033},
    language = {en},
    number = {4},
    urldate = {2026-03-03},
    journal = {MNRAS, Volume 533, Issue 4, pp.3878-3892},
    author = {Mai, Yifan and Croom, Scott M. and Wisnioski, Emily and Vaughan, Sam P. and Varidel, Mathew R. and Battisti, Andrew J. and Mendel, J. Trevor and Mun, Marcie and Tsukui, Takafumi and Foster, Caroline and Harborne, Katherine E. and Lagos, Claudia D. P. and Wang, Di and Bellstedt, Sabine and Bland-Hawthorn, Joss and Colless, Matthew and D'Eugenio, Francesco and Grasha, Kathryn and Peng, Yingjie and Santucci, Giulia and Sweet, Sarah M. and Thater, Sabine and Valenzuela, Lucas M. and Ziegler, Bodo},
    month = oct,
    year = {2024},
    pages = {3878},
}

@ARTICLE{weng-gasflows-2023,
       author = {{Weng}, Simon and {P{\'e}roux}, C{\'e}line and {Karki}, Arjun and {Augustin}, Ramona and {Kulkarni}, Varsha P. and {Hamanowicz}, Aleksandra and {Zwaan}, Martin and {Sadler}, Elaine M. and {Nelson}, Dylan and {Hayes}, Matthew J. and {Kacprzak}, Glenn G. and {Fox}, Andrew J. and {Bollo}, Victoria and {Casavecchia}, Benedetta and {Szakacs}, Roland},
        title = "{MUSE-ALMA Haloes XI: gas flows in the circumgalactic medium}",
      journal = {\mnras},
         year = 2023,
        month = jul,
       volume = {523},
       number = {1},
        pages = {676-700},
          doi = {10.1093/mnras/stad1462},
archivePrefix = {arXiv},
       eprint = {2305.11219},
 primaryClass = {astro-ph.GA},
       adsurl = {https://ui.adsabs.harvard.edu/abs/2023MNRAS.523..676W}
}

@article{turner_kmos_2017,
    title = {The {KMOS} {Deep} {Survey} ({KDS}) - {I}. {Dynamical} measurements of typical star-forming galaxies at z ≃ 3.5},
    volume = {471},
    issn = {0035-8711},
    url = {https://ui.adsabs.harvard.edu/abs/2017MNRAS.471.1280T/abstract},
    doi = {10.1093/mnras/stx1366},
    language = {en},
    number = {2},
    urldate = {2025-06-23},
    journal = {MNRAS, Volume 471, Issue 2, p.1280-1320},
    author = {Turner, O. J. and Cirasuolo, M. and Harrison, C. M. and McLure, R. J. and Dunlop, J. S. and Swinbank, A. M. and Johnson, H. L. and Sobral, D. and Matthee, J. and Sharples, R. M.},
    month = oct,
    year = {2017},
    pages = {1280},
}

@article{price_mosdef_2020,
    title = {The {MOSDEF} {Survey}: {Kinematic} and {Structural} {Evolution} of {Star}-forming {Galaxies} at 1.4 ≤ z ≤ 3.8},
    volume = {894},
    issn = {0004-637X},
    shorttitle = {The {MOSDEF} {Survey}},
    url = {https://ui.adsabs.harvard.edu/abs/2020ApJ...894...91P/abstract},
    doi = {10.3847/1538-4357/ab7990},
    language = {en},
    number = {2},
    urldate = {2026-03-03},
    journal = {ApJ},
    author = {Price, Sedona H. and Kriek, Mariska and Barro, Guillermo and Shapley, Alice E. and Reddy, Naveen A. and Freeman, William R. and Coil, Alison L. and Shivaei, Irene and Azadi, Mojegan and de Groot, Laura and Siana, Brian and Mobasher, Bahram and Sanders, Ryan L. and Leung, Gene C. K. and Fetherolf, Tara and Zick, Tom O. and Übler, Hannah and Förster Schreiber, Natascha M.},
    month = may,
    year = {2020},
    pages = {91},
}

@article{rizzo_alma-alpaka_2023,
    title = {The {ALMA}-{ALPAKA} survey. {I}. {High}-resolution {CO} and [{CI}] kinematics of star-forming galaxies at z = 0.5-3.5},
    volume = {679},
    issn = {0004-6361},
    url = {https://ui.adsabs.harvard.edu/abs/2023A&A...679A.129R},
    doi = {10.1051/0004-6361/202346444},
    urldate = {2025-04-07},
    journal = {A\&A},
    author = {Rizzo, F. and Roman-Oliveira, F. and Fraternali, F. and Frickmann, D. and Valentino, F. M. and Brammer, G. and Zanella, A. and Kokorev, V. and Popping, G. and Whitaker, K. E. and Kohandel, M. and Magdis, G. E. and Di Mascolo, L. and Ikeda, R. and Jin, S. and Toft, S.},
    month = nov,
    year = {2023},
    note = {ADS Bibcode: 2023A\&A...679A.129R},
    pages = {A129},
}

@article{danhaive_dawn_2025,
    title = {The dawn of discs: unveiling the turbulent ionized gas kinematics of the galaxy population at z ∼ 4─6 with {JWST}/{NIRCam} grism spectroscopy},
    volume = {543},
    issn = {0035-8711},
    shorttitle = {The dawn of discs},
    url = {https://ui.adsabs.harvard.edu/abs/2025MNRAS.543.3249D/abstract},
    doi = {10.1093/mnras/staf1540},
    language = {en},
    number = {4},
    urldate = {2026-03-03},
    journal = {MNRAS, Volume 543, Issue 4, pp. 3249-3302, 54 pp.},
    author = {Danhaive, A. Lola and Tacchella, Sandro and Übler, Hannah and de Graaff, Anna and Egami, Eiichi and Johnson, Benjamin D. and Sun, Fengwu and Arribas, Santiago and Bunker, Andrew J. and Carniani, Stefano and Jones, Gareth C. and Maiolino, Roberto and McClymont, William and Parlanti, Eleonora and Simmonds, Charlotte and Villanueva, Natalia C. and Baker, William M. and Jaffe, Daniel T. and Eisenstein, Daniel and Hainline, Kevin and Helton, Jakob M. and Ji, Zhiyuan and Lin, Xiaojing and Liu, Yichen and Puskás, Dávid and Rieke, Marcia and Rinaldi, Pierluigi and Robertson, Brant and Scholz, Jan and Williams, Christina C. and Willmer, Christopher N. A.},
    month = nov,
    year = {2025},
    pages = {3249},
}

@article{de_graaff_ionised_2024,
    title = {Ionised gas kinematics and dynamical masses of z ≳ 6 galaxies from {JADES}/{NIRSpec} high-resolution spectroscopy},
    volume = {684},
    issn = {0004-6361},
    url = {https://ui.adsabs.harvard.edu/abs/2024A%26A...684A..87D/abstract},
    doi = {10.1051/0004-6361/202347755},
    language = {en},
    urldate = {2026-03-03},
    journal = {A\&A, Volume 684, id.A87, pp.},
    author = {de Graaff, Anna and Rix, Hans-Walter and Carniani, Stefano and Suess, Katherine A. and Charlot, Stéphane and Curtis-Lake, Emma and Arribas, Santiago and Baker, William M. and Boyett, Kristan and Bunker, Andrew J. and Cameron, Alex J. and Chevallard, Jacopo and Curti, Mirko and Eisenstein, Daniel J. and Franx, Marijn and Hainline, Kevin and Hausen, Ryan and Ji, Zhiyuan and Johnson, Benjamin D. and Jones, Gareth C. and Maiolino, Roberto and Maseda, Michael V. and Nelson, Erica and Parlanti, Eleonora and Rawle, Tim and Robertson, Brant and Tacchella, Sandro and Übler, Hannah and Williams, Christina C. and Willmer, Christopher N. A. and Willott, Chris},
    month = apr,
    year = {2024},
    pages = {A87},
}

@article{nestor_shachar_rc100_2023,
    title = {{RC100}: {Rotation} {Curves} of 100 {Massive} {Star}-forming {Galaxies} at z = 0.6-2.5 {Reveal} {Little} {Dark} {Matter} on {Galactic} {Scales}},
    volume = {944},
    issn = {0004-637X},
    shorttitle = {{RC100}},
    url = {https://ui.adsabs.harvard.edu/abs/2023ApJ...944...78N/abstract},
    doi = {10.3847/1538-4357/aca9cf},
    language = {en},
    number = {1},
    urldate = {2025-06-24},
    journal = {ApJ},
    author = {Nestor Shachar, A. and Price, S. H. and Förster Schreiber, N. M. and Genzel, R. and Shimizu, T. T. and Tacconi, L. J. and Übler, H. and Burkert, A. and Davies, R. I. and Dekel, A. and Herrera-Camus, R. and Lee, L. L. and Liu, D. and Lutz, D. and Naab, T. and Neri, R. and Renzini, A. and Saglia, R. and Schuster, K. F. and Sternberg, A. and Wisnioski, E. and Wuyts, S.},
    month = feb,
    year = {2023},
    pages = {78},
}

@article{sharma_magpi_2026,
    title = {The {MAGPI} {Survey}: co-evolution of baryons and dark matter in star-forming disk-like galaxies at \$0.1 {\textbackslash}lesssim z {\textbackslash}lesssim 0.85\$},
    shorttitle = {The {MAGPI} {Survey}},
    url = {https://ui.adsabs.harvard.edu/abs/2026arXiv260112315S/abstract},
    doi = {10.48550/arXiv.2601.12315},
    language = {en},
    urldate = {2026-03-03},
    journal = {JCAP subm.},
    author = {Sharma, Gauri and Battisti, Andrew J. and Wisnioski, Emily and Mendel, J. Trevor and Bellstedt, Sabine and Lagos, Claudia Del P. and Foster, Caroline and Poci, Adriano and Harborne, Katherine E. and Bagge, Ryan and Barsanti, Stefania and Bland-Hawthorn, Joss and Breda, Iris and Croom, Scott M. and Glazebrook, Karl and Mai, Yifan and Sweet, Sarah M. and Thater, Sabine and Valenzuela, Lucas M. and van de Ven, Glenn and Yi, Sukyoung and Zafar, Tayyaba and Ziegler, Bodo},
    month = jan,
    year = {2026},
    pages = {arXiv:2601.12315},
}

@article{forster_schreiber_sinszc-sinf_2018,
    title = {The {SINS}/{zC}-{SINF} {Survey} of z ∼ 2 {Galaxy} {Kinematics}: {SINFONI} {Adaptive} {Optics}-assisted {Data} and {Kiloparsec}-scale {Emission}-line {Properties}},
    volume = {238},
    issn = {0067-0049},
    shorttitle = {The {SINS}/{zC}-{SINF} {Survey} of z ∼ 2 {Galaxy} {Kinematics}},
    url = {https://ui.adsabs.harvard.edu/abs/2018ApJS..238...21F},
    doi = {10.3847/1538-4365/aadd49},
    urldate = {2025-02-27},
    journal = {ApJ Supplement Series},
    publisher = {IOP},
    author = {Förster Schreiber, N. M. and Renzini, A. and Mancini, C. and Genzel, R. and Bouché, N. and Cresci, G. and Hicks, E. K. S. and Lilly, S. J. and Peng, Y. and Burkert, A. and Carollo, C. M. and Cimatti, A. and Daddi, E. and Davies, R. I. and Genel, S. and Kurk, J. D. and Lang, P. and Lutz, D. and Mainieri, V. and McCracken, H. J. and Mignoli, M. and Naab, T. and Oesch, P. and Pozzetti, L. and Scodeggio, M. and Shapiro Griffin, K. and Shapley, A. E. and Sternberg, A. and Tacchella, S. and Tacconi, L. J. and Wuyts, S. and Zamorani, G.},
    month = oct,
    year = {2018},
    note = {ADS Bibcode: 2018ApJS..238...21F},
    pages = {21},
}

@article{forster_schreiber_sins_2009,
    title = {The {SINS} {Survey}: {SINFONI} {Integral} {Field} {Spectroscopy} of z {\textasciitilde} 2 {Star}-forming {Galaxies}},
    volume = {706},
    issn = {0004-637X},
    shorttitle = {The {SINS} {Survey}},
    url = {https://ui.adsabs.harvard.edu/abs/2009ApJ...706.1364F},
    doi = {10.1088/0004-637X/706/2/1364},
    urldate = {2025-04-07},
    journal = {ApJ},
    publisher = {IOP},
    author = {Förster Schreiber, N. M. and Genzel, R. and Bouché, N. and Cresci, G. and Davies, R. and Buschkamp, P. and Shapiro, K. and Tacconi, L. J. and Hicks, E. K. S. and Genel, S. and Shapley, A. E. and Erb, D. K. and Steidel, C. C. and Lutz, D. and Eisenhauer, F. and Gillessen, S. and Sternberg, A. and Renzini, A. and Cimatti, A. and Daddi, E. and Kurk, J. and Lilly, S. and Kong, X. and Lehnert, M. D. and Nesvadba, N. and Verma, A. and McCracken, H. and Arimoto, N. and Mignoli, M. and Onodera, M.},
    month = dec,
    year = {2009},
    note = {ADS Bibcode: 2009ApJ...706.1364F},
    pages = {1364--1428},
}

@article{kassin_epoch_2012,
    title = {The {Epoch} of {Disk} {Settling}: z {\textasciitilde} 1 to {Now}},
    volume = {758},
    issn = {0004-637X},
    shorttitle = {The {Epoch} of {Disk} {Settling}},
    url = {https://ui.adsabs.harvard.edu/abs/2012ApJ...758..106K/abstract},
    doi = {10.1088/0004-637X/758/2/106},
    language = {en},
    number = {2},
    urldate = {2026-03-17},
    journal = {ApJ},
    author = {Kassin, Susan A. and Weiner, Benjamin J. and Faber, S. M. and Gardner, Jonathan P. and Willmer, C. N. A. and Coil, Alison L. and Cooper, Michael C. and Devriendt, Julien and Dutton, Aaron A. and Guhathakurta, Puragra and Koo, David C. and Metevier, A. J. and Noeske, Kai G. and Primack, Joel R.},
    month = oct,
    year = {2012},
    pages = {106},
}

@article{wisnioski_kmos3d_2015,
    title = {The {KMOS}{\textless}{SUP}{\textgreater}{3D}{\textless}/{SUP}{\textgreater} {Survey}: {Design}, {First} {Results}, and the {Evolution} of {Galaxy} {Kinematics} from 0.7 \&lt;= z \&lt;= 2.7},
    volume = {799},
    issn = {0004-637X},
    shorttitle = {The {KMOS}{\textless}{SUP}{\textgreater}{3D}{\textless}/{SUP}{\textgreater} {Survey}},
    url = {https://ui.adsabs.harvard.edu/abs/2015ApJ...799..209W/abstract},
    doi = {10.1088/0004-637X/799/2/209},
    language = {en},
    number = {2},
    urldate = {2026-03-17},
    journal = {ApJ},
    author = {Wisnioski, E. and Förster Schreiber, N. M. and Wuyts, S. and Wuyts, E. and Bandara, K. and Wilman, D. and Genzel, R. and Bender, R. and Davies, R. and Fossati, M. and Lang, P. and Mendel, J. T. and Beifiori, A. and Brammer, G. and Chan, J. and Fabricius, M. and Fudamoto, Y. and Kulkarni, S. and Kurk, J. and Lutz, D. and Nelson, E. J. and Momcheva, I. and Rosario, D. and Saglia, R. and Seitz, S. and Tacconi, L. J. and van Dokkum, P. G.},
    month = feb,
    year = {2015},
    pages = {209},
}

@article{wisnioski_kmos3d_2019,
    title = {The {KMOS}{\textless}{SUP}{\textgreater}{3D}{\textless}/{SUP}{\textgreater} {Survey}: {Data} {Release} and {Final} {Survey} {Paper}},
    volume = {886},
    issn = {0004-637X},
    shorttitle = {The {KMOS}{\textless}{SUP}{\textgreater}{3D}{\textless}/{SUP}{\textgreater} {Survey}},
    url = {https://ui.adsabs.harvard.edu/abs/2019ApJ...886..124W/abstract},
    doi = {10.3847/1538-4357/ab4db8},
    language = {en},
    number = {2},
    urldate = {2026-03-17},
    journal = {ApJ},
    author = {Wisnioski, E. and Förster Schreiber, N. M. and Fossati, M. and Mendel, J. T. and Wilman, D. and Genzel, R. and Bender, R. and Wuyts, S. and Davies, R. L. and Übler, H. and Bandara, K. and Beifiori, A. and Belli, S. and Brammer, G. and Chan, J. and Davies, R. I. and Fabricius, M. and Galametz, A. and Lang, P. and Lutz, D. and Nelson, E. J. and Momcheva, I. and Price, S. and Rosario, D. and Saglia, R. and Seitz, S. and Shimizu, T. and Tacconi, L. J. and Tadaki, K. and van Dokkum, P. G. and Wuyts, E.},
    month = dec,
    year = {2019},
    pages = {124},
}

@article{forster_schreiber_sinfoni_2006,
    title = {{SINFONI} {Integral} {Field} {Spectroscopy} of z {\textasciitilde} 2 {UV}-selected {Galaxies}: {Rotation} {Curves} and {Dynamical} {Evolution}},
    volume = {645},
    issn = {0004-637X},
    shorttitle = {{SINFONI} {Integral} {Field} {Spectroscopy} of z {\textasciitilde} 2 {UV}-selected {Galaxies}},
    url = {https://ui.adsabs.harvard.edu/abs/2006ApJ...645.1062F},
    doi = {10.1086/504403},
    urldate = {2025-04-07},
    journal = {ApJ},
    publisher = {IOP},
    author = {Förster Schreiber, N. M. and Genzel, R. and Lehnert, M. D. and Bouché, N. and Verma, A. and Erb, D. K. and Shapley, A. E. and Steidel, C. C. and Davies, R. and Lutz, D. and Nesvadba, N. and Tacconi, L. J. and Eisenhauer, F. and Abuter, R. and Gilbert, A. and Gillessen, S. and Sternberg, A.},
    month = jul,
    year = {2006},
    note = {ADS Bibcode: 2006ApJ...645.1062F},
    pages = {1062--1075},
}

@article{tacconi_evolution_2020,
    title = {The {Evolution} of the {Star}-{Forming} {Interstellar} {Medium} {Across} {Cosmic} {Time}},
    volume = {58},
    issn = {0066-4146},
    url = {https://ui.adsabs.harvard.edu/abs/2020ARA&A..58..157T/abstract},
    doi = {10.1146/annurev-astro-082812-141034},
    language = {en},
    urldate = {2026-03-17},
    journal = {ARA\&A},
    author = {Tacconi, Linda J. and Genzel, Reinhard and Sternberg, Amiel},
    month = aug,
    year = {2020},
    pages = {157--203},
}

@article{dekel_mass_2020,
    title = {A mass threshold for galactic gas discs by spin flips},
    volume = {493},
    issn = {0035-8711},
    url = {https://ui.adsabs.harvard.edu/abs/2020MNRAS.493.4126D/abstract},
    doi = {10.1093/mnras/staa470},
    language = {en},
    number = {3},
    urldate = {2026-03-17},
    journal = {MNRAS},
    author = {Dekel, Avishai and Ginzburg, Omri and Jiang, Fangzhou and Freundlich, Jonathan and Lapiner, Sharon and Ceverino, Daniel and Primack, Joel},
    month = apr,
    year = {2020},
    pages = {4126--4142},
}

@article{levy_edge-califa_2018,
    title = {The {EDGE}-{CALIFA} {Survey}: {Molecular} and {Ionized} {Gas} {Kinematics} in {Nearby} {Galaxies}},
    volume = {860},
    issn = {0004-637X},
    shorttitle = {The {EDGE}-{CALIFA} {Survey}},
    url = {https://ui.adsabs.harvard.edu/abs/2018ApJ...860...92L/abstract},
    doi = {10.3847/1538-4357/aac2e5},
    language = {en},
    number = {2},
    urldate = {2026-03-09},
    journal = {ApJ},
    author = {Levy, Rebecca C. and Bolatto, Alberto D. and Teuben, Peter and Sánchez, Sebastián F. and Barrera-Ballesteros, Jorge K. and Blitz, Leo and Colombo, Dario and García-Benito, Rubén and Herrera-Camus, Rodrigo and Husemann, Bernd and Kalinova, Veselina and Lan, Tian and Leung, Gigi Y. C. and Mast, Damián and Utomo, Dyas and van de Ven, Glenn and Vogel, Stuart N. and Wong, Tony},
    month = jun,
    year = {2018},
    pages = {92},
}

@article{rizzo_dynamically_2020,
    title = {A dynamically cold disk galaxy in the early {Universe}},
    volume = {584},
    issn = {0028-0836},
    url = {https://ui.adsabs.harvard.edu/abs/2020Natur.584..201R/abstract},
    doi = {10.1038/s41586-020-2572-6},
    language = {en},
    number = {7820},
    urldate = {2026-03-17},
    journal = {Nature},
    author = {Rizzo, F. and Vegetti, S. and Powell, D. and Fraternali, F. and McKean, J. P. and Stacey, H. R. and White, S. D. M.},
    month = aug,
    year = {2020},
    pages = {201--204},
}

@article{rowland_rebels-25_2024,
    title = {{REBELS}-25: discovery of a dynamically cold disc galaxy at z = 7.31},
    volume = {535},
    issn = {0035-8711},
    shorttitle = {{REBELS}-25},
    url = {https://ui.adsabs.harvard.edu/abs/2024MNRAS.535.2068R/abstract},
    doi = {10.1093/mnras/stae2217},
    language = {en},
    number = {3},
    urldate = {2026-03-17},
    journal = {MNRAS},
    author = {Rowland, Lucie E. and Hodge, Jacqueline and Bouwens, Rychard and Mancera Piña, Pavel E. and Hygate, Alexander and Algera, Hiddo and Aravena, Manuel and Bowler, Rebecca and da Cunha, Elisabete and Dayal, Pratika and Ferrara, Andrea and Herard-Demanche, Thomas and Inami, Hanae and van Leeuwen, Ivana and de Looze, Ilse and Oesch, Pascal and Pallottini, Andrea and Phillips, Siân and Rybak, Matus and Schouws, Sander and Smit, Renske and Sommovigo, Laura and Stefanon, Mauro and van der Werf, Paul},
    month = dec,
    year = {2024},
    pages = {2068--2091},
}

@article{norris_discovery_2021,
    title = {Discovery of extremely low-metallicity circumgalactic gas at z = 0.5 towards {Q0454}-220},
    volume = {506},
    issn = {0035-8711},
    url = {https://ui.adsabs.harvard.edu/abs/2021MNRAS.506.5640N/abstract},
    doi = {10.1093/mnras/stab1812},
    language = {en},
    number = {4},
    urldate = {2025-04-09},
    journal = {MNRAS},
    author = {Norris, Jackson M. and Muzahid, Sowgat and Charlton, Jane C. and Kacprzak, Glenn G. and Wakker, Bart P. and Churchill, Christopher W.},
    month = oct,
    year = {2021},
    pages = {5640--5657},
}

@article{cappellari_full_2023,
    title = {Full spectrum fitting with photometry in {PPXF}: stellar population versus dynamical masses, non-parametric star formation history and metallicity for 3200 {LEGA}-{C} galaxies at redshift z ≈ 0.8},
    volume = {526},
    issn = {0035-8711},
    shorttitle = {Full spectrum fitting with photometry in {PPXF}},
    url = {https://ui.adsabs.harvard.edu/abs/2023MNRAS.526.3273C/abstract},
    doi = {10.1093/mnras/stad2597},
    language = {en},
    number = {3},
    urldate = {2026-03-18},
    journal = {MNRAS},
    author = {Cappellari, Michele},
    month = dec,
    year = {2023},
    pages = {3273--3300},
}

@article{verro_modelling_2022,
    title = {Modelling simple stellar populations in the near-ultraviolet to near-infrared with the {X}-shooter {Spectral} {Library} ({XSL})},
    volume = {661},
    issn = {0004-6361},
    url = {https://ui.adsabs.harvard.edu/abs/2022A&A...661A..50V/abstract},
    doi = {10.1051/0004-6361/202142387},
    language = {en},
    urldate = {2026-03-18},
    journal = {A\&A},
    author = {Verro, K. and Trager, S. C. and Peletier, R. F. and Lançon, A. and Arentsen, A. and Chen, Y.-P. and Coelho, P. R. T. and Dries, M. and Falcón-Barroso, J. and Gonneau, A. and Lyubenova, M. and Martins, L. and Prugniel, P. and Sánchez-Blázquez, P. and Vazdekis, A.},
    month = may,
    year = {2022},
    pages = {A50},
}

@article{bouche_muse_2022,
    title = {The {MUSE} {Extremely} {Deep} {Field}: {Evidence} for {SFR}-induced cores in dark-matter dominated galaxies at z ≃ 1},
    volume = {658},
    issn = {0004-6361},
    shorttitle = {The {MUSE} {Extremely} {Deep} {Field}},
    url = {https://ui.adsabs.harvard.edu/abs/2022A&A...658A..76B/abstract},
    doi = {10.1051/0004-6361/202141762},
    language = {en},
    urldate = {2026-03-23},
    journal = {A\&A},
    author = {Bouché, Nicolas F. and Bera, Samuel and Krajnović, Davor and Emsellem, Eric and Mercier, Wilfried and Schaye, Joop and Epinat, Benoît and Richard, Johan and Zoutendijk, Sebastiaan L. and Abril-Melgarejo, Valentina and Brinchmann, Jarle and Bacon, Roland and Contini, Thierry and Boogaard, Leindert and Wisotzki, Lutz and Maseda, Michael and Steinmetz, Matthias},
    month = feb,
    year = {2022},
    pages = {A76},
}

@article{martinsson_diskmass_2013,
    title = {The {DiskMass} {Survey}. {VI}. {Gas} and stellar kinematics in spiral galaxies from {PPak} integral-field spectroscopy},
    volume = {557},
    issn = {0004-6361},
    url = {https://ui.adsabs.harvard.edu/abs/2013A&A...557A.130M/abstract},
    doi = {10.1051/0004-6361/201220515},
    language = {en},
    urldate = {2026-03-23},
    journal = {A\&A},
    author = {Martinsson, Thomas P. K. and Verheijen, Marc A. W. and Westfall, Kyle B. and Bershady, Matthew A. and Schechtman-Rook, Andrew and Andersen, David R. and Swaters, Rob A.},
    month = sep,
    year = {2013},
    pages = {A130},
}

@article{barnabe_swells_2012,
    title = {The {SWELLS} survey - {IV}. {Precision} measurements of the stellar and dark matter distributions in a spiral lens galaxy},
    volume = {423},
    issn = {0035-8711},
    url = {https://ui.adsabs.harvard.edu/abs/2012MNRAS.423.1073B/abstract},
    doi = {10.1111/j.1365-2966.2012.20934.x},
    language = {en},
    number = {2},
    urldate = {2026-03-23},
    journal = {MNRAS},
    author = {Barnabè, Matteo and Dutton, Aaron A. and Marshall, Philip J. and Auger, Matthew W. and Brewer, Brendon J. and Treu, Tommaso and Bolton, Adam S. and Koo, David C. and Koopmans, Léon V. E.},
    month = jun,
    year = {2012},
    pages = {1073--1088},
}

@article{yttergren_kinematics_2025,
    title = {Kinematics of synthetically observed high-z rotating discs: reliability and biases of {3D} fitting tools},
    volume = {543},
    issn = {0035-8711},
    shorttitle = {Kinematics of synthetically observed high-z rotating discs},
    url = {https://ui.adsabs.harvard.edu/abs/2025MNRAS.543.3103Y/abstract},
    doi = {10.1093/mnras/staf1489},
    language = {en},
    number = {4},
    urldate = {2026-03-23},
    journal = {MNRAS},
    author = {Yttergren, M. and Knudsen, K. K. and Molina, J. and Jones, G. C. and Kade, K. and Scholtz, J. and Bewketu Belete, A.},
    month = nov,
    year = {2025},
    pages = {3103--3122},
}

@article{bacchini_3d_2024,
    title = {A {3D} view on the local gravitational instability of cold gas discs in star-forming galaxies at 0 ≲ z ≲ 5},
    volume = {687},
    issn = {0004-6361},
    url = {https://ui.adsabs.harvard.edu/abs/2024A&A...687A.115B/abstract},
    doi = {10.1051/0004-6361/202449925},
    language = {en},
    urldate = {2026-03-24},
    journal = {A\&A},
    author = {Bacchini, C. and Nipoti, C. and Iorio, G. and Roman-Oliveira, F. and Rizzo, F. and Mancera Piña, P. E. and Marasco, A. and Zanella, A. and Lelli, F.},
    month = jul,
    year = {2024},
    pages = {A115},
}

@article{genzel_rotation_2020,
    title = {Rotation {Curves} in z ∼ 1-2 {Star}-forming {Disks}: {Evidence} for {Cored} {Dark} {Matter} {Distributions}},
    volume = {902},
    issn = {0004-637X},
    shorttitle = {Rotation {Curves} in z ∼ 1-2 {Star}-forming {Disks}},
    url = {https://ui.adsabs.harvard.edu/abs/2020ApJ...902...98G/abstract},
    doi = {10.3847/1538-4357/abb0ea},
    language = {en},
    number = {2},
    urldate = {2026-02-09},
    journal = {ApJ},
    author = {Genzel, R. and Price, S. H. and Übler, H. and Förster Schreiber, N. M. and Shimizu, T. T. and Tacconi, L. J. and Bender, R. and Burkert, A. and Contursi, A. and Coogan, R. and Davies, R. L. and Davies, R. I. and Dekel, A. and Herrera-Camus, R. and Lee, M.-J. and Lutz, D. and Naab, T. and Neri, R. and Nestor, A. and Renzini, A. and Saglia, R. and Schuster, K. and Sternberg, A. and Wisnioski, E. and Wuyts, S.},
    month = oct,
    year = {2020},
    pages = {98},
}

@article{bland-hawthorn_turbulent_2024,
    title = {Turbulent {Gas}-rich {Disks} at {High} {Redshift}: {Bars} and {Bulges} in a {Radial} {Shear} {Flow}},
    volume = {968},
    issn = {0004-637X},
    shorttitle = {Turbulent {Gas}-rich {Disks} at {High} {Redshift}},
    url = {https://ui.adsabs.harvard.edu/abs/2024ApJ...968...86B/abstract},
    doi = {10.3847/1538-4357/ad4118},
    language = {en},
    number = {2},
    urldate = {2026-03-25},
    journal = {ApJ},
    author = {Bland-Hawthorn, Joss and Tepper-Garcia, Thor and Agertz, Oscar and Federrath, Christoph},
    month = jun,
    year = {2024},
    pages = {86},
}

@article{newman_sinszc-sinf_2013,
    title = {The {SINS}/{zC}-{SINF} {Survey} of z {\textasciitilde} {2Galaxy} {Kinematics}: {The} {Nature} of {Dispersion}-dominated {Galaxies}},
    volume = {767},
    issn = {0004-637X},
    shorttitle = {The {SINS}/{zC}-{SINF} {Survey} of z {\textasciitilde} {2Galaxy} {Kinematics}},
    url = {https://ui.adsabs.harvard.edu/abs/2013ApJ...767..104N/abstract},
    doi = {10.1088/0004-637X/767/2/104},
    language = {en},
    number = {2},
    urldate = {2026-03-26},
    journal = {ApJ},
    author = {Newman, Sarah F. and Genzel, Reinhard and Förster Schreiber, Natascha M. and Shapiro Griffin, Kristen and Mancini, Chiara and Lilly, Simon J. and Renzini, Alvio and Bouché, Nicolas and Burkert, Andreas and Buschkamp, Peter and Carollo, C. Marcella and Cresci, Giovanni and Davies, Ric and Eisenhauer, Frank and Genel, Shy and Hicks, Erin K. S. and Kurk, Jaron and Lutz, Dieter and Naab, Thorsten and Peng, Yingjie and Sternberg, Amiel and Tacconi, Linda J. and Wuyts, Stijn and Zamorani, Gianni and Vergani, Daniela},
    month = apr,
    year = {2013},
    pages = {104},
}

@article{law_kiloparsec-scale_2009,
    title = {The {Kiloparsec}-scale {Kinematics} of {High}-redshift {Star}-forming {Galaxies}},
    volume = {697},
    issn = {0004-637X},
    url = {https://ui.adsabs.harvard.edu/abs/2009ApJ...697.2057L/abstract},
    doi = {10.1088/0004-637X/697/2/2057},
    language = {en},
    number = {2},
    urldate = {2026-03-26},
    journal = {ApJ},
    author = {Law, David R. and Steidel, Charles C. and Erb, Dawn K. and Larkin, James E. and Pettini, Max and Shapley, Alice E. and Wright, Shelley A.},
    month = jun,
    year = {2009},
    pages = {2057--2082},
}

@article{pastras_noema3d_2025,
    title = {{NOEMA}{\textless}{SUP}{\textgreater}{3D}{\textless}/{SUP}{\textgreater}: {A} first kiloparsec resolution study of a z ∼ 1.5 main sequence barred galaxy channeling gas into a growing bulge},
    volume = {704},
    issn = {0004-6361},
    shorttitle = {{NOEMA}{\textless}{SUP}{\textgreater}{3D}{\textless}/{SUP}{\textgreater}},
    url = {https://ui.adsabs.harvard.edu/abs/2025A&A...704A.329P/abstract},
    doi = {10.1051/0004-6361/202555430},
    language = {en},
    urldate = {2026-03-26},
    journal = {A\&A},
    author = {Pastras, Stavros and Genzel, Reinhard and Tacconi, Linda J. and Schuster, Karl and Neri, Roberto and Förster Schreiber, Natascha M. and Naab, Thorsten and Barfety, Capucine and Burkert, Andreas and Cao, Yixian and Chen, Jianhang and Combes, Françoise and Davies, Ric and Eisenhauer, Frank and Espejo Salcedo, Juan M. and García-Burillo, Santiago and Herrera-Camus, Rodrigo and Jolly, Jean-Baptiste and Lee, Lilian L. and Lee, Minju M. and Liu, Daizhong and Lutz, Dieter and Nestor Shachar, Amit and Parlanti, Eleonora and Price, Sedona H. and Pulsoni, Claudia and Renzini, Alvio and Scaloni, Letizia and Shimizu, Taro T. and Springel, Volker and Sternberg, Amiel and Sturm, Eckhard and Tozzi, Giulia and Wuyts, Stijn and Übler, Hannah},
    month = dec,
    year = {2025},
    pages = {A329},
}

@article{wuyts_kmos3d_2016,
    title = {{KMOS3D}: {Dynamical} {Constraints} on the {Mass} {Budget} in {Early} {Star}-forming {Disks}},
    volume = {831},
    issn = {0004-637X},
    shorttitle = {{KMOS3D}},
    url = {https://ui.adsabs.harvard.edu/abs/2016ApJ...831..149W/abstract},
    doi = {10.3847/0004-637X/831/2/149},
    language = {en},
    number = {2},
    urldate = {2026-03-26},
    journal = {ApJ},
    author = {Wuyts, Stijn and Förster Schreiber, Natascha M. and Wisnioski, Emily and Genzel, Reinhard and Burkert, Andreas and Bandara, Kaushala and Beifiori, Alessandra and Belli, Sirio and Bender, Ralf and Brammer, Gabriel B. and Chan, Jeffrey and Davies, Ric and Fossati, Matteo and Galametz, Audrey and Kulkarni, Sandesh K. and Lang, Philipp and Lutz, Dieter and Mendel, J. Trevor and Momcheva, Ivelina G. and Naab, Thorsten and Nelson, Erica J. and Saglia, Roberto P. and Seitz, Stella and Tacconi, Linda J. and Tadaki, Ken-ichi and Übler, Hannah and van Dokkum, Pieter G. and Wilman, David J. and Wuyts, Eva},
    month = nov,
    year = {2016},
    pages = {149},
}

@article{bouche_impact_2010,
    title = {The {Impact} of {Cold} {Gas} {Accretion} {Above} a {Mass} {Floor} on {Galaxy} {Scaling} {Relations}},
    volume = {718},
    issn = {0004-637X},
    url = {https://ui.adsabs.harvard.edu/abs/2010ApJ...718.1001B/abstract},
    doi = {10.1088/0004-637X/718/2/1001},
    language = {en},
    number = {2},
    urldate = {2026-03-26},
    journal = {ApJ},
    author = {Bouché, N. and Dekel, A. and Genzel, R. and Genel, S. and Cresci, G. and Förster Schreiber, N. M. and Shapiro, K. L. and Davies, R. I. and Tacconi, L.},
    month = aug,
    year = {2010},
    pages = {1001--1018},
}

@article{dave_analytic_2012,
    title = {An analytic model for the evolution of the stellar, gas and metal content of galaxies},
    volume = {421},
    issn = {0035-8711},
    url = {https://ui.adsabs.harvard.edu/abs/2012MNRAS.421...98D/abstract},
    doi = {10.1111/j.1365-2966.2011.20148.x},
    language = {en},
    number = {1},
    urldate = {2026-03-26},
    journal = {MNRAS},
    author = {Davé, Romeel and Finlator, Kristian and Oppenheimer, Benjamin D.},
    month = mar,
    year = {2012},
    pages = {98--107},
}

@article{lilly_gas_2013,
    title = {Gas {Regulation} of {Galaxies}: {The} {Evolution} of the {Cosmic} {Specific} {Star} {Formation} {Rate}, the {Metallicity}-{Mass}-{Star}-formation {Rate} {Relation}, and the {Stellar} {Content} of {Halos}},
    volume = {772},
    issn = {0004-637X},
    shorttitle = {Gas {Regulation} of {Galaxies}},
    url = {https://ui.adsabs.harvard.edu/abs/2013ApJ...772..119L/abstract},
    doi = {10.1088/0004-637X/772/2/119},
    language = {en},
    number = {2},
    urldate = {2026-03-26},
    journal = {ApJ},
    author = {Lilly, Simon J. and Carollo, C. Marcella and Pipino, Antonio and Renzini, Alvio and Peng, Yingjie},
    month = aug,
    year = {2013},
    pages = {119},
}

@article{cortese_alma_2017,
    title = {{ALMA} {Shows} that {Gas} {Reservoirs} of {Star}-forming {Disks} over the {Past} 3 {Billion} {Years} {Are} {Not} {Predominantly} {Molecular}},
    volume = {848},
    issn = {0004-637X},
    url = {https://ui.adsabs.harvard.edu/abs/2017ApJ...848L...7C/abstract},
    doi = {10.3847/2041-8213/aa8cc3},
    language = {en},
    number = {1},
    urldate = {2026-03-26},
    journal = {ApJ},
    author = {Cortese, Luca and Catinella, Barbara and Janowiecki, Steven},
    month = oct,
    year = {2017},
    pages = {L7},
}

@article{rathjen_silcc_2023,
    title = {{SILCC} - {VII}. {Gas} kinematics and multiphase outflows of the simulated {ISM} at high gas surface densities},
    volume = {522},
    issn = {0035-8711},
    url = {https://ui.adsabs.harvard.edu/abs/2023MNRAS.522.1843R/abstract},
    doi = {10.1093/mnras/stad1104},
    language = {en},
    number = {2},
    urldate = {2026-03-26},
    journal = {MNRAS},
    author = {Rathjen, Tim-Eric and Naab, Thorsten and Walch, Stefanie and Seifried, Daniel and Girichidis, Philipp and Wünsch, Richard},
    month = jun,
    year = {2023},
    pages = {1843--1862},
}

@article{peng_detailed_2010,
    title = {Detailed {Decomposition} of {Galaxy} {Images}. {II}. {Beyond} {Axisymmetric} {Models}},
    volume = {139},
    issn = {0004-6256},
    url = {https://ui.adsabs.harvard.edu/abs/2010AJ....139.2097P/abstract},
    doi = {10.1088/0004-6256/139/6/2097},
    language = {en},
    number = {6},
    urldate = {2026-03-26},
    journal = {AJ},
    author = {Peng, Chien Y. and Ho, Luis C. and Impey, Chris D. and Rix, Hans-Walter},
    month = jun,
    year = {2010},
    pages = {2097--2129},
}

@article{arnouts_measuring_1999,
    title = {Measuring and modelling the redshift evolution of clustering: the {Hubble} {Deep} {Field} {North}},
    volume = {310},
    issn = {0035-8711},
    shorttitle = {Measuring and modelling the redshift evolution of clustering},
    url = {https://ui.adsabs.harvard.edu/abs/1999MNRAS.310..540A/abstract},
    doi = {10.1046/j.1365-8711.1999.02978.x},
    language = {en},
    number = {2},
    urldate = {2026-03-26},
    journal = {MNRAS},
    author = {Arnouts, S. and Cristiani, S. and Moscardini, L. and Matarrese, S. and Lucchin, F. and Fontana, A. and Giallongo, E.},
    month = dec,
    year = {1999},
    pages = {540--556},
}

@article{ilbert_accurate_2006,
    title = {Accurate photometric redshifts for the {CFHT} legacy survey calibrated using the {VIMOS} {VLT} deep survey},
    volume = {457},
    issn = {0004-6361},
    url = {https://ui.adsabs.harvard.edu/abs/2006A&A...457..841I/abstract},
    doi = {10.1051/0004-6361:20065138},
    language = {en},
    number = {3},
    urldate = {2026-03-26},
    journal = {A\&A},
    author = {Ilbert, O. and Arnouts, S. and McCracken, H. J. and Bolzonella, M. and Bertin, E. and Le Fèvre, O. and Mellier, Y. and Zamorani, G. and Pellò, R. and Iovino, A. and Tresse, L. and Le Brun, V. and Bottini, D. and Garilli, B. and Maccagni, D. and Picat, J. P. and Scaramella, R. and Scodeggio, M. and Vettolani, G. and Zanichelli, A. and Adami, C. and Bardelli, S. and Cappi, A. and Charlot, S. and Ciliegi, P. and Contini, T. and Cucciati, O. and Foucaud, S. and Franzetti, P. and Gavignaud, I. and Guzzo, L. and Marano, B. and Marinoni, C. and Mazure, A. and Meneux, B. and Merighi, R. and Paltani, S. and Pollo, A. and Pozzetti, L. and Radovich, M. and Zucca, E. and Bondi, M. and Bongiorno, A. and Busarello, G. and de La Torre, S. and Gregorini, L. and Lamareille, F. and Mathez, G. and Merluzzi, P. and Ripepi, V. and Rizzo, D. and Vergani, D.},
    month = oct,
    year = {2006},
    pages = {841--856},
}

@article{davies_how_2011,
    title = {How {Well} {Can} {We} {Measure} the {Intrinsic} {Velocity} {Dispersion} of {Distant} {Disk} {Galaxies}?},
    volume = {741},
    issn = {0004-637X},
    url = {https://ui.adsabs.harvard.edu/abs/2011ApJ...741...69D/abstract},
    doi = {10.1088/0004-637X/741/2/69},
    language = {en},
    number = {2},
    urldate = {2026-03-26},
    journal = {ApJ},
    author = {Davies, R. and Förster Schreiber, N. M. and Cresci, G. and Genzel, R. and Bouché, N. and Burkert, A. and Buschkamp, P. and Genel, S. and Hicks, E. and Kurk, J. and Lutz, D. and Newman, S. and Shapiro, K. and Sternberg, A. and Tacconi, L. J. and Wuyts, S.},
    month = nov,
    year = {2011},
    pages = {69},
}

@article{davies_nuclear_2004,
    title = {The {Nuclear} {Gasdynamics} and {Star} {Formation} of {NGC} 7469},
    volume = {602},
    issn = {0004-637X},
    url = {https://ui.adsabs.harvard.edu/abs/2004ApJ...602..148D/abstract},
    doi = {10.1086/380995},
    language = {en},
    number = {1},
    urldate = {2026-03-26},
    journal = {ApJ},
    author = {Davies, R. I. and Tacconi, L. J. and Genzel, R.},
    month = feb,
    year = {2004},
    pages = {148--161},
}

@article{cresci_sins_2009,
    title = {The {SINS} {Survey}: {Modeling} the {Dynamics} of z {\textasciitilde} 2 {Galaxies} and the {High}-z {Tully}-{Fisher} {Relation}},
    volume = {697},
    issn = {0004-637X},
    shorttitle = {The {SINS} {Survey}},
    url = {https://ui.adsabs.harvard.edu/abs/2009ApJ...697..115C/abstract},
    doi = {10.1088/0004-637X/697/1/115},
    language = {en},
    number = {1},
    urldate = {2026-03-26},
    journal = {ApJ},
    author = {Cresci, G. and Hicks, E. K. S. and Genzel, R. and Förster Schreiber, N. M. and Davies, R. and Bouché, N. and Buschkamp, P. and Genel, S. and Shapiro, K. and Tacconi, L. and Sommer-Larsen, J. and Burkert, A. and Eisenhauer, F. and Gerhard, O. and Lutz, D. and Naab, T. and Sternberg, A. and Cimatti, A. and Daddi, E. and Erb, D. K. and Kurk, J. and Lilly, S. L. and Renzini, A. and Shapley, A. and Steidel, C. C. and Caputi, K.},
    month = may,
    year = {2009},
    pages = {115--132},
}

@article{lang_falling_2017,
    title = {Falling {Outer} {Rotation} {Curves} of {Star}-forming {Galaxies} at 0.6 ≲ z ≲ 2.6 {Probed} with {KMOS}{\textless}{SUP}{\textgreater}{3D}{\textless}/{SUP}{\textgreater} and {SINS}/{zC}-{SINF}},
    volume = {840},
    issn = {0004-637X},
    url = {https://ui.adsabs.harvard.edu/abs/2017ApJ...840...92L/abstract},
    doi = {10.3847/1538-4357/aa6d82},
    language = {en},
    number = {2},
    urldate = {2026-03-26},
    journal = {ApJ},
    author = {Lang, Philipp and Förster Schreiber, Natascha M. and Genzel, Reinhard and Wuyts, Stijn and Wisnioski, Emily and Beifiori, Alessandra and Belli, Sirio and Bender, Ralf and Brammer, Gabe and Burkert, Andreas and Chan, Jeffrey and Davies, Ric and Fossati, Matteo and Galametz, Audrey and Kulkarni, Sandesh K. and Lutz, Dieter and Mendel, J. Trevor and Momcheva, Ivelina G. and Naab, Thorsten and Nelson, Erica J. and Saglia, Roberto P. and Seitz, Stella and Tacchella, Sandro and Tacconi, Linda J. and Tadaki, Ken-ichi and Übler, Hannah and van Dokkum, Pieter G. and Wilman, David J.},
    month = may,
    year = {2017},
    pages = {92},
}

@article{bouche_galpak_2015,
    title = {{GalPaK} {3D}: {Galaxy} parameters and kinematics extraction from {3D} data},
    shorttitle = {{GalPaK} {3D}},
    url = {https://ui.adsabs.harvard.edu/abs/2015ascl.soft01014B/abstract},
    language = {en},
    urldate = {2026-03-26},
    journal = {Astrophysics Source Code Library},
    author = {Bouché, N. and Carfantan, H. and Schroetter, I. and Michel-Dansac, L. and Contini, T.},
    month = jan,
    year = {2015},
    pages = {ascl:1501.014},
}

@article{bouche_galpak3d_2015,
    title = {{GalPak}{\textless}{SUP}{\textgreater}{3D}{\textless}/{SUP}{\textgreater}: {A} {Bayesian} {Parametric} {Tool} for {Extracting} {Morphokinematics} of {Galaxies} from {3D} {Data}},
    volume = {150},
    issn = {0004-6256},
    shorttitle = {{GalPak}{\textless}{SUP}{\textgreater}{3D}{\textless}/{SUP}{\textgreater}},
    url = {https://ui.adsabs.harvard.edu/abs/2015AJ....150...92B/abstract},
    doi = {10.1088/0004-6256/150/3/92},
    language = {en},
    number = {3},
    urldate = {2026-03-26},
    journal = {AJ},
    author = {Bouché, N. and Carfantan, H. and Schroetter, I. and Michel-Dansac, L. and Contini, T.},
    month = sep,
    year = {2015},
    pages = {92},
}

@article{neeleman_qubefit_2020,
    title = {qubefit: {MCMC} kinematic modeling},
    shorttitle = {qubefit},
    url = {https://ui.adsabs.harvard.edu/abs/2020ascl.soft05013N/abstract},
    language = {en},
    urldate = {2026-03-26},
    journal = {Astrophysics Source Code Library},
    author = {Neeleman, Marcel and Prochaska, J. Xavier and Kanekar, Nissim and Rafelski, Marc},
    month = may,
    year = {2020},
    pages = {ascl:2005.013},
}

@article{di_teodoro_3d_2015,
    title = {{\textless}{SUP}{\textgreater}{3D}{\textless}/{SUP}{\textgreater} {BAROLO}: a new {3D} algorithm to derive rotation curves of galaxies},
    volume = {451},
    issn = {0035-8711},
    shorttitle = {{\textless}{SUP}{\textgreater}{3D}{\textless}/{SUP}{\textgreater} {BAROLO}},
    url = {https://ui.adsabs.harvard.edu/abs/2015MNRAS.451.3021D/abstract},
    doi = {10.1093/mnras/stv1213},
    language = {en},
    number = {3},
    urldate = {2026-03-26},
    journal = {MNRAS},
    author = {Di Teodoro, E. M. and Fraternali, F.},
    month = aug,
    year = {2015},
    pages = {3021--3033},
}

@article{genzel_sins_2011,
    title = {The {Sins} {Survey} of z {\textasciitilde} 2 {Galaxy} {Kinematics}: {Properties} of the {Giant} {Star}-forming {Clumps}},
    volume = {733},
    issn = {0004-637X},
    shorttitle = {The {Sins} {Survey} of z {\textasciitilde} 2 {Galaxy} {Kinematics}},
    url = {https://ui.adsabs.harvard.edu/abs/2011ApJ...733..101G/abstract},
    doi = {10.1088/0004-637X/733/2/101},
    language = {en},
    number = {2},
    urldate = {2026-03-26},
    journal = {ApJ},
    author = {Genzel, R. and Newman, S. and Jones, T. and Förster Schreiber, N. M. and Shapiro, K. and Genel, S. and Lilly, S. J. and Renzini, A. and Tacconi, L. J. and Bouché, N. and Burkert, A. and Cresci, G. and Buschkamp, P. and Carollo, C. M. and Ceverino, D. and Davies, R. and Dekel, A. and Eisenhauer, F. and Hicks, E. and Kurk, J. and Lutz, D. and Mancini, C. and Naab, T. and Peng, Y. and Sternberg, A. and Vergani, D. and Zamorani, G.},
    month = jun,
    year = {2011},
    pages = {101},
}

@article{casavecchia_atomic_2025,
    title = {Atomic and molecular gas as traced by [{C} {II}] emission},
    volume = {693},
    issn = {0004-6361},
    url = {https://ui.adsabs.harvard.edu/abs/2025A&A...693A.119C/abstract},
    doi = {10.1051/0004-6361/202452282},
    language = {en},
    urldate = {2026-03-26},
    journal = {A\&A},
    author = {Casavecchia, Benedetta and Maio, Umberto and Péroux, Céline and Ciardi, Benedetta},
    month = jan,
    year = {2025},
    pages = {A119},
}

@article{binney_galactic_2008,
    title = {Galactic {Dynamics}: {Second} {Edition}},
    shorttitle = {Galactic {Dynamics}},
    url = {https://ui.adsabs.harvard.edu/abs/2008gady.book.....B/abstract},
    language = {en},
    urldate = {2026-03-26},
    journal = {Galactic Dynamics: Second Edition},
    author = {Binney, James and Tremaine, Scott},
    year = {2008},
    note = {ISBN: 9780691130262},
}

@article{behrendt_structure_2015,
    title = {Structure formation in gas-rich galactic discs with finite thickness: from discs to rings},
    volume = {448},
    issn = {0035-8711},
    shorttitle = {Structure formation in gas-rich galactic discs with finite thickness},
    url = {https://ui.adsabs.harvard.edu/abs/2015MNRAS.448.1007B/abstract},
    doi = {10.1093/mnras/stv027},
    language = {en},
    number = {1},
    urldate = {2026-03-26},
    journal = {MNRAS},
    author = {Behrendt, M. and Burkert, A. and Schartmann, M.},
    month = mar,
    year = {2015},
    pages = {1007--1019},
}

@article{fujimoto_primordial_2025,
    title = {Primordial rotating disk composed of at least 15 dense star-forming clumps at cosmic dawn},
    volume = {9},
    issn = {2397-3366},
    url = {https://ui.adsabs.harvard.edu/abs/2025NatAs...9.1553F/abstract},
    doi = {10.1038/s41550-025-02592-w},
    language = {en},
    urldate = {2026-03-26},
    journal = {Nature Astronomy},
    author = {Fujimoto, S. and Ouchi, M. and Kohno, K. and Valentino, F. and Giménez-Arteaga, C. and Brammer, G. B. and Furtak, L. J. and Kohandel, M. and Oguri, M. and Pallottini, A. and Richard, J. and Zitrin, A. and Bauer, F. E. and Boylan-Kolchin, M. and Dessauges-Zavadsky, M. and Egami, E. and Finkelstein, S. L. and Ma, Z. and Smail, I. and Watson, D. and Hutchison, T. A. and Rigby, J. R. and Welch, B. D. and Ao, Y. and Bradley, L. D. and Caminha, G. B. and Caputi, K. I. and Espada, D. and Endsley, R. and Fudamoto, Y. and González-López, J. and Hatsukade, B. and Koekemoer, A. M. and Kokorev, V. and Laporte, N. and Lee, M. and Magdis, G. E. and Ono, Y. and Rizzo, F. and Shibuya, T. and Shimasaku, K. and Sun, F. and Toft, S. and Umehata, H. and Wang, T. and Yajima, H.},
    month = aug,
    year = {2025},
    pages = {1553--1567},
}

@article{bacon_mpdaf_2016,
    title = {{MPDAF}: {MUSE} {Python} {Data} {Analysis} {Framework}},
    shorttitle = {{MPDAF}},
    url = {https://ui.adsabs.harvard.edu/abs/2016ascl.soft11003B/abstract},
    language = {en},
    urldate = {2026-03-26},
    journal = {Astrophysics Source Code Library},
    author = {Bacon, Roland and Piqueras, Laure and Conseil, Simon and Richard, Johan and Shepherd, Martin},
    month = nov,
    year = {2016},
    pages = {ascl:1611.003},
}

@article{piqueras_mpdaf_2019,
    title = {{MPDAF} - {A} {Python} {Package} for the {Analysis} of {VLT}/{MUSE} {Data}},
    volume = {521},
    issn = {1050-3390},
    url = {https://ui.adsabs.harvard.edu/abs/2019ASPC..521..545P/abstract},
    language = {en},
    urldate = {2026-03-26},
    journal = {Astronomical Data Analysis Software and Systems XXVI},
    author = {Piqueras, L. and Conseil, S. and Shepherd, M. and Bacon, R. and Leclercq, F. and Richard, J.},
    month = oct,
    year = {2019},
    pages = {545},
}

@misc{hogg_data_2010,
    title = {Data analysis recipes: {Fitting} a model to data},
    shorttitle = {Data analysis recipes},
    url = {https://ui.adsabs.harvard.edu/abs/2010arXiv1008.4686H},
    doi = {10.48550/arXiv.1008.4686},
    urldate = {2025-01-21},
    author = {Hogg, David W. and Bovy, Jo and Lang, Dustin},
    month = aug,
    year = {2010},
    note = {arXiv e-prints
ADS Bibcode: 2010arXiv1008.4686H},
}

@article{schwarz_estimating_1978,
    title = {Estimating the {Dimension} of a {Model}},
    volume = {6},
    url = {https://ui.adsabs.harvard.edu/abs/1978AnSta...6..461S},
    urldate = {2025-01-21},
    journal = {Annals of Statistics},
    author = {Schwarz, Gideon},
    month = jul,
    year = {1978},
    note = {ADS Bibcode: 1978AnSta...6..461S},
    pages = {461--464},
}

@article{liddle_information_2007,
    title = {Information criteria for astrophysical model selection},
    volume = {377},
    issn = {1745-3933, 1745-3925},
    url = {https://academic.oup.com/mnrasl/article/377/1/L74/1210361},
    doi = {10.1111/j.1745-3933.2007.00306.x},
    language = {en},
    number = {1},
    urldate = {2025-01-21},
    journal = {MNRAS: Letters},
    author = {Liddle, Andrew R.},
    month = may,
    year = {2007},
    pages = {L74--L78},
}

@article{foreman-mackey_emcee_2013,
    title = {emcee: {The} {MCMC} {Hammer}},
    volume = {125},
    issn = {0004-6280},
    shorttitle = {emcee},
    url = {https://ui.adsabs.harvard.edu/abs/2013PASP..125..306F/abstract},
    doi = {10.1086/670067},
    language = {en},
    number = {925},
    urldate = {2026-03-26},
    journal = {PASP},
    author = {Foreman-Mackey, Daniel and Hogg, David W. and Lang, Dustin and Goodman, Jonathan},
    month = mar,
    year = {2013},
    pages = {306},
}

@article{concas_being_2022,
    title = {Being {KLEVER} at cosmic noon: {Ionized} gas outflows are inconspicuous in low-mass star-forming galaxies but prominent in massive {AGN} hosts},
    volume = {513},
    issn = {0035-8711},
    shorttitle = {Being {KLEVER} at cosmic noon},
    url = {https://ui.adsabs.harvard.edu/abs/2022MNRAS.513.2535C/abstract},
    doi = {10.1093/mnras/stac1026},
    language = {en},
    number = {2},
    urldate = {2026-03-26},
    journal = {MNRAS},
    author = {Concas, Alice and Maiolino, Roberto and Curti, Mirko and Hayden-Pawson, Connor and Cirasuolo, Michele and Jones, Gareth C. and Mercurio, Amata and Belfiore, Francesco and Cresci, Giovanni and Cullen, Fergus and Mannucci, Filippo and Marconi, Alessandro and Cappellari, Michele and Cicone, Claudia and Peng, Yingjie and Troncoso, Paulina},
    month = jun,
    year = {2022},
    pages = {2535--2562},
}

@article{cappellari_powerbin_2025,
    title = {{PowerBin}: fast adaptive data binning with {Centroidal} {Power} {Diagrams}},
    volume = {544},
    issn = {0035-8711},
    shorttitle = {{PowerBin}},
    url = {https://ui.adsabs.harvard.edu/abs/2025MNRAS.544.1432C/abstract},
    doi = {10.1093/mnras/staf1726},
    language = {en},
    number = {2},
    urldate = {2026-03-26},
    journal = {MNRAS},
    author = {Cappellari, Michele},
    month = dec,
    year = {2025},
    pages = {1432--1446},
}

@article{van_de_sande_fundamental_2014,
    title = {The {Fundamental} {Plane} of {Massive} {Quiescent} {Galaxies} {Out} to z {\textasciitilde} 2},
    volume = {793},
    issn = {0004-637X},
    url = {https://ui.adsabs.harvard.edu/abs/2014ApJ...793L..31V/abstract},
    doi = {10.1088/2041-8205/793/2/L31},
    language = {en},
    number = {2},
    urldate = {2026-03-26},
    journal = {ApJ},
    author = {van de Sande, Jesse and Kriek, Mariska and Franx, Marijn and Bezanson, Rachel and van Dokkum, Pieter G.},
    month = oct,
    year = {2014},
    pages = {L31},
}

@article{deugenio_fast-rotator_2024,
    title = {A fast-rotator post-starburst galaxy quenched by supermassive black-hole feedback at z = 3},
    volume = {8},
    issn = {2397-3366},
    url = {https://ui.adsabs.harvard.edu/abs/2024NatAs...8.1443D/abstract},
    doi = {10.1038/s41550-024-02345-1},
    language = {en},
    urldate = {2026-03-19},
    journal = {Nature Astronomy},
    author = {D'Eugenio, Francesco and Pérez-González, Pablo G. and Maiolino, Roberto and Scholtz, Jan and Perna, Michele and Circosta, Chiara and Übler, Hannah and Arribas, Santiago and Böker, Torsten and Bunker, Andrew J. and Carniani, Stefano and Charlot, Stephane and Chevallard, Jacopo and Cresci, Giovanni and Curtis-Lake, Emma and Jones, Gareth C. and Kumari, Nimisha and Lamperti, Isabella and Looser, Tobias J. and Parlanti, Eleonora and Rix, Hans-Walter and Robertson, Brant and Rodríguez Del Pino, Bruno and Tacchella, Sandro and Venturi, Giacomo and Willott, Chris J.},
    month = nov,
    year = {2024},
    pages = {1443--1456},
}

@article{bacon_muse_2017,
    title = {The {MUSE} {Hubble} {Ultra} {Deep} {Field} {Survey}. {I}. {Survey} description, data reduction, and source detection},
    volume = {608},
    issn = {0004-6361},
    url = {https://ui.adsabs.harvard.edu/abs/2017A&A...608A...1B/abstract},
    doi = {10.1051/0004-6361/201730833},
    language = {en},
    urldate = {2026-03-26},
    journal = {A\&A},
    author = {Bacon, Roland and Conseil, Simon and Mary, David and Brinchmann, Jarle and Shepherd, Martin and Akhlaghi, Mohammad and Weilbacher, Peter M. and Piqueras, Laure and Wisotzki, Lutz and Lagattuta, David and Epinat, Benoit and Guerou, Adrien and Inami, Hanae and Cantalupo, Sebastiano and Courbot, Jean Baptiste and Contini, Thierry and Richard, Johan and Maseda, Michael and Bouwens, Rychard and Bouché, Nicolas and Kollatschny, Wolfram and Schaye, Joop and Marino, Raffaella Anna and Pello, Roser and Herenz, Christian and Guiderdoni, Bruno and Carollo, Marcella},
    month = dec,
    year = {2017},
    pages = {A1},
}

@article{navarro_structure_1996,
    title = {The {Structure} of {Cold} {Dark} {Matter} {Halos}},
    volume = {462},
    issn = {0004-637X},
    url = {https://ui.adsabs.harvard.edu/abs/1996ApJ...462..563N/abstract},
    doi = {10.1086/177173},
    language = {en},
    urldate = {2026-03-26},
    journal = {ApJ},
    author = {Navarro, Julio F. and Frenk, Carlos S. and White, Simon D. M.},
    month = may,
    year = {1996},
    pages = {563},
}

@article{sofue_rotation_2001,
    title = {Rotation {Curves} of {Spiral} {Galaxies}},
    volume = {39},
    issn = {0066-4146},
    url = {https://ui.adsabs.harvard.edu/abs/2001ARA&A..39..137S/abstract},
    doi = {10.1146/annurev.astro.39.1.137},
    language = {en},
    urldate = {2026-03-26},
    journal = {ARA\&A},
    author = {Sofue, Yoshiaki and Rubin, Vera},
    year = {2001},
    pages = {137--174},
}

@article{lelli_wisdom_2022,
    title = {{WISDOM} {Project} - {XIII}. {Feeding} molecular gas to the supermassive black hole in the starburst {AGN}-host galaxy {Fairall} 49},
    volume = {516},
    issn = {0035-8711},
    url = {https://ui.adsabs.harvard.edu/abs/2022MNRAS.516.4066L/abstract},
    doi = {10.1093/mnras/stac2493},
    language = {en},
    number = {3},
    urldate = {2026-03-26},
    journal = {MNRAS},
    author = {Lelli, Federico and Davis, Timothy A. and Bureau, Martin and Cappellari, Michele and Liu, Lijie and Ruffa, Ilaria and Smith, Mark D. and Williams, Thomas G.},
    month = nov,
    year = {2022},
    pages = {4066--4083},
}

@article{peroux_cosmic_2020,
    title = {The {Cosmic} {Baryon} and {Metal} {Cycles}},
    volume = {58},
    issn = {0066-4146},
    url = {https://ui.adsabs.harvard.edu/abs/2020ARA&A..58..363P/abstract},
    doi = {10.1146/annurev-astro-021820-120014},
    language = {en},
    urldate = {2026-04-13},
    journal = {ARA\&A},
    author = {Péroux, Céline and Howk, J. Christopher},
    month = aug,
    year = {2020},
    pages = {363--406},
}

@article{peroux_multiphase_2019,
    title = {Multiphase circumgalactic medium probed with {MUSE} and {ALMA}},
    volume = {485},
    issn = {0035-8711},
    url = {https://ui.adsabs.harvard.edu/abs/2019MNRAS.485.1595P/abstract},
    doi = {10.1093/mnras/stz202},
    language = {en},
    number = {2},
    urldate = {2026-04-13},
    journal = {MNRAS},
    author = {Péroux, Céline and Zwaan, Martin A. and Klitsch, Anne and Augustin, Ramona and Hamanowicz, Aleksandra and Rahmani, Hadi and Pettini, Max and Kulkarni, Varsha and Straka, Lorrie A. and Biggs, Andy D. and York, Donald G. and Milliard, Bruno},
    month = may,
    year = {2019},
    pages = {1595--1613},
}

@article{klitsch_almacal_2018,
    title = {{ALMACAL} - {III}. {A} combined {ALMA} and {MUSE} survey for neutral, molecular, and ionized gas in an {H} {I}-absorption-selected system},
    volume = {475},
    issn = {0035-8711},
    url = {https://ui.adsabs.harvard.edu/abs/2018MNRAS.475..492K/abstract},
    doi = {10.1093/mnras/stx3184},
    language = {en},
    number = {1},
    urldate = {2026-04-13},
    journal = {MNRAS},
    author = {Klitsch, A. and Péroux, C. and Zwaan, M. A. and Smail, I. and Oteo, I. and Biggs, A. D. and Popping, G. and Swinbank, A. M.},
    month = mar,
    year = {2018},
    pages = {492--507},
}

@article{szakacs_muse-alma_2021,
    title = {{MUSE}-{ALMA} haloes {VI}: coupling atomic, ionized, and molecular gas kinematics of galaxies},
    volume = {505},
    issn = {0035-8711},
    shorttitle = {{MUSE}-{ALMA} haloes {VI}},
    url = {https://ui.adsabs.harvard.edu/abs/2021MNRAS.505.4746S/abstract},
    doi = {10.1093/mnras/stab1434},
    language = {en},
    number = {4},
    urldate = {2026-04-13},
    journal = {MNRAS},
    author = {Szakacs, Roland and Péroux, Céline and Zwaan, Martin and Hamanowicz, Aleksandra and Klitsch, Anne and Fresco, Alejandra Y. and Augustin, Ramona and Biggs, Andrew and Kulkarni, Varsha and Rahmani, Hadi},
    month = aug,
    year = {2021},
    pages = {4746--4761},
}

@article{simons_z_2017,
    title = {z ∼ 2: {An} {Epoch} of {Disk} {Assembly}},
    volume = {843},
    issn = {0004-637X},
    shorttitle = {z ∼ 2},
    url = {https://ui.adsabs.harvard.edu/abs/2017ApJ...843...46S/abstract},
    doi = {10.3847/1538-4357/aa740c},
    language = {en},
    number = {1},
    urldate = {2026-04-21},
    journal = {ApJ},
    author = {Simons, Raymond C. and Kassin, Susan A. and Weiner, Benjamin J. and Faber, Sandra M. and Trump, Jonathan R. and Heckman, Timothy M. and Koo, David C. and Pacifici, Camilla and Primack, Joel R. and Snyder, Gregory F. and de la Vega, Alexander},
    month = jul,
    year = {2017},
    pages = {46},
}

@article{forster_schreiber_star-forming_2020,
    title = {Star-{Forming} {Galaxies} at {Cosmic} {Noon}},
    volume = {58},
    issn = {0066-4146},
    url = {https://ui.adsabs.harvard.edu/abs/2020ARA&A..58..661F/abstract},
    doi = {10.1146/annurev-astro-032620-021910},
    language = {en},
    urldate = {2026-03-26},
    journal = {ARA\&A},
    author = {Förster Schreiber, Natascha M. and Wuyts, Stijn},
    month = aug,
    year = {2020},
    pages = {661--725},
}

@article{bryant_sami_2015,
    title = {The {SAMI} {Galaxy} {Survey}: instrument specification and target selection},
    volume = {447},
    issn = {0035-8711},
    shorttitle = {The {SAMI} {Galaxy} {Survey}},
    url = {https://ui.adsabs.harvard.edu/abs/2015MNRAS.447.2857B/abstract},
    doi = {10.1093/mnras/stu2635},
    language = {en},
    number = {3},
    urldate = {2026-04-21},
    journal = {MNRAS},
    author = {Bryant, J. J. and Owers, M. S. and Robotham, A. S. G. and Croom, S. M. and Driver, S. P. and Drinkwater, M. J. and Lorente, N. P. F. and Cortese, L. and Scott, N. and Colless, M. and Schaefer, A. and Taylor, E. N. and Konstantopoulos, I. S. and Allen, J. T. and Baldry, I. and Barnes, L. and Bauer, A. E. and Bland-Hawthorn, J. and Bloom, J. V. and Brooks, A. M. and Brough, S. and Cecil, G. and Couch, W. and Croton, D. and Davies, R. and Ellis, S. and Fogarty, L. M. R. and Foster, C. and Glazebrook, K. and Goodwin, M. and Green, A. and Gunawardhana, M. L. and Hampton, E. and Ho, I.-T. and Hopkins, A. M. and Kewley, L. and Lawrence, J. S. and Leon-Saval, S. G. and Leslie, S. and McElroy, R. and Lewis, G. and Liske, J. and López-Sánchez, Á R. and Mahajan, S. and Medling, A. M. and Metcalfe, N. and Meyer, M. and Mould, J. and Obreschkow, D. and O'Toole, S. and Pracy, M. and Richards, S. N. and Shanks, T. and Sharp, R. and Sweet, S. M. and Thomas, A. D. and Tonini, C. and Walcher, C. J.},
    month = mar,
    year = {2015},
    pages = {2857--2879},
}

@article{sanchez_califa_2012,
    title = {{CALIFA}, the {Calar} {Alto} {Legacy} {Integral} {Field} {Area} survey. {I}. {Survey} presentation},
    volume = {538},
    issn = {0004-6361},
    url = {https://ui.adsabs.harvard.edu/abs/2012A&A...538A...8S/abstract},
    doi = {10.1051/0004-6361/201117353},
    language = {en},
    urldate = {2026-04-21},
    journal = {A\&A},
    author = {Sánchez, S. F. and Kennicutt, R. C. and Gil de Paz, A. and van de Ven, G. and Vílchez, J. M. and Wisotzki, L. and Walcher, C. J. and Mast, D. and Aguerri, J. a. L. and Albiol-Pérez, S. and Alonso-Herrero, A. and Alves, J. and Bakos, J. and Bartáková, T. and Bland-Hawthorn, J. and Boselli, A. and Bomans, D. J. and Castillo-Morales, A. and Cortijo-Ferrero, C. and de Lorenzo-Cáceres, A. and Del Olmo, A. and Dettmar, R.-J. and Díaz, A. and Ellis, S. and Falcón-Barroso, J. and Flores, H. and Gallazzi, A. and García-Lorenzo, B. and González Delgado, R. and Gruel, N. and Haines, T. and Hao, C. and Husemann, B. and Iglésias-Páramo, J. and Jahnke, K. and Johnson, B. and Jungwiert, B. and Kalinova, V. and Kehrig, C. and Kupko, D. and López-Sánchez, Á R. and Lyubenova, M. and Marino, R. A. and Mármol-Queraltó, E. and Márquez, I. and Masegosa, J. and Meidt, S. and Mendez-Abreu, J. and Monreal-Ibero, A. and Montijo, C. and Mourão, A. M. and Palacios-Navarro, G. and Papaderos, P. and Pasquali, A. and Peletier, R. and Pérez, E. and Pérez, I. and Quirrenbach, A. and Relaño, M. and Rosales-Ortega, F. F. and Roth, M. M. and Ruiz-Lara, T. and Sánchez-Blázquez, P. and Sengupta, C. and Singh, R. and Stanishev, V. and Trager, S. C. and Vazdekis, A. and Viironen, K. and Wild, V. and Zibetti, S. and Ziegler, B.},
    month = feb,
    year = {2012},
    pages = {A8},
}

@article{bundy_overview_2015,
    title = {Overview of the {SDSS}-{IV} {MaNGA} {Survey}: {Mapping} nearby {Galaxies} at {Apache} {Point} {Observatory}},
    volume = {798},
    issn = {0004-637X},
    shorttitle = {Overview of the {SDSS}-{IV} {MaNGA} {Survey}},
    url = {https://ui.adsabs.harvard.edu/abs/2015ApJ...798....7B/abstract},
    doi = {10.1088/0004-637X/798/1/7},
    language = {en},
    number = {1},
    urldate = {2026-04-21},
    journal = {ApJ},
    author = {Bundy, Kevin and Bershady, Matthew A. and Law, David R. and Yan, Renbin and Drory, Niv and MacDonald, Nicholas and Wake, David A. and Cherinka, Brian and Sánchez-Gallego, José R. and Weijmans, Anne-Marie and Thomas, Daniel and Tremonti, Christy and Masters, Karen and Coccato, Lodovico and Diamond-Stanic, Aleksandar M. and Aragón-Salamanca, Alfonso and Avila-Reese, Vladimir and Badenes, Carles and Falcón-Barroso, Jésus and Belfiore, Francesco and Bizyaev, Dmitry and Blanc, Guillermo A. and Bland-Hawthorn, Joss and Blanton, Michael R. and Brownstein, Joel R. and Byler, Nell and Cappellari, Michele and Conroy, Charlie and Dutton, Aaron A. and Emsellem, Eric and Etherington, James and Frinchaboy, Peter M. and Fu, Hai and Gunn, James E. and Harding, Paul and Johnston, Evelyn J. and Kauffmann, Guinevere and Kinemuchi, Karen and Klaene, Mark A. and Knapen, Johan H. and Leauthaud, Alexie and Li, Cheng and Lin, Lihwai and Maiolino, Roberto and Malanushenko, Viktor and Malanushenko, Elena and Mao, Shude and Maraston, Claudia and McDermid, Richard M. and Merrifield, Michael R. and Nichol, Robert C. and Oravetz, Daniel and Pan, Kaike and Parejko, John K. and Sanchez, Sebastian F. and Schlegel, David and Simmons, Audrey and Steele, Oliver and Steinmetz, Matthias and Thanjavur, Karun and Thompson, Benjamin A. and Tinker, Jeremy L. and van den Bosch, Remco C. E. and Westfall, Kyle B. and Wilkinson, David and Wright, Shelley and Xiao, Ting and Zhang, Kai},
    month = jan,
    year = {2015},
    pages = {7},
}

@article{stott_kmos_2016,
    title = {The {KMOS} {Redshift} {One} {Spectroscopic} {Survey} ({KROSS}): dynamical properties, gas and dark matter fractions of typical z ∼ 1 star-forming galaxies},
    volume = {457},
    issn = {0035-8711},
    shorttitle = {The {KMOS} {Redshift} {One} {Spectroscopic} {Survey} ({KROSS})},
    url = {https://ui.adsabs.harvard.edu/abs/2016MNRAS.457.1888S/abstract},
    doi = {10.1093/mnras/stw129},
    language = {en},
    number = {2},
    urldate = {2026-04-21},
    journal = {MNRAS},
    author = {Stott, John P. and Swinbank, A. M. and Johnson, Helen L. and Tiley, Alfie and Magdis, Georgios and Bower, Richard and Bunker, Andrew J. and Bureau, Martin and Harrison, Chris M. and Jarvis, Matt J. and Sharples, Ray and Smail, Ian and Sobral, David and Best, Philip and Cirasuolo, Michele},
    month = apr,
    year = {2016},
    pages = {1888--1904},
}

@article{tiley_kmos_2021,
    title = {The {KMOS} galaxy evolution survey ({KGES}): the angular momentum of star-forming galaxies over the last ≍10 {Gyr}},
    volume = {506},
    issn = {0035-8711},
    shorttitle = {The {KMOS} galaxy evolution survey ({KGES})},
    url = {https://ui.adsabs.harvard.edu/abs/2021MNRAS.506..323T/abstract},
    doi = {10.1093/mnras/stab1692},
    language = {en},
    number = {1},
    urldate = {2026-04-21},
    journal = {MNRAS},
    author = {Tiley, Alfred L. and Gillman, S. and Cortese, L. and Swinbank, A. M. and Dudzevičiūtė, U. and Harrison, C. M. and Smail, I. and Obreschkow, D. and Croom, S. M. and Sharples, R. M. and Puglisi, A.},
    month = sep,
    year = {2021},
    pages = {323--342},
}

@article{peroux_muse-alma_2026,
    title = {{MUSE}-{ALMA} {Haloes}: {XIV}. {The} {ALMA} {Large} {Programme} {Data} {Release}},
    volume = {708},
    issn = {0004-6361},
    shorttitle = {{MUSE}-{ALMA} {Haloes}},
    url = {https://ui.adsabs.harvard.edu/abs/2026A&A...708A.146P/abstract},
    doi = {10.1051/0004-6361/202558645},
    language = {en},
    urldate = {2026-04-21},
    journal = {A\&A},
    author = {Péroux, Céline and Chen, Jianhang and Bollo, Victoria and O'Beirne, Tamsyn and Barfety, Capucine and Weng, Simon and Augustin, Ramona and Casavecchia, Benedetta and Zwaan, Martin and Tacconi, Linda and Föster Schreiber, Natascha and Chemin, Laurent and Kulkarni, Varsha},
    month = apr,
    year = {2026},
    pages = {A146},
}

@article{van_der_kruit_kinematics_1978,
    title = {The kinematics of spiral and irregular galaxies.},
    volume = {16},
    issn = {0066-4146},
    url = {https://ui.adsabs.harvard.edu/abs/1978ARA&A..16..103V/abstract},
    doi = {10.1146/annurev.aa.16.090178.000535},
    language = {en},
    urldate = {2026-04-21},
    journal = {ARA\&A},
    author = {van der Kruit, P. C. and Allen, R. J.},
    year = {1978},
    pages = {103--139},
}

@article{forster_schreiber_sinszc-sinf_2014,
    title = {The {Sins}/{zC}-{Sinf} {Survey} of z {\textasciitilde} 2 {Galaxy} {Kinematics}: {Evidence} for {Powerful} {Active} {Galactic} {Nucleus}-{Driven} {Nuclear} {Outflows} in {Massive} {Star}-{Forming} {Galaxies}},
    volume = {787},
    issn = {0004-637X},
    shorttitle = {The {Sins}/{zC}-{Sinf} {Survey} of z {\textasciitilde} 2 {Galaxy} {Kinematics}},
    url = {https://ui.adsabs.harvard.edu/abs/2014ApJ...787...38F},
    doi = {10.1088/0004-637X/787/1/38},
    urldate = {2025-04-07},
    journal = {ApJ},
    publisher = {IOP},
    author = {Förster Schreiber, N. M. and Genzel, R. and Newman, S. F. and Kurk, J. D. and Lutz, D. and Tacconi, L. J. and Wuyts, S. and Bandara, K. and Burkert, A. and Buschkamp, P. and Carollo, C. M. and Cresci, G. and Daddi, E. and Davies, R. and Eisenhauer, F. and Hicks, E. K. S. and Lang, P. and Lilly, S. J. and Mainieri, V. and Mancini, C. and Naab, T. and Peng, Y. and Renzini, A. and Rosario, D. and Shapiro Griffin, K. and Shapley, A. E. and Sternberg, A. and Tacchella, S. and Vergani, D. and Wisnioski, E. and Wuyts, E. and Zamorani, G.},
    month = may,
    year = {2014},
    note = {ADS Bibcode: 2014ApJ...787...38F},
    pages = {38},
}

@article{jolly_noema3d_2026,
    title = {{NOEMA3D}: {Resolving} radial gas flows in disk galaxies at z{\textasciitilde}1.1-1.6 with high-resolution {CO} observations},
    shorttitle = {{NOEMA3D}},
    url = {https://ui.adsabs.harvard.edu/abs/2026arXiv260418503J/abstract},
    language = {en},
    urldate = {2026-04-23},
    journal = {A\&A subm.},
    author = {Jolly, Jean-Baptiste and Tacconi, Linda J. and Genzel, Reinhard and Neri, Roberto and Schuster, Karl and Chen, Jianhang and Förster Schreiber, Natascha M. and Pastras, Stavros and Scaloni, Letizia and Tozzi, Giulia and Barféty, Capucine and Bolatto, Alberto and Burker, Andreas and Combes, Françoise and Cox, Pierre and Davies, Ric and Eisenhauer, Frank and Espejo Salcedo, Juan Manuel and Herrera-Camu, Rodrigo and García-Burillo, Santi and Kodama, Tadayuki and Lee, Lilian and Lee, Minju M. and Liu, Daizhong and Lutz, Dieter and Mazzolari, Giovanni and Naab, Thorsten and Nestor Shachar, Amit and Pulsoni, Claudia and Renzini, Alvio and Rubio, Monica and Shimizu, Taro T. and Sternberg, Amiel and Sturm, Eckhard and Übler, Hannah and Usero, Antonio and Wuyts, Stijn},
    month = apr,
    year = {2026},
    pages = {arXiv:2604.18503},
}

@article{genzel_evidence_2014,
    title = {Evidence for {Wide}-spread {Active} {Galactic} {Nucleus}-driven {Outflows} in the {Most} {Massive} z {\textasciitilde} 1-2 {Star}-forming {Galaxies}},
    volume = {796},
    issn = {0004-637X},
    url = {https://ui.adsabs.harvard.edu/abs/2014ApJ...796....7G/abstract},
    doi = {10.1088/0004-637X/796/1/7},
    language = {en},
    number = {1},
    urldate = {2026-04-23},
    journal = {ApJ},
    author = {Genzel, R. and Förster Schreiber, N. M. and Rosario, D. and Lang, P. and Lutz, D. and Wisnioski, E. and Wuyts, E. and Wuyts, S. and Bandara, K. and Bender, R. and Berta, S. and Kurk, J. and Mendel, J. T. and Tacconi, L. J. and Wilman, D. and Beifiori, A. and Brammer, G. and Burkert, A. and Buschkamp, P. and Chan, J. and Carollo, C. M. and Davies, R. and Eisenhauer, F. and Fabricius, M. and Fossati, M. and Kriek, M. and Kulkarni, S. and Lilly, S. J. and Mancini, C. and Momcheva, I. and Naab, T. and Nelson, E. J. and Renzini, A. and Saglia, R. and Sharples, R. M. and Sternberg, A. and Tacchella, S. and van Dokkum, P.},
    month = nov,
    year = {2014},
    pages = {7},
}

@ARTICLE{planck-collaboration_2020,
       author = {{Planck Collaboration} and {Aghanim}, N. and {Akrami}, Y. and {Ashdown}, M. and {Aumont}, J. and {Baccigalupi}, C. and {Ballardini}, M. and {Banday}, A.~J. and {Barreiro}, R.~B. and {Bartolo}, N. and {Basak}, S. and {Battye}, R. and {Benabed}, K. and {Bernard}, J.-P. and {Bersanelli}, M. and {Bielewicz}, P. and {Bock}, J.~J. and {Bond}, J.~R. and {Borrill}, J. and {Bouchet}, F.~R. and {Boulanger}, F. and {Bucher}, M. and {Burigana}, C. and {Butler}, R.~C. and {Calabrese}, E. and {Cardoso}, J.-F. and {Carron}, J. and {Challinor}, A. and {Chiang}, H.~C. and {Chluba}, J. and {Colombo}, L.~P.~L. and {Combet}, C. and {Contreras}, D. and {Crill}, B.~P. and {Cuttaia}, F. and {de Bernardis}, P. and {de Zotti}, G. and {Delabrouille}, J. and {Delouis}, J.-M. and {Di Valentino}, E. and {Diego}, J.~M. and {Dor{\'e}}, O. and {Douspis}, M. and {Ducout}, A. and {Dupac}, X. and {Dusini}, S. and {Efstathiou}, G. and {Elsner}, F. and {En{\ss}lin}, T.~A. and {Eriksen}, H.~K. and {Fantaye}, Y. and {Farhang}, M. and {Fergusson}, J. and {Fernandez-Cobos}, R. and {Finelli}, F. and {Forastieri}, F. and {Frailis}, M. and {Fraisse}, A.~A. and {Franceschi}, E. and {Frolov}, A. and {Galeotta}, S. and {Galli}, S. and {Ganga}, K. and {G{\'e}nova-Santos}, R.~T. and {Gerbino}, M. and {Ghosh}, T. and {Gonz{\'a}lez-Nuevo}, J. and {G{\'o}rski}, K.~M. and {Gratton}, S. and {Gruppuso}, A. and {Gudmundsson}, J.~E. and {Hamann}, J. and {Handley}, W. and {Hansen}, F.~K. and {Herranz}, D. and {Hildebrandt}, S.~R. and {Hivon}, E. and {Huang}, Z. and {Jaffe}, A.~H. and {Jones}, W.~C. and {Karakci}, A. and {Keih{\"a}nen}, E. and {Keskitalo}, R. and {Kiiveri}, K. and {Kim}, J. and {Kisner}, T.~S. and {Knox}, L. and {Krachmalnicoff}, N. and {Kunz}, M. and {Kurki-Suonio}, H. and {Lagache}, G. and {Lamarre}, J.-M. and {Lasenby}, A. and {Lattanzi}, M. and {Lawrence}, C.~R. and {Le Jeune}, M. and {Lemos}, P. and {Lesgourgues}, J. and {Levrier}, F. and {Lewis}, A. and {Liguori}, M. and {Lilje}, P.~B. and {Lilley}, M. and {Lindholm}, V. and {L{\'o}pez-Caniego}, M. and {Lubin}, P.~M. and {Ma}, Y.-Z. and {Mac{\'\i}as-P{\'e}rez}, J.~F. and {Maggio}, G. and {Maino}, D. and {Mandolesi}, N. and {Mangilli}, A. and {Marcos-Caballero}, A. and {Maris}, M. and {Martin}, P.~G. and {Martinelli}, M. and {Mart{\'\i}nez-Gonz{\'a}lez}, E. and {Matarrese}, S. and {Mauri}, N. and {McEwen}, J.~D. and {Meinhold}, P.~R. and {Melchiorri}, A. and {Mennella}, A. and {Migliaccio}, M. and {Millea}, M. and {Mitra}, S. and {Miville-Desch{\^e}nes}, M.-A. and {Molinari}, D. and {Montier}, L. and {Morgante}, G. and {Moss}, A. and {Natoli}, P. and {N{\o}rgaard-Nielsen}, H.~U. and {Pagano}, L. and {Paoletti}, D. and {Partridge}, B. and {Patanchon}, G. and {Peiris}, H.~V. and {Perrotta}, F. and {Pettorino}, V. and {Piacentini}, F. and {Polastri}, L. and {Polenta}, G. and {Puget}, J.-L. and {Rachen}, J.~P. and {Reinecke}, M. and {Remazeilles}, M. and {Renzi}, A. and {Rocha}, G. and {Rosset}, C. and {Roudier}, G. and {Rubi{\~n}o-Mart{\'\i}n}, J.~A. and {Ruiz-Granados}, B. and {Salvati}, L. and {Sandri}, M. and {Savelainen}, M. and {Scott}, D. and {Shellard}, E.~P.~S. and {Sirignano}, C. and {Sirri}, G. and {Spencer}, L.~D. and {Sunyaev}, R. and {Suur-Uski}, A.-S. and {Tauber}, J.~A. and {Tavagnacco}, D. and {Tenti}, M. and {Toffolatti}, L. and {Tomasi}, M. and {Trombetti}, T. and {Valenziano}, L. and {Valiviita}, J. and {Van Tent}, B. and {Vibert}, L. and {Vielva}, P. and {Villa}, F. and {Vittorio}, N. and {Wandelt}, B.~D. and {Wehus}, I.~K. and {White}, M. and {White}, S.~D.~M. and {Zacchei}, A. and {Zonca}, A.},
        title = "{Planck 2018 results. VI. Cosmological parameters}",
      journal = {\aap},
         year = 2020,
        month = sep,
       volume = {641},
          eid = {A6},
        pages = {A6},
          doi = {10.1051/0004-6361/201833910},
archivePrefix = {arXiv},
       eprint = {1807.06209},
 primaryClass = {astro-ph.CO},
       adsurl = {https://ui.adsabs.harvard.edu/abs/2020A&A...641A...6P}
}

@article{derkenne_magpi_2024,
    title = {The {MAGPI} survey: evidence against the bulge-halo conspiracy},
    volume = {533},
    issn = {0035-8711},
    shorttitle = {The {MAGPI} survey},
    url = {https://ui.adsabs.harvard.edu/abs/2024MNRAS.533.1300D/abstract},
    doi = {10.1093/mnras/stae1836},
    language = {en},
    number = {2},
    urldate = {2026-03-23},
    journal = {MNRAS},
    author = {Derkenne, C. and McDermid, R. M. and Santucci, G. and Poci, A. and Thater, S. and Bellstedt, S. and Mendel, J. T. and Foster, C. and Harborne, K. E. and Lagos, C. D. P. and Wisnioski, E. and Croom, S. and Remus, R.-S. and Valenzuela, L. M. and van de Sande, J. and Sweet, S. M. and Ziegler, B.},
    month = sep,
    year = {2024},
    pages = {1300--1320},
}

@article{genzel_strongly_2017,
    title = {Strongly baryon-dominated disk galaxies at the peak of galaxy formation ten billion years ago},
    volume = {543},
    issn = {0028-0836},
    url = {https://ui.adsabs.harvard.edu/abs/2017Natur.543..397G/abstract},
    doi = {10.1038/nature21685},
    language = {en},
    number = {7645},
    urldate = {2026-04-27},
    journal = {Nature},
    author = {Genzel, R. and Förster Schreiber, N. M. and Übler, H. and Lang, P. and Naab, T. and Bender, R. and Tacconi, L. J. and Wisnioski, E. and Wuyts, S. and Alexander, T. and Beifiori, A. and Belli, S. and Brammer, G. and Burkert, A. and Carollo, C. M. and Chan, J. and Davies, R. and Fossati, M. and Galametz, A. and Genel, S. and Gerhard, O. and Lutz, D. and Mendel, J. T. and Momcheva, I. and Nelson, E. J. and Renzini, A. and Saglia, R. and Sternberg, A. and Tacchella, S. and Tadaki, K. and Wilman, D.},
    month = mar,
    year = {2017},
    pages = {397--401},
}

@article{price_kinematics_2022,
    title = {Kinematics and mass distributions for non-spherical deprojected {Sérsic} density profiles and applications to multi-component galactic systems},
    volume = {665},
    issn = {0004-6361},
    url = {https://ui.adsabs.harvard.edu/abs/2022A&A...665A.159P/abstract},
    doi = {10.1051/0004-6361/202244143},
    language = {en},
    urldate = {2026-04-27},
    journal = {A\&A},
    author = {Price, S. H. and Übler, H. and Förster Schreiber, N. M. and de Zeeuw, P. T. and Burkert, A. and Genzel, R. and Tacconi, L. J. and Davies, R. I. and Price, C. P.},
    month = sep,
    year = {2022},
    pages = {A159},
}

@article{erb_stellar_2006,
    title = {The {Stellar}, {Gas}, and {Dynamical} {Masses} of {Star}-forming {Galaxies} at z {\textasciitilde} 2},
    volume = {646},
    issn = {0004-637X},
    url = {https://ui.adsabs.harvard.edu/abs/2006ApJ...646..107E/abstract},
    doi = {10.1086/504891},
    language = {en},
    number = {1},
    urldate = {2026-04-27},
    journal = {ApJ},
    author = {Erb, Dawn K. and Steidel, Charles C. and Shapley, Alice E. and Pettini, Max and Reddy, Naveen A. and Adelberger, Kurt L.},
    month = jul,
    year = {2006},
    pages = {107--132},
}

@article{sanders_luminous_1996,
    title = {Luminous {Infrared} {Galaxies}},
    volume = {34},
    issn = {0066-4146},
    url = {https://ui.adsabs.harvard.edu/abs/1996ARA&A..34..749S/abstract},
    doi = {10.1146/annurev.astro.34.1.749},
    language = {en},
    urldate = {2026-04-28},
    journal = {ARA\&A},
    author = {Sanders, D. B. and Mirabel, I. F.},
    year = {1996},
    pages = {749},
}

@article{puech_3d_2007,
    title = {{3D} spectroscopy with {VLT}/{GIRAFFE}. {IV}. {Angular} momentum and dynamical support of intermediate redshift galaxies},
    volume = {466},
    issn = {0004-6361},
    url = {https://ui.adsabs.harvard.edu/abs/2007A&A...466...83P/abstract},
    doi = {10.1051/0004-6361:20065978},
    language = {en},
    number = {1},
    urldate = {2026-04-28},
    journal = {A\&A},
    author = {Puech, M. and Hammer, F. and Lehnert, M. D. and Flores, H.},
    month = apr,
    year = {2007},
    pages = {83--92},
}

@article{belfiore_sdss_2018,
    title = {{SDSS} {IV} {MaNGA} - {sSFR} profiles and the slow quenching of discs in green valley galaxies},
    volume = {477},
    issn = {0035-8711},
    url = {https://ui.adsabs.harvard.edu/abs/2018MNRAS.477.3014B/abstract},
    doi = {10.1093/mnras/sty768},
    language = {en},
    number = {3},
    urldate = {2026-04-28},
    journal = {MNRAS},
    author = {Belfiore, Francesco and Maiolino, Roberto and Bundy, Kevin and Masters, Karen and Bershady, Matthew and Oyarzún, Grecco A. and Lin, Lihwai and Cano-Diaz, Mariana and Wake, David and Spindler, Ashley and Thomas, Daniel and Brownstein, Joel R. and Drory, Niv and Yan, Renbin},
    month = jul,
    year = {2018},
    pages = {3014--3029},
}

@article{tsukui_emergence_2025,
    title = {The emergence of galactic thin and thick discs across cosmic history},
    volume = {540},
    issn = {0035-8711},
    url = {https://ui.adsabs.harvard.edu/abs/2025MNRAS.540.3493T/abstract},
    doi = {10.1093/mnras/staf604},
    language = {en},
    number = {4},
    urldate = {2026-05-18},
    journal = {MNRAS},
    author = {Tsukui, Takafumi and Wisnioski, Emily and Bland-Hawthorn, Joss and Freeman, Ken},
    month = jul,
    year = {2025},
    pages = {3493--3522},
}

@article{tsukui_spatially_2023,
    title = {Spatially resolved dust properties and quasar-galaxy decomposition of a hyper-luminous infrared galaxy at z = 4.4},
    volume = {523},
    issn = {0035-8711},
    url = {https://ui.adsabs.harvard.edu/abs/2023MNRAS.523.4654T/abstract},
    doi = {10.1093/mnras/stad1464},
    language = {en},
    number = {3},
    urldate = {2026-05-18},
    journal = {MNRAS},
    author = {Tsukui, Takafumi and Wisnioski, Emily and Krumholz, Mark R. and Battisti, Andrew},
    month = aug,
    year = {2023},
    pages = {4654--4679},
}

@ARTICLE{collaboration-gaia-2016,
       author = {{Gaia Collaboration} and {Prusti}, T. and {de Bruijne}, J.~H.~J. and {Brown}, A.~G.~A. and {Vallenari}, A. and {Babusiaux}, C. and {Bailer-Jones}, C.~A.~L. and {Bastian}, U. and {Biermann}, M. and {Evans}, D.~W. and {Eyer}, L. and {Jansen}, F. and {Jordi}, C. and {Klioner}, S.~A. and {Lammers}, U. and {Lindegren}, L. and {Luri}, X. and {Mignard}, F. and {Milligan}, D.~J. and {Panem}, C. and {Poinsignon}, V. and {Pourbaix}, D. and {Randich}, S. and {Sarri}, G. and {Sartoretti}, P. and {Siddiqui}, H.~I. and {Soubiran}, C. and {Valette}, V. and {van Leeuwen}, F. and {Walton}, N.~A. and {Aerts}, C. and {Arenou}, F. and {Cropper}, M. and {Drimmel}, R. and {H{\o}g}, E. and {Katz}, D. and {Lattanzi}, M.~G. and {O'Mullane}, W. and {Grebel}, E.~K. and {Holland}, A.~D. and {Huc}, C. and {Passot}, X. and {Bramante}, L. and {Cacciari}, C. and {Casta{\~n}eda}, J. and {Chaoul}, L. and {Cheek}, N. and {De Angeli}, F. and {Fabricius}, C. and {Guerra}, R. and {Hern{\'a}ndez}, J. and {Jean-Antoine-Piccolo}, A. and {Masana}, E. and {Messineo}, R. and {Mowlavi}, N. and {Nienartowicz}, K. and {Ord{\'o}{\~n}ez-Blanco}, D. and {Panuzzo}, P. and {Portell}, J. and {Richards}, P.~J. and {Riello}, M. and {Seabroke}, G.~M. and {Tanga}, P. and {Th{\'e}venin}, F. and {Torra}, J. and {Els}, S.~G. and {Gracia-Abril}, G. and {Comoretto}, G. and {Garcia-Reinaldos}, M. and {Lock}, T. and {Mercier}, E. and {Altmann}, M. and {Andrae}, R. and {Astraatmadja}, T.~L. and {Bellas-Velidis}, I. and {Benson}, K. and {Berthier}, J. and {Blomme}, R. and {Busso}, G. and {Carry}, B. and {Cellino}, A. and {Clementini}, G. and {Cowell}, S. and {Creevey}, O. and {Cuypers}, J. and {Davidson}, M. and {De Ridder}, J. and {de Torres}, A. and {Delchambre}, L. and {Dell'Oro}, A. and {Ducourant}, C. and {Fr{\'e}mat}, Y. and {Garc{\'\i}a-Torres}, M. and {Gosset}, E. and {Halbwachs}, J.-L. and {Hambly}, N.~C. and {Harrison}, D.~L. and {Hauser}, M. and {Hestroffer}, D. and {Hodgkin}, S.~T. and {Huckle}, H.~E. and {Hutton}, A. and {Jasniewicz}, G. and {Jordan}, S. and {Kontizas}, M. and {Korn}, A.~J. and {Lanzafame}, A.~C. and {Manteiga}, M. and {Moitinho}, A. and {Muinonen}, K. and {Osinde}, J. and {Pancino}, E. and {Pauwels}, T. and {Petit}, J.-M. and {Recio-Blanco}, A. and {Robin}, A.~C. and {Sarro}, L.~M. and {Siopis}, C. and {Smith}, M. and {Smith}, K.~W. and {Sozzetti}, A. and {Thuillot}, W. and {van Reeven}, W. and {Viala}, Y. and {Abbas}, U. and {Abreu Aramburu}, A. and {Accart}, S. and {Aguado}, J.~J. and {Allan}, P.~M. and {Allasia}, W. and {Altavilla}, G. and {{\'A}lvarez}, M.~A. and {Alves}, J. and {Anderson}, R.~I. and {Andrei}, A.~H. and {Anglada Varela}, E. and {Antiche}, E. and {Antoja}, T. and {Ant{\'o}n}, S. and {Arcay}, B. and {Atzei}, A. and {Ayache}, L. and {Bach}, N. and {Baker}, S.~G. and {Balaguer-N{\'u}{\~n}ez}, L. and {Barache}, C. and {Barata}, C. and {Barbier}, A. and {Barblan}, F. and {Baroni}, M. and {Barrado y Navascu{\'e}s}, D. and {Barros}, M. and {Barstow}, M.~A. and {Becciani}, U. and {Bellazzini}, M. and {Bellei}, G. and {Bello Garc{\'\i}a}, A. and {Belokurov}, V. and {Bendjoya}, P. and {Berihuete}, A. and {Bianchi}, L. and {Bienaym{\'e}}, O. and {Billebaud}, F. and {Blagorodnova}, N. and {Blanco-Cuaresma}, S. and {Boch}, T. and {Bombrun}, A. and {Borrachero}, R. and {Bouquillon}, S. and {Bourda}, G. and {Bouy}, H. and {Bragaglia}, A. and {Breddels}, M.~A. and {Brouillet}, N. and {Br{\"u}semeister}, T. and {Bucciarelli}, B. and {Budnik}, F. and {Burgess}, P. and {Burgon}, R. and {Burlacu}, A. and {Busonero}, D. and {Buzzi}, R. and {Caffau}, E. and {Cambras}, J. and {Campbell}, H. and {Cancelliere}, R. and {Cantat-Gaudin}, T. and {Carlucci}, T. and {Carrasco}, J.~M. and {Castellani}, M. and {Charlot}, P. and {Charnas}, J. and {Charvet}, P. and {Chassat}, F. and {Chiavassa}, A. and {Clotet}, M. and {Cocozza}, G. and {Collins}, R.~S. and {Collins}, P. and {Costigan}, G.},
        title = "{The Gaia mission}",
      journal = {\aap},
         year = 2016,
        month = nov,
       volume = {595},
          eid = {A1},
        pages = {A1},
          doi = {10.1051/0004-6361/201629272},
archivePrefix = {arXiv},
       eprint = {1609.04153},
 primaryClass = {astro-ph.IM},
       adsurl = {https://ui.adsabs.harvard.edu/abs/2016A&A...595A...1G}
}

@ARTICLE{collaboration-gaia-2023,
       author = {{Gaia Collaboration} and {Vallenari}, A. and {Brown}, A.~G.~A. and {Prusti}, T. and {de Bruijne}, J.~H.~J. and {Arenou}, F. and {Babusiaux}, C. and {Biermann}, M. and {Creevey}, O.~L. and {Ducourant}, C. and {Evans}, D.~W. and {Eyer}, L. and {Guerra}, R. and {Hutton}, A. and {Jordi}, C. and {Klioner}, S.~A. and {Lammers}, U.~L. and {Lindegren}, L. and {Luri}, X. and {Mignard}, F. and {Panem}, C. and {Pourbaix}, D. and {Randich}, S. and {Sartoretti}, P. and {Soubiran}, C. and {Tanga}, P. and {Walton}, N.~A. and {Bailer-Jones}, C.~A.~L. and {Bastian}, U. and {Drimmel}, R. and {Jansen}, F. and {Katz}, D. and {Lattanzi}, M.~G. and {van Leeuwen}, F. and {Bakker}, J. and {Cacciari}, C. and {Casta{\~n}eda}, J. and {De Angeli}, F. and {Fabricius}, C. and {Fouesneau}, M. and {Fr{\'e}mat}, Y. and {Galluccio}, L. and {Guerrier}, A. and {Heiter}, U. and {Masana}, E. and {Messineo}, R. and {Mowlavi}, N. and {Nicolas}, C. and {Nienartowicz}, K. and {Pailler}, F. and {Panuzzo}, P. and {Riclet}, F. and {Roux}, W. and {Seabroke}, G.~M. and {Sordo}, R. and {Th{\'e}venin}, F. and {Gracia-Abril}, G. and {Portell}, J. and {Teyssier}, D. and {Altmann}, M. and {Andrae}, R. and {Audard}, M. and {Bellas-Velidis}, I. and {Benson}, K. and {Berthier}, J. and {Blomme}, R. and {Burgess}, P.~W. and {Busonero}, D. and {Busso}, G. and {C{\'a}novas}, H. and {Carry}, B. and {Cellino}, A. and {Cheek}, N. and {Clementini}, G. and {Damerdji}, Y. and {Davidson}, M. and {de Teodoro}, P. and {Nu{\~n}ez Campos}, M. and {Delchambre}, L. and {Dell'Oro}, A. and {Esquej}, P. and {Fern{\'a}ndez-Hern{\'a}ndez}, J. and {Fraile}, E. and {Garabato}, D. and {Garc{\'\i}a-Lario}, P. and {Gosset}, E. and {Haigron}, R. and {Halbwachs}, J.-L. and {Hambly}, N.~C. and {Harrison}, D.~L. and {Hern{\'a}ndez}, J. and {Hestroffer}, D. and {Hodgkin}, S.~T. and {Holl}, B. and {Jan{\ss}en}, K. and {Jevardat de Fombelle}, G. and {Jordan}, S. and {Krone-Martins}, A. and {Lanzafame}, A.~C. and {L{\"o}ffler}, W. and {Marchal}, O. and {Marrese}, P.~M. and {Moitinho}, A. and {Muinonen}, K. and {Osborne}, P. and {Pancino}, E. and {Pauwels}, T. and {Recio-Blanco}, A. and {Reyl{\'e}}, C. and {Riello}, M. and {Rimoldini}, L. and {Roegiers}, T. and {Rybizki}, J. and {Sarro}, L.~M. and {Siopis}, C. and {Smith}, M. and {Sozzetti}, A. and {Utrilla}, E. and {van Leeuwen}, M. and {Abbas}, U. and {{\'A}brah{\'a}m}, P. and {Abreu Aramburu}, A. and {Aerts}, C. and {Aguado}, J.~J. and {Ajaj}, M. and {Aldea-Montero}, F. and {Altavilla}, G. and {{\'A}lvarez}, M.~A. and {Alves}, J. and {Anders}, F. and {Anderson}, R.~I. and {Anglada Varela}, E. and {Antoja}, T. and {Baines}, D. and {Baker}, S.~G. and {Balaguer-N{\'u}{\~n}ez}, L. and {Balbinot}, E. and {Balog}, Z. and {Barache}, C. and {Barbato}, D. and {Barros}, M. and {Barstow}, M.~A. and {Bartolom{\'e}}, S. and {Bassilana}, J.-L. and {Bauchet}, N. and {Becciani}, U. and {Bellazzini}, M. and {Berihuete}, A. and {Bernet}, M. and {Bertone}, S. and {Bianchi}, L. and {Binnenfeld}, A. and {Blanco-Cuaresma}, S. and {Blazere}, A. and {Boch}, T. and {Bombrun}, A. and {Bossini}, D. and {Bouquillon}, S. and {Bragaglia}, A. and {Bramante}, L. and {Breedt}, E. and {Bressan}, A. and {Brouillet}, N. and {Brugaletta}, E. and {Bucciarelli}, B. and {Burlacu}, A. and {Butkevich}, A.~G. and {Buzzi}, R. and {Caffau}, E. and {Cancelliere}, R. and {Cantat-Gaudin}, T. and {Carballo}, R. and {Carlucci}, T. and {Carnerero}, M.~I. and {Carrasco}, J.~M. and {Casamiquela}, L. and {Castellani}, M. and {Castro-Ginard}, A. and {Chaoul}, L. and {Charlot}, P. and {Chemin}, L. and {Chiaramida}, V. and {Chiavassa}, A. and {Chornay}, N. and {Comoretto}, G. and {Contursi}, G. and {Cooper}, W.~J. and {Cornez}, T. and {Cowell}, S. and {Crifo}, F. and {Cropper}, M. and {Crosta}, M. and {Crowley}, C. and {Dafonte}, C. and {Dapergolas}, A. and {David}, M. and {David}, P. and {de Laverny}, P. and {De Luise}, F. and {De March}, R.},
        title = "{Gaia Data Release 3. Summary of the content and survey properties}",
      journal = {\aap},
         year = 2023,
        month = jun,
       volume = {674},
          eid = {A1},
        pages = {A1},
          doi = {10.1051/0004-6361/202243940},
archivePrefix = {arXiv},
       eprint = {2208.00211},
 primaryClass = {astro-ph.GA},
       adsurl = {https://ui.adsabs.harvard.edu/abs/2023A&A...674A...1G}
}

@article{danhaive_dark_2026,
    title = {The dark side of early galaxies: {GEKO} uncovers dark-matter fractions at z ∼ 4 − 6},
    volume = {546},
    issn = {0035-8711},
    shorttitle = {The dark side of early galaxies},
    url = {https://ui.adsabs.harvard.edu/abs/2026MNRAS.546ag119D/abstract},
    doi = {10.1093/mnras/stag119},
    language = {en},
    number = {3},
    urldate = {2026-05-19},
    journal = {MNRAS},
    author = {Danhaive, A. Lola and Tacchella, Sandro and Bunker, Andrew J. and Curtis-Lake, Emma and de Graaff, Anna and D'Eugenio, Francesco and Duan, Qiao and Egami, Eiichi and Eisenstein, Daniel J. and Johnson, Benjamin D. and Maiolino, Roberto and McClymont, William and Rieke, Marcia and Robertson, Brant and Sun, Fengwu and Willmer, Christopher N. A. and Wu, Zihao and Zhu, Yongda},
    month = mar,
    year = {2026},
    pages = {stag119},
}

@article{kohandel_dynamically_2024,
    title = {Dynamically cold disks in the early {Universe}: {Myth} or reality?},
    volume = {685},
    issn = {0004-6361},
    shorttitle = {Dynamically cold disks in the early {Universe}},
    url = {https://ui.adsabs.harvard.edu/abs/2024A&A...685A..72K/abstract},
    doi = {10.1051/0004-6361/202348209},
    language = {en},
    urldate = {2026-05-20},
    journal = {A\&A},
    author = {Kohandel, M. and Pallottini, A. and Ferrara, A. and Zanella, A. and Rizzo, F. and Carniani, S.},
    month = may,
    year = {2024},
    pages = {A72},
}

@article{bhagwat_stellar_2026,
    title = {Stellar feedback {SPICEs} up [{C} {II}] emission in the first galaxies},
    url = {https://ui.adsabs.harvard.edu/abs/2026arXiv260418731B/abstract},
    doi = {10.48550/arXiv.2604.18731},
    language = {en},
    urldate = {2026-05-20},
    journal = {MNRAS subm.},
    author = {Bhagwat, Aniket and Costa, Tiago and Ciardi, Benedetta and Arrigoni Battaia, Fabrizio},
    month = apr,
    year = {2026},
    pages = {arXiv:2604.18731},
}

\begin{appendix}

\FloatBarrier 
\onecolumn
\section{Kinematic extraction with {\scshape linefit}}\label{ap:linefit}
In this section, we describe the {\scshape linefit} fitting code and the extraction of the 2D kinematics from data cubes. The code takes as input the data cube, a noise cube used both for the fitting and for uncertainty estimate, as well as a template line spread function (LSF) to be convolved with the fit (which in the case of interferometric data is inferior to the channel width). For each of the MUSE-ALMA Haloes galaxies used in the analysis, the input noise cube is constructed by taking the RMS of regions of the original cube with no detected sources. {\scshape linefit} extracts the flux, velocity and velocity dispersion maps from the data cubes by performing a pixel-by-pixel Gaussian fit to the spectrum, after subtraction of the continuum and 2$\sigma$-clipping. The Gaussian fit is performed within a given wavelength range which contains the emission line, while the continuum is estimated through a first-order polynomial fit within two wavelength intervals framing the emission line. The Gaussian fit is weighted by the given input noise cube. The area, centroid and standard deviation of the Gaussian fit are then used to reconstruct the pixel-by-pixel flux, velocity and velocity dispersion maps respectively. The uncertainties are computed through 100 Monte Carlo (MC) simulations where each point of the spectra is perturbed using a Gaussian distribution with standard deviation given by the input noise cube.

\begin{figure*}[ht!]
\section{{\scshape DysmalPy} residuals and corner plots}\label{ap:dysmal}
In this section, we present the kinematic fit to the cold molecular and ionised gas for the remaining three galaxies: Q2131-120\_5, Q0152-2001\_12 and Q1345-0023\_13. The methodology is described in detail in Sec. \ref{sec:dysmalpy}, and the results are discussed in Sec. \ref{sec:disc}. 

\subsection{Q2131-120\_5}
    \centering
    \includegraphics[width=\textwidth]{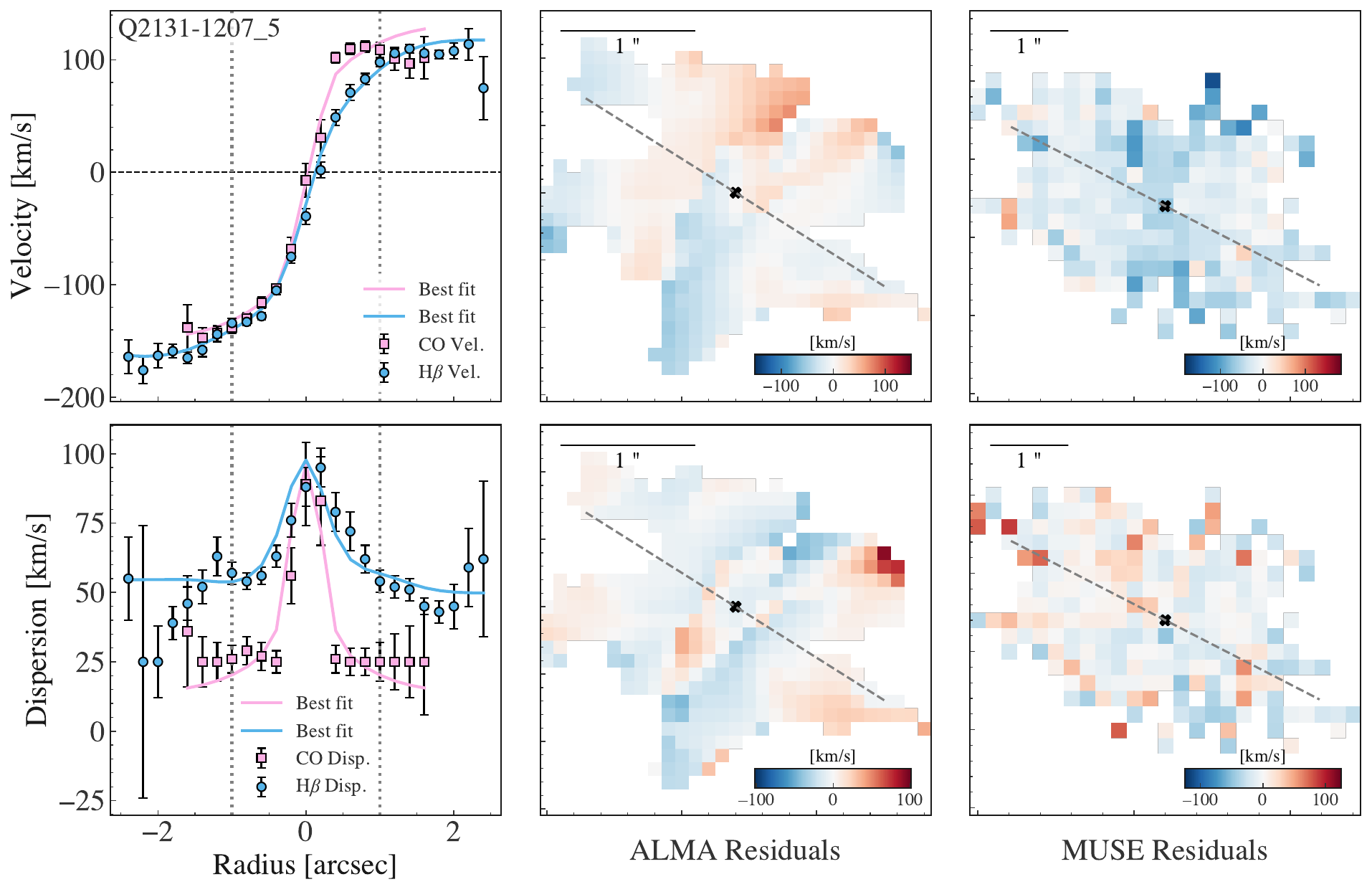}
    \caption{Fit from the kinematic modelling with {\scshape DysmalPy} for Q2131-1207\_5. The top row shows the fit results and residuals to the velocity profile, whereas the bottom row shows the fit and residuals for the velocity dispersion profile. The left column displays the 1D profiles extracted from the ALMA (pink data points) and MUSE (blue data points) data, as well as the best-fit profiles from {\scshape DysmalPy}. The grey dotted lines delimit the effective radius of the galaxy from \cite{karki_muse-alma_2023}. The middle and right column show the reconstructed 2D velocity and velocity dispersion residual maps, for the ALMA and MUSE data respectively.}
    \label{fig:2131_dysmal}
\end{figure*}

\begin{figure*}
    \centering
    \includegraphics[width=\textwidth]{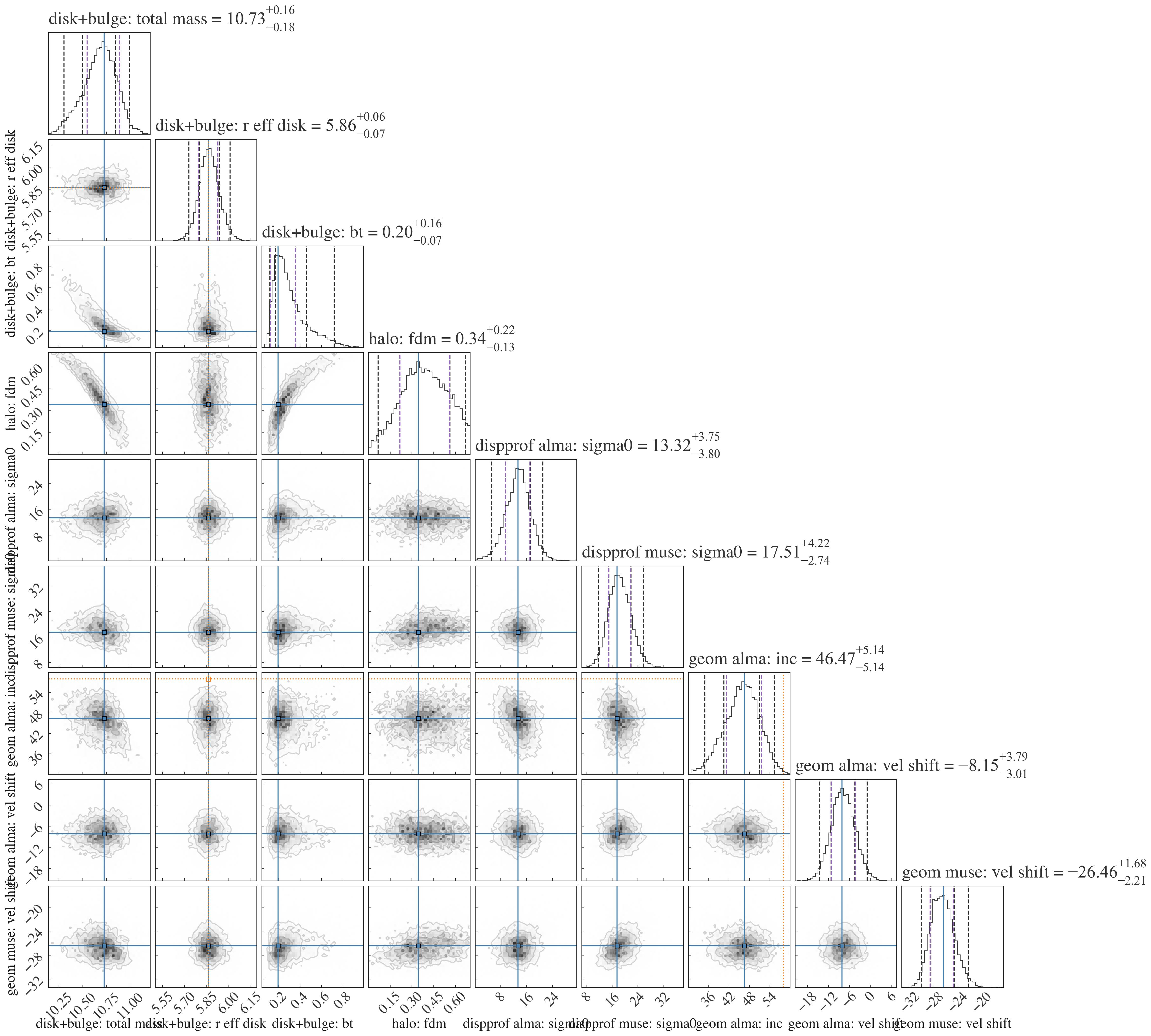}
    \caption{{\scshape DysmalPy} corner plots for the best-fit of the CO(2-1) and H$\beta$ data of Q2131-1207\_5.}
    \label{fig:2131_corner_co}
\end{figure*}

\begin{figure*}
\subsection{Q0152-2001\_12}
    \centering
    \includegraphics[width=\textwidth]{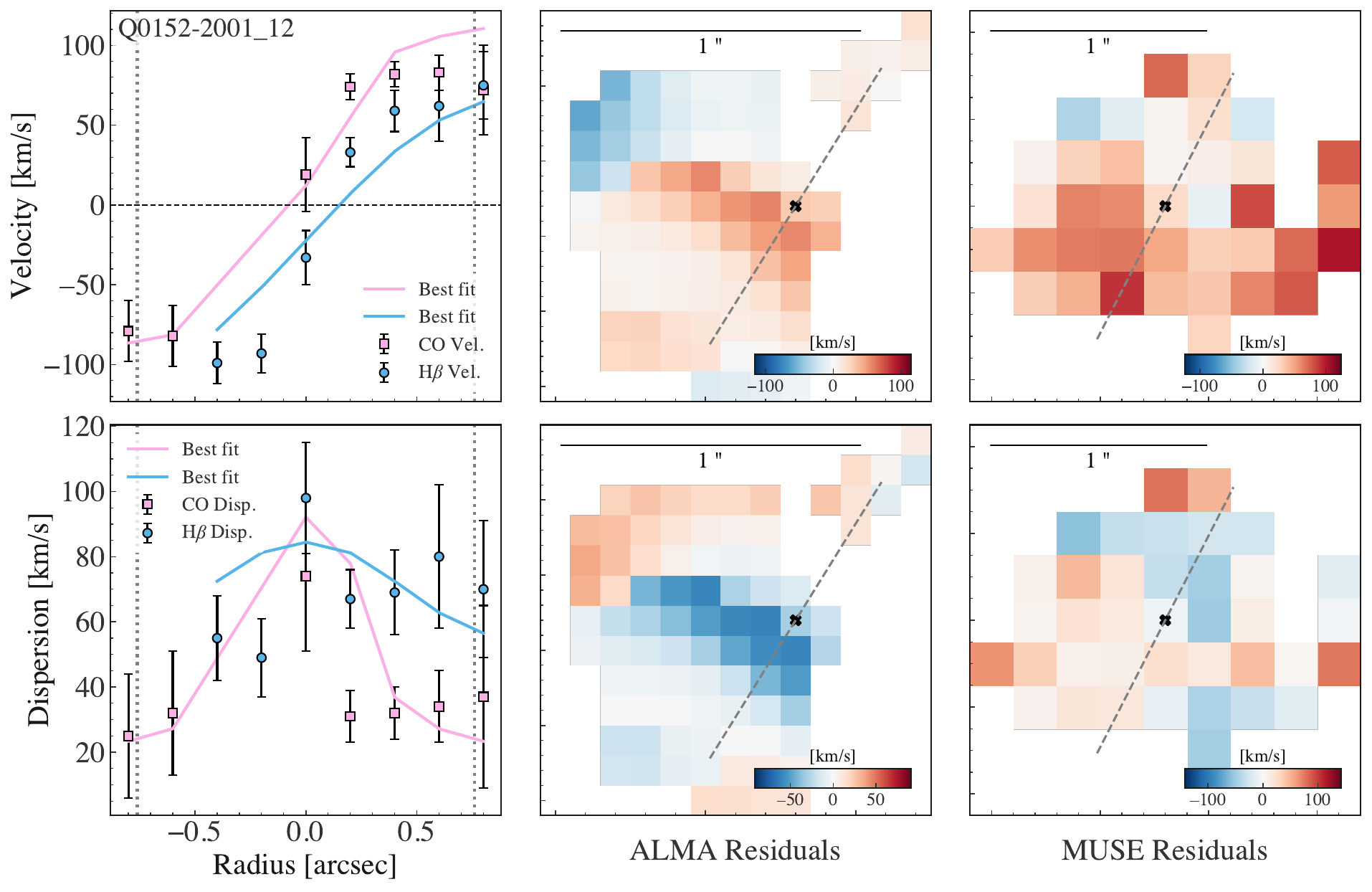}
    \caption{Fit from the kinematic modelling with {\scshape DysmalPy} for Q0152-2001\_12. The top row shows the fit and residuals to the velocity profile, whereas the bottom row shows the fit and residuals for the velocity dispersion profile. The left column displays the 1D profiles extracted from the ALMA (pink data points) and MUSE (blue data points) data, as well as the best-fit profiles from {\scshape DysmalPy}. The grey dotted lines delimit the effective radius of the galaxy from \cite{karki_muse-alma_2023}. The middle and right column show the reconstructed 2D velocity and velocity dispersion residual maps, for the ALMA and MUSE data respectively.}
    \label{fig:0152_dysmal}
\end{figure*}

\begin{figure*}
    \centering
    \includegraphics[width=\textwidth]{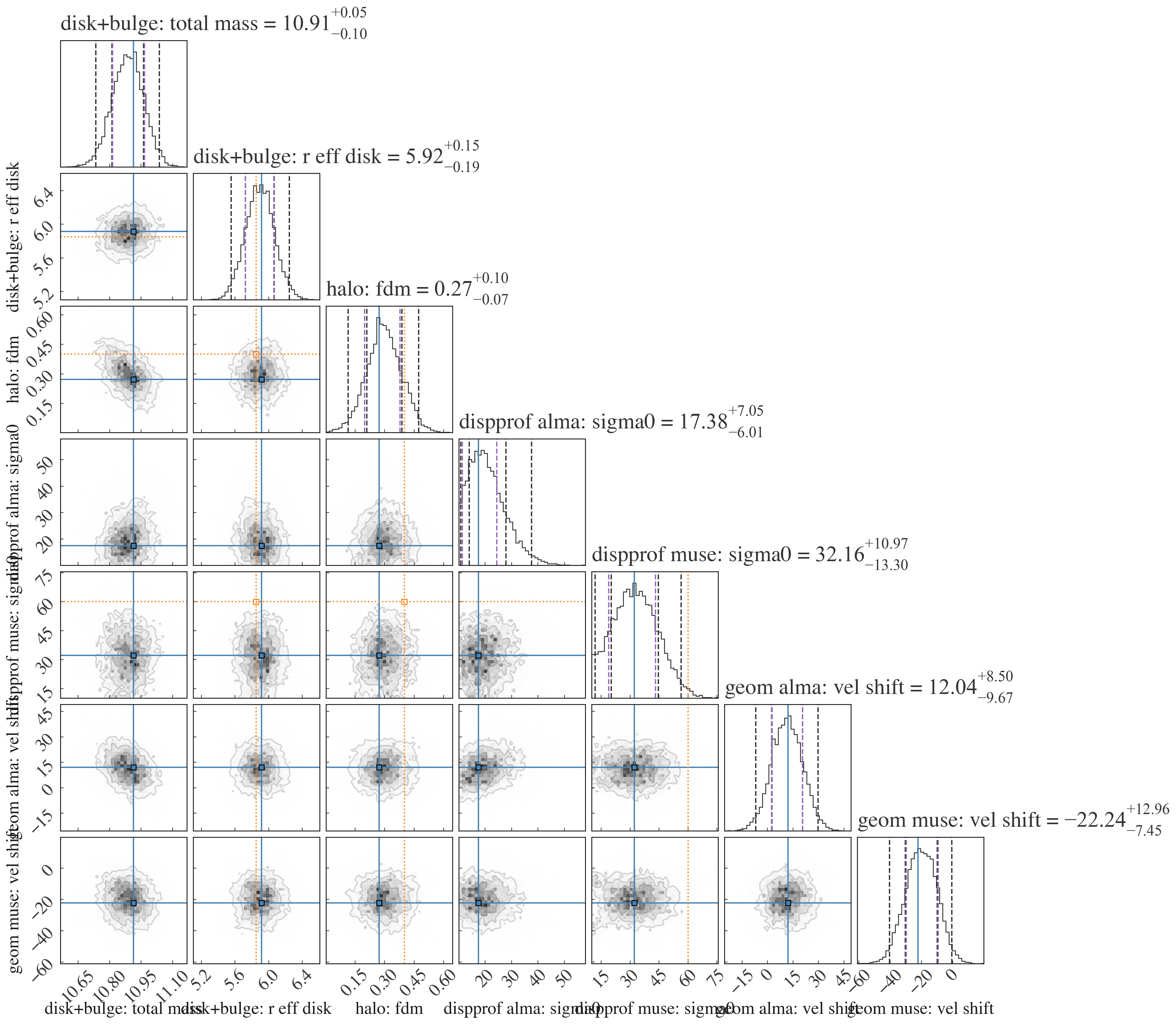}
    \caption{{\scshape DysmalPy} corner plots for the best-fit kinematics of the CO(2-1) and H$\beta$ data of Q0152-2001\_12.}
    \label{fig:0152_corner_co}
\end{figure*}

\begin{figure*}
\subsection{Q1345-0023\_13}
    \centering
    \includegraphics[width=\textwidth]{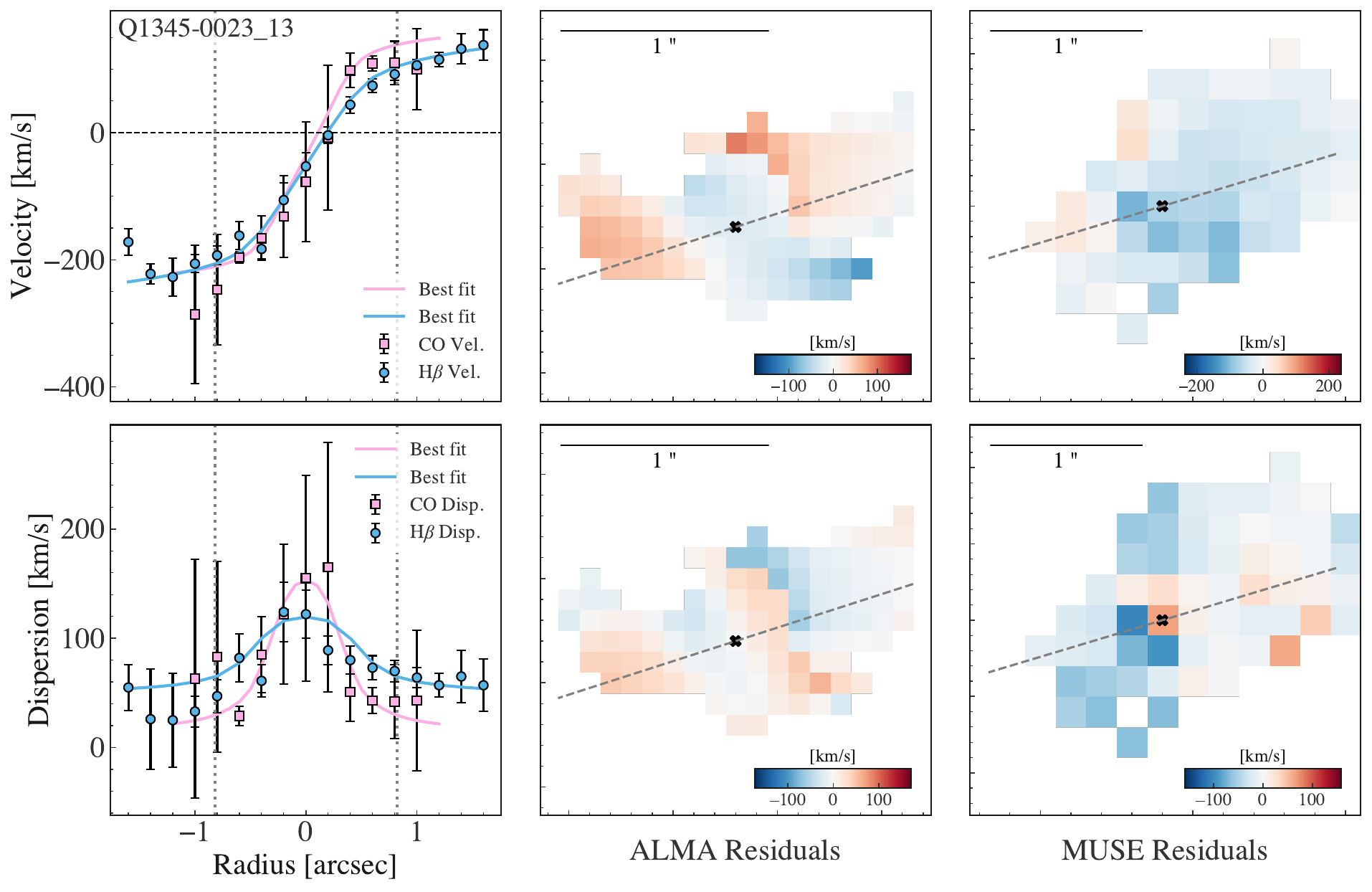}
    \caption{Fit results from the kinematic modelling with {\scshape DysmalPy} for Q1345-0023\_13. The top row shows the fit and residuals of the velocity profile, whereas the bottom row shows the fit and residuals for the velocity dispersion profile. The left column displays the 1D profiles extracted from the ALMA (pink data points) and MUSE (blue data points) data, as well as the best-fit profiles from {\scshape DysmalPy}. The grey dotted lines delimit the effective radius of the galaxy from \cite{karki_muse-alma_2023}. The middle and right column show the reconstructed 2D velocity and velocity dispersion residual maps, for the ALMA and MUSE data respectively.}
    \label{fig:1345_dysmal}
\end{figure*}

\begin{figure*}
    \centering
    \includegraphics[width=\textwidth]{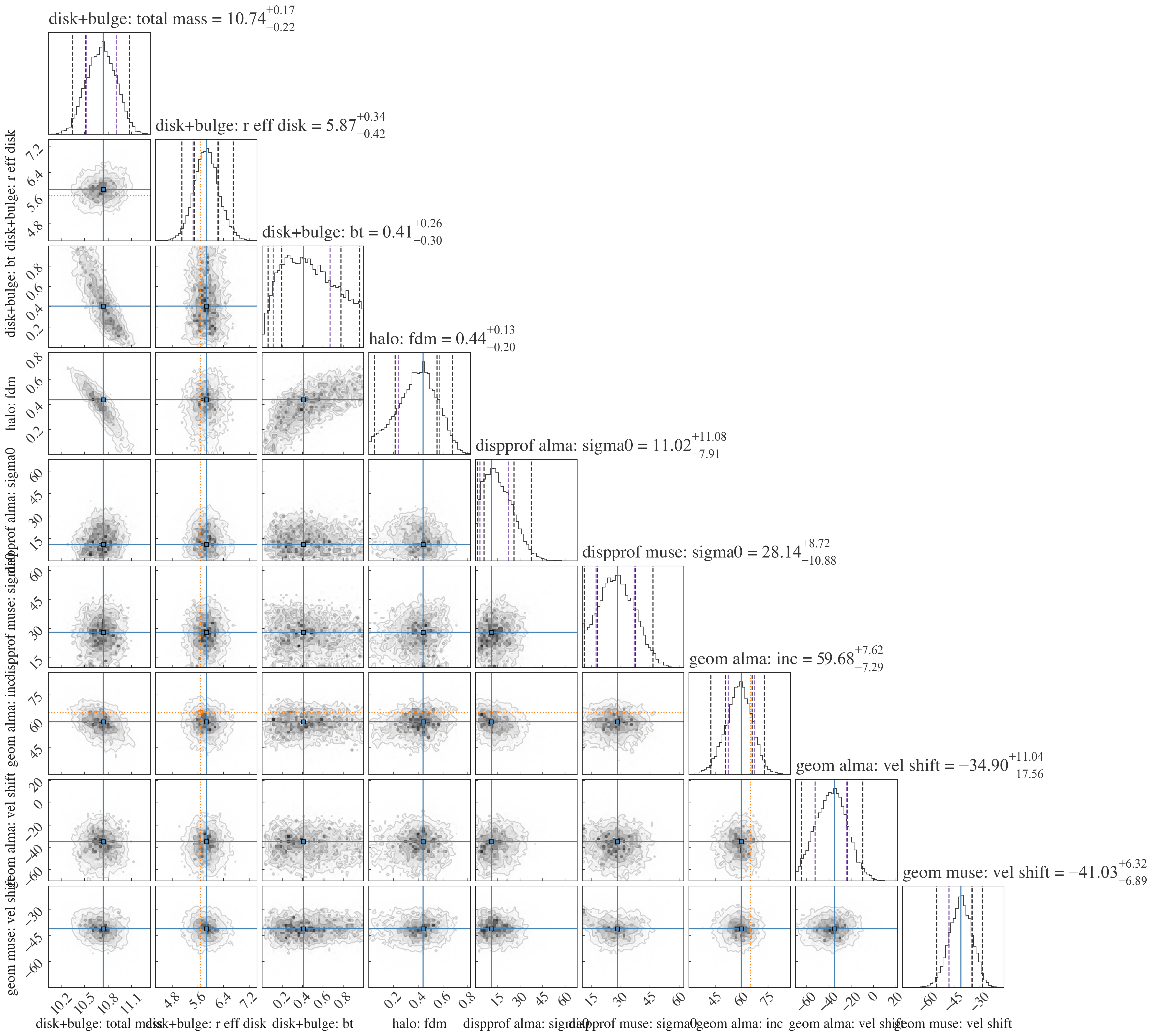}
    \caption{{\scshape DysmalPy} corner plots for the best-fit kinematics of the CO(2-1) and H$\beta$ data of Q1345-0023\_13.}
    \label{fig:1345_corner_co}
\end{figure*}

\begin{figure*}
\subsection{Q0454-220\_4}
    \centering
    \includegraphics[width=\textwidth]{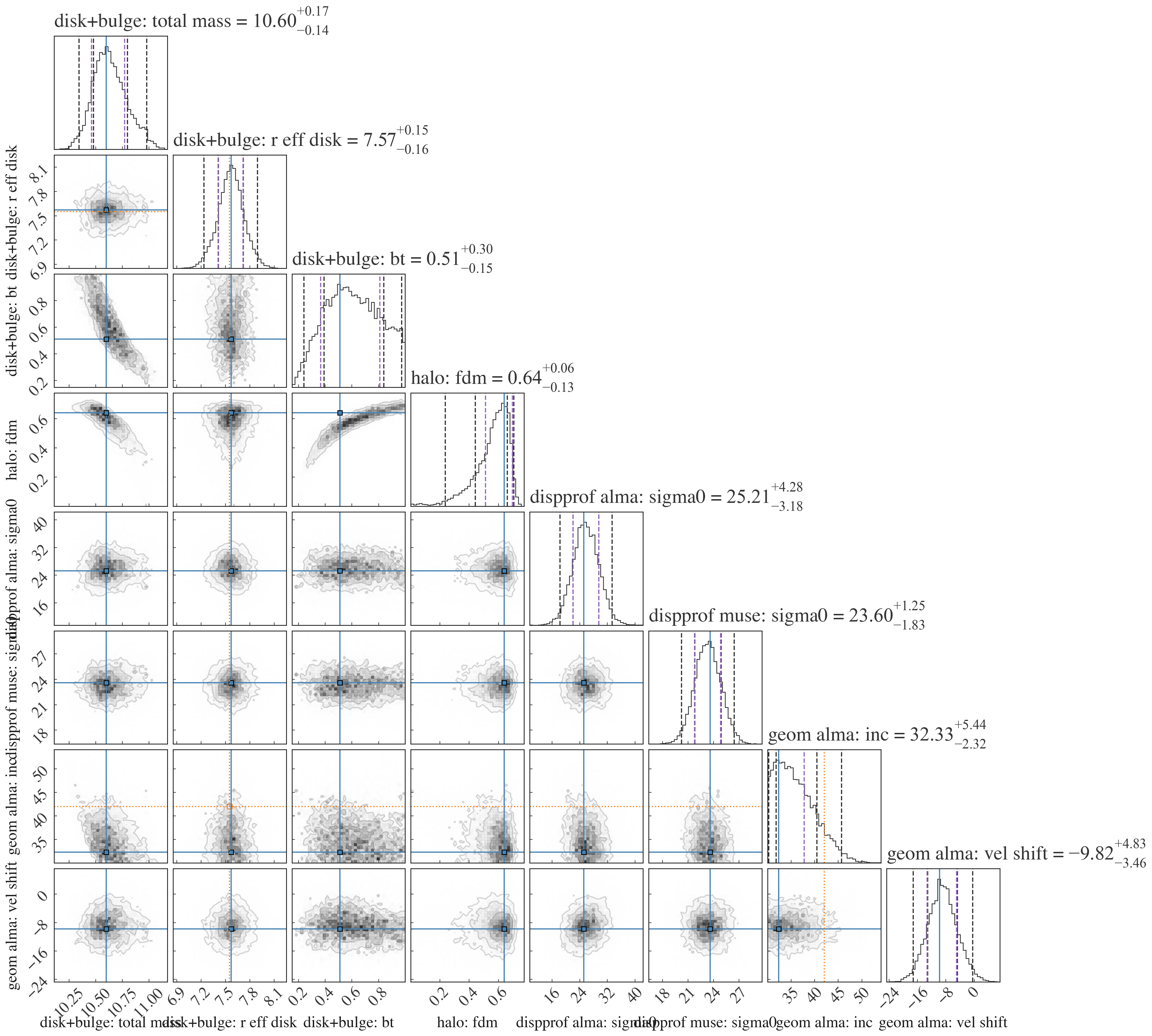}
    \caption{{\scshape DysmalPy} corner plots for the best-fit kinematics of the CO(2-1) and H$\beta$ data of Q0454-220\_4.}
    \label{fig:0454_corner_co}
\end{figure*}

\onecolumn
\begin{figure*}
\section{Outflow search}\label{ap:outflow}
\subsection{Integrated spectra}

In this section, we present the remaining six integrated spectra from the analysis performed in Sec. \ref{sec:outflows} for Q0454-200\_4, Q1345-0023\_13 and Q0152-2001\_12. As for Q2131-1207\_5, the CO spectrum of Q0454-200\_4 shows signatures of a broad component; however, we show in the following section that these can be attributed to instrumental effects. 

    \centering
    \begin{subfigure}[b]{0.5\textwidth}
        \centering
        \includegraphics[width=\textwidth]{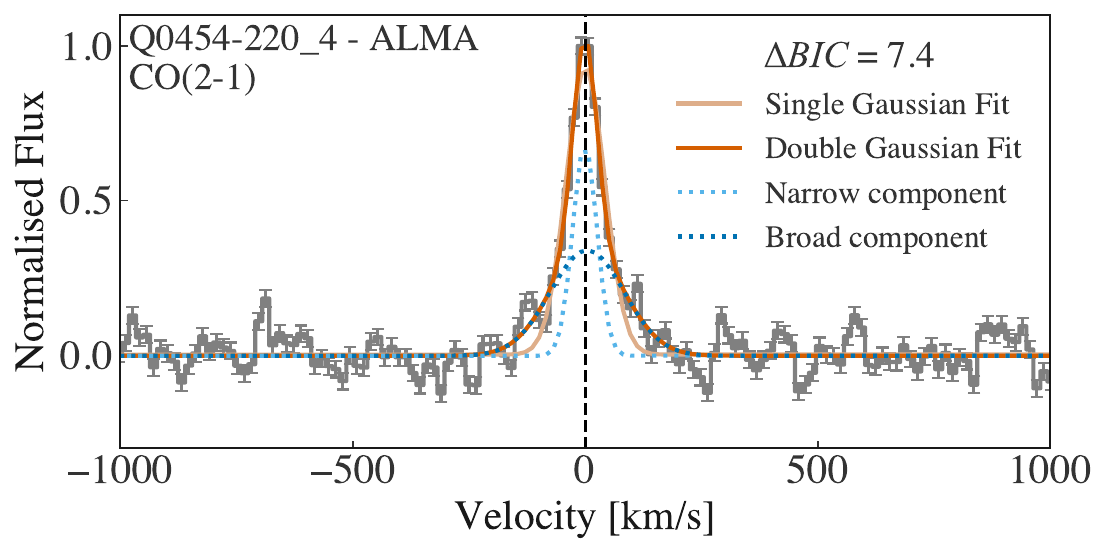}
    \end{subfigure}%
    ~ 
    \begin{subfigure}[b]{0.5\textwidth}
        \centering
        \includegraphics[width=\textwidth]{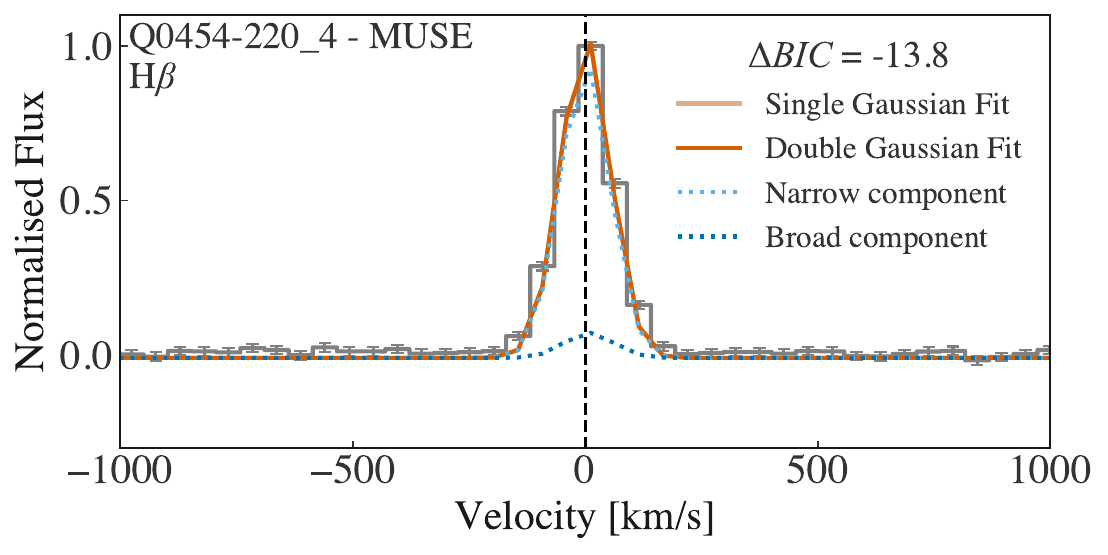}
    \end{subfigure}
    \caption{Integrated spectra of the CO(2-1) (left) and H$\beta$ (right) emission of Q0454-200\_4. In both plots, the yellow line shows the best-fit single Gaussian, the red line shows the double Gaussian fit, and the light and dark blue dotted lines show the narrow and broad components of the double Gaussian component respectively.}
    \label{fig:0454_outflows}
\end{figure*}

\begin{figure*}
    \centering
    \begin{subfigure}[b]{0.5\textwidth}
        \centering
        \includegraphics[width=\textwidth]{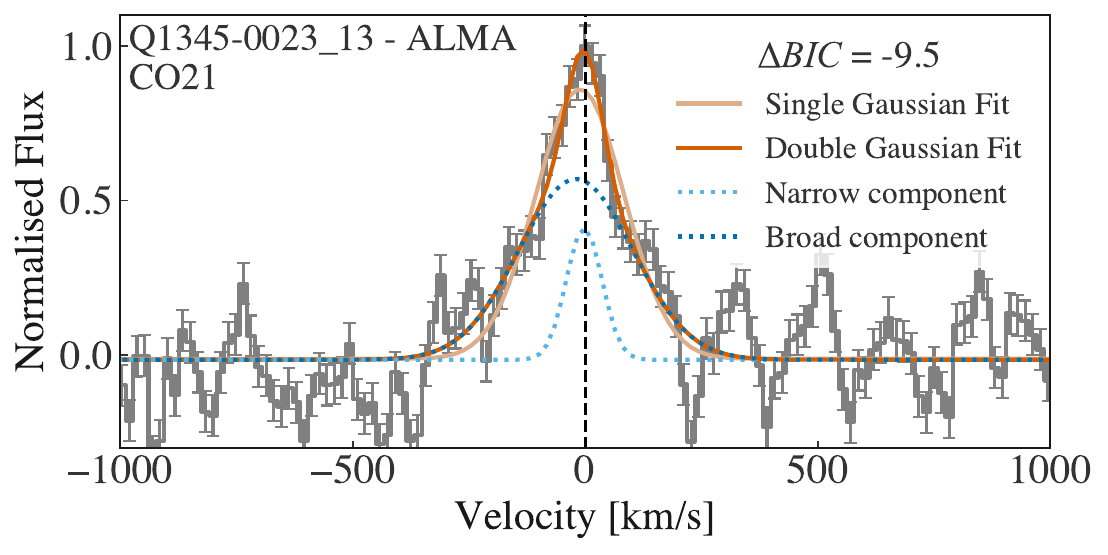}
    \end{subfigure}%
    ~ 
    \begin{subfigure}[b]{0.5\textwidth}
        \centering
        \includegraphics[width=\textwidth]{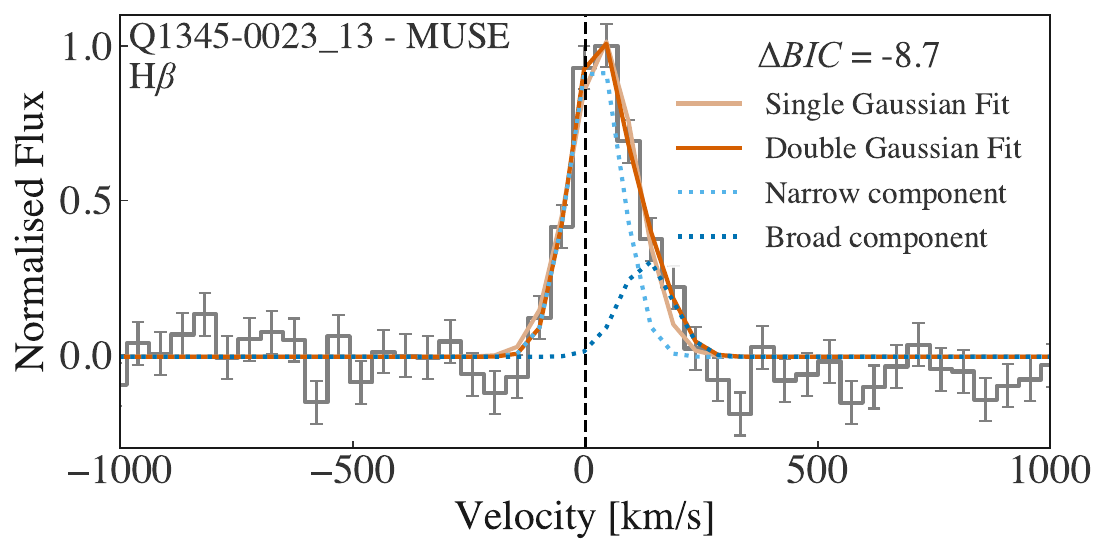}
    \end{subfigure}
    \caption{Integrated spectra of the CO(2-1) (left) and H$\beta$ (right) emission of Q1345-0023\_13. In both plots, the yellow line shows the best-fit single Gaussian, the red line shows the double Gaussian fit, and the light and dark blue dotted lines show the narrow and broad components of the double Gaussian component respectively.}
    \label{fig:1345_outflows}
\end{figure*}

\begin{figure*}
    \centering
    \begin{subfigure}[b]{0.5\textwidth}
        \centering
        \includegraphics[width=\textwidth]{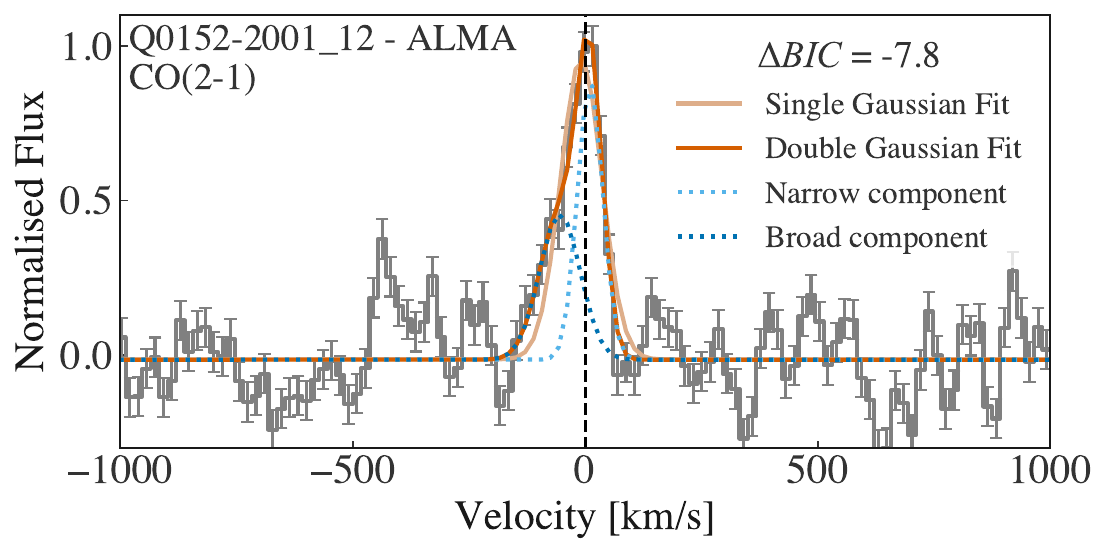}
    \end{subfigure}%
    ~ 
    \begin{subfigure}[b]{0.5\textwidth}
        \centering
        \includegraphics[width=\textwidth]{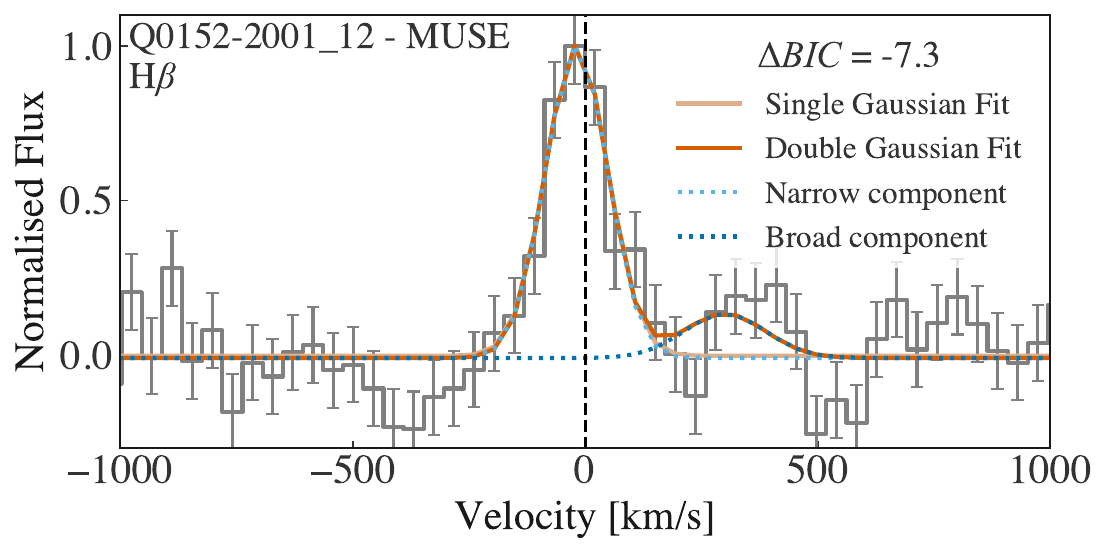}
    \end{subfigure}
    \caption{Integrated spectra of the CO(2-1) (left) and H$\beta$ (right) emission of Q0152-2001\_12. In both plots, the yellow line shows the best-fit single Gaussian, the red line shows the double Gaussian fit, and the light and dark blue dotted lines show the narrow and broad components of the double Gaussian component respectively.}
    \label{fig:0152_outflows}
\end{figure*}

\begin{figure*}
\subsection{Comparison to {\scshape DysmalPy} model}
As mentioned in Sec. \ref{sec:outflows}, for the integrated spectra of the CO emission in Q0454-220\_4 and Q2131-1207\_5, the $\Delta \mathrm{BIC}$ indicates a moderate preference for a double Gaussian fit rather than a single Gaussian fit. In order to investigate if this results from a physical process or instrumental effect, we use the bestfit {\scshape DysmalPy} models obtained for both galaxies. The software produces data cubes with the model galaxy (following the best fit mass profile and geometry) and convolves it with instrumental effects such as beam smearing and LSF. As a result, since we have constructed the model of a purely rotating galaxy, if we extract an integrated spectrum over the same area as the data, shifting each pixel by the best-fit velocity, we have an integrated galaxy line convolved with the instrumental effects. For both galaxies, we performed this analysis and used the resulting integrated line as a new "model" for our data. We computed $\Delta\mathrm{BIC} = \mathrm{BIC_{double}} - \mathrm{BIC_{DysmalPy}}$, and in both cases, we found that the fit strongly favours the {\scshape DysmalPy} model to the double Gaussian component model indicating that our line is well characterised by systemic emission convolved with beam smearing (see Fig. \ref{fig:dysmalpy_outflows}), and thus shows no indication of outflows.

    \centering
    \begin{subfigure}[b]{0.5\textwidth}
        \centering
        \includegraphics[width=\textwidth]{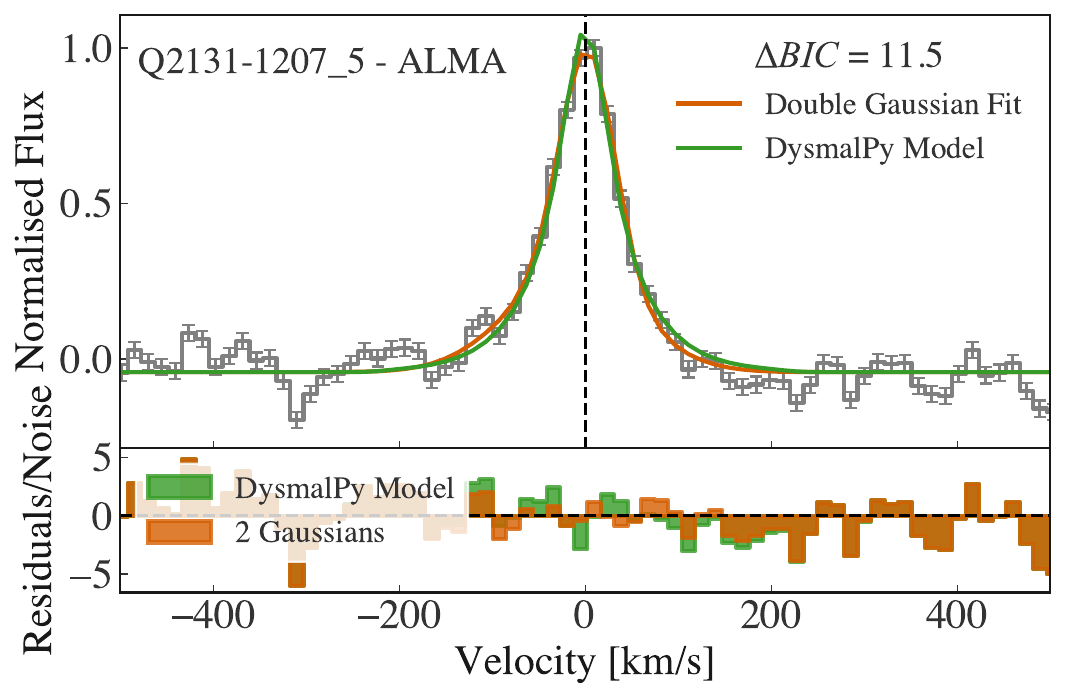}
    \end{subfigure}%
    ~ 
    \begin{subfigure}[b]{0.5\textwidth}
        \centering
        \includegraphics[width=\textwidth]{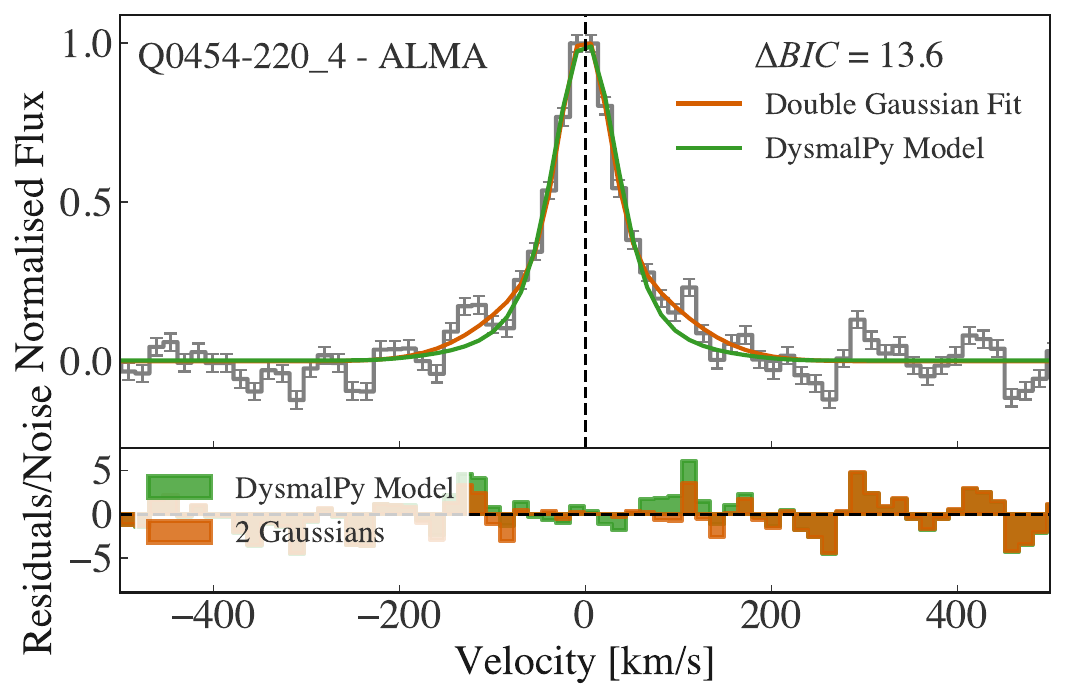}
    \end{subfigure}
    \caption{Comparison of the two Gaussian best-fit model (red line) with the model extracted from {\scshape DysmalPy} (green line) for Q2131-1207\_5 (left) and Q0454-220\_4 (right). The bottom sub-plots show the fit residuals.}
    \label{fig:dysmalpy_outflows}
\end{figure*}

\onecolumn
\begin{figure*}
\section{{\scshape pPXF} fitting}\label{ap:ppxf}

To fit the galaxies using {\scshape pPXF}, we use the highest spectral resolution stellar templates (R $\sim 10,000$), the X-shooter Spectral Library (XSL) which covers 3000 $\AA$ to 25,000 $\AA$ \citep{verro_modelling_2022}. As a sanity check, we also used the e-MILES library and observed that it affects our velocity dispersion estimates by $< 10\%$. To account for instrumental broadening of our lines, we create a wavelength-dependent LSF, using Eq. 7 from \citealt{bacon_muse_2017}, which we convolve with both our stellar templates and emission lines templates. In this way, the resulting velocity dispersions are the intrinsic line-of-sight velocity dispersions of the galaxies. We fit both the stellar continuum and the emission lines of our spectra ([O\,{\sc ii}]3726,3728; [Ne\,{\sc iii}]3968; [Ne\,{\sc iii}]3869; H$\gamma$; He {\sc ii}4687; H$\beta$; [O\,{\sc iii}]4959,5007) over the range [3600; 5100] $\AA$, where all emission lines are tied together in wavelength and width.

Here we present the remaining three fits to the integrated spectra of Q0454-220\_4, Q0152-2001\_12 and Q1345-0023\_13. For Q0454-220\_4, we remove spaxels of the galaxy due to the presence of a companion south-east of the galaxy, as well as a diffraction spike from the nearby quasar. However, some leftover contamination to the stellar continuum can be seen in the spatially resolved dispersion map of the galaxy, and it adds some uncertainty to the estimate of the stellar velocity dispersion.

        \centering
        \includegraphics[width=\textwidth]{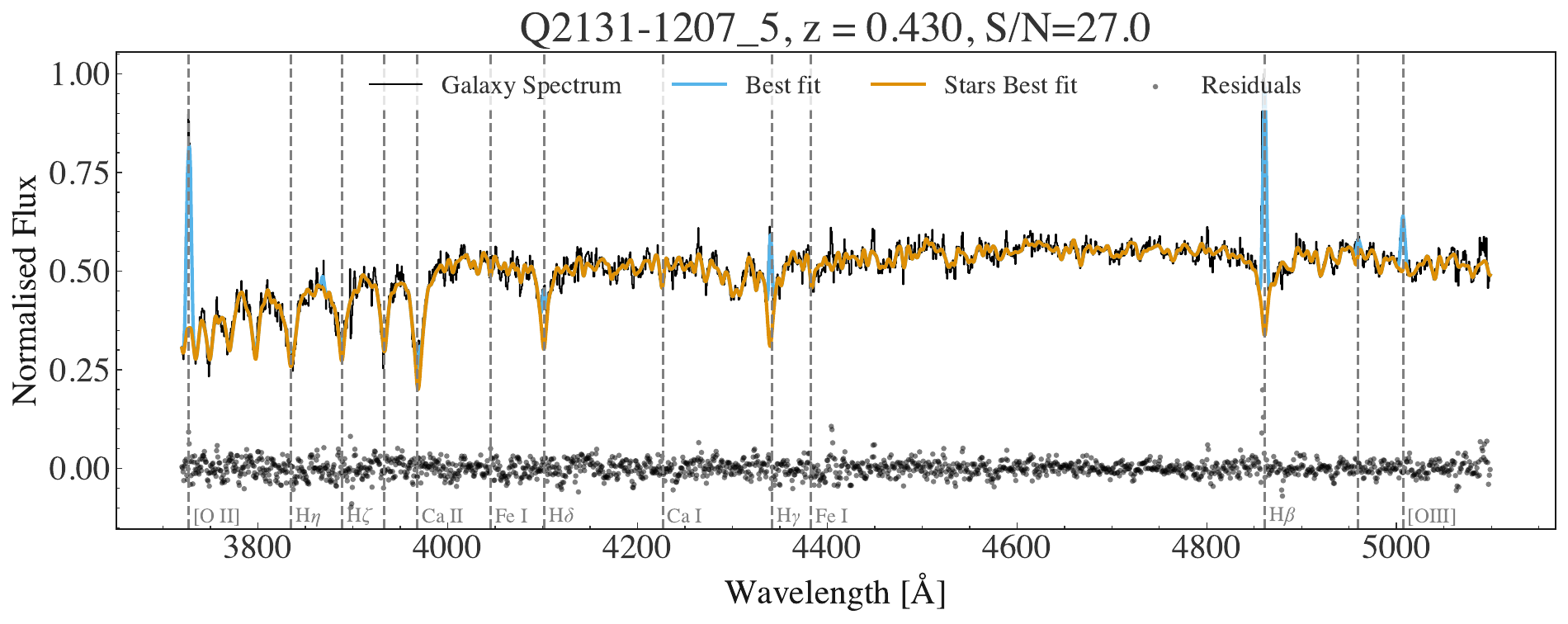}
        \caption{Best-fit to the integrated spectrum of Q2131-1207\_5 from {\scshape pPXF}. The black solid line represents the data, the orange solid line represents the fit to the stellar continuum, and the blue solid line represents the fit to the stellar continuum. The grey dots are the fit residuals.}
        \label{fig:2131_ppxf}
    \end{figure*}

\begin{figure*}
        \centering
        \includegraphics[width=\textwidth]{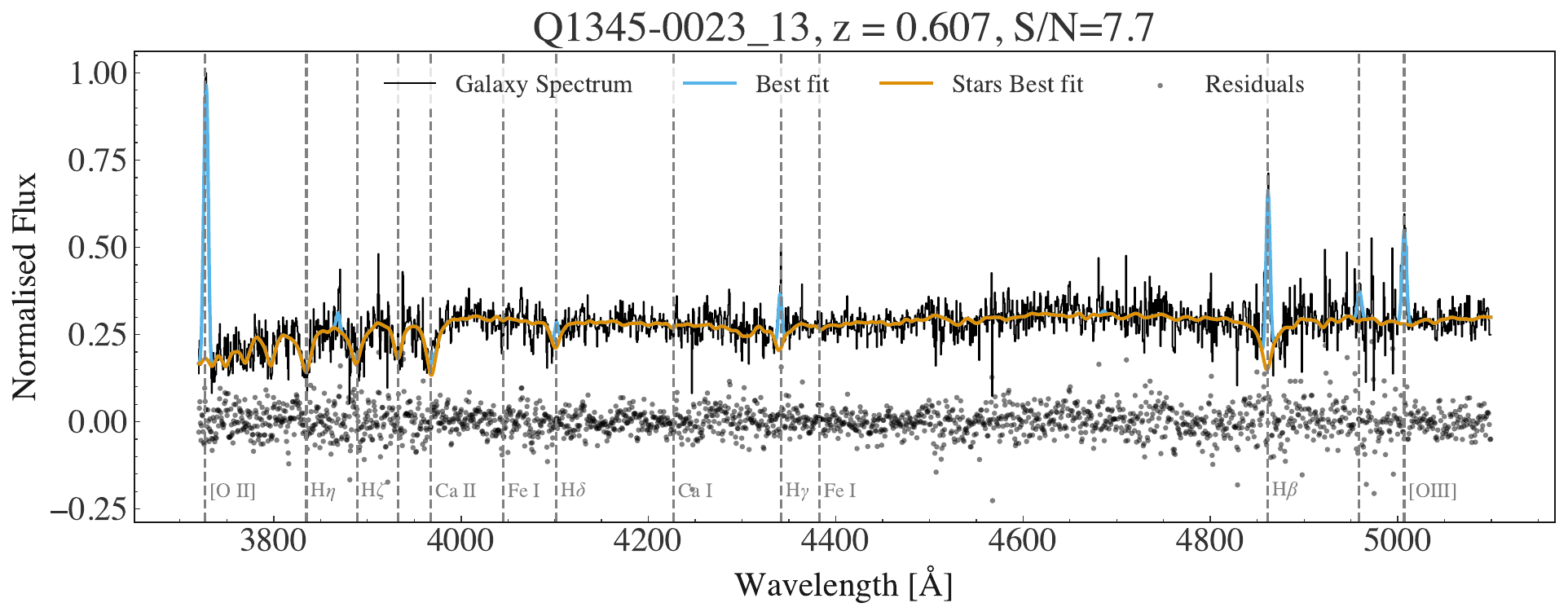}
        \caption{Best-fit to the integrated spectrum of Q1345-0023\_13 from {\scshape pPXF}. The legend is the same as for Figures \ref{fig:2131_ppxf} \& \ref{fig:0454_ppxf}.}
        \label{fig:1345_ppxf}
    \end{figure*}

\begin{figure*}
        \centering
        \includegraphics[width=\textwidth]{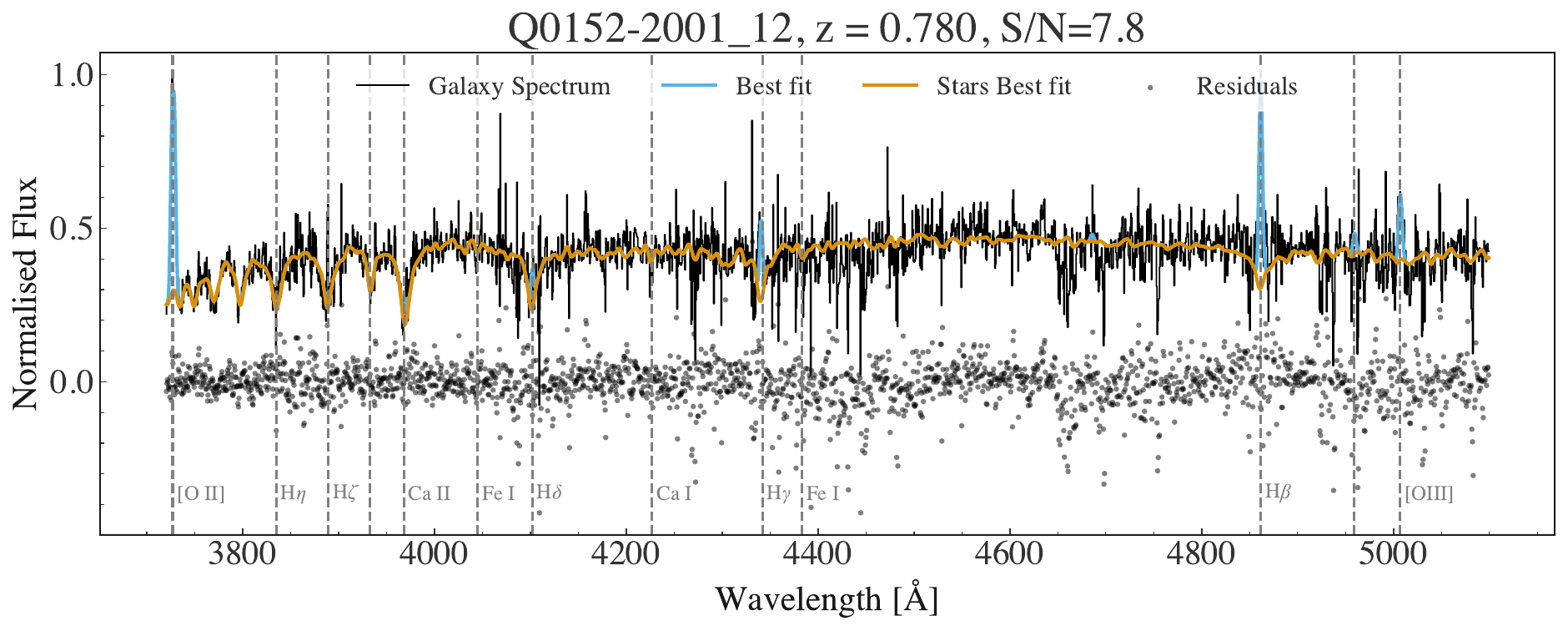}
        \caption{Best-fit to the integrated spectrum of Q0152-2001\_12 from {\scshape pPXF}. The legend is the same as for Figures \ref{fig:2131_ppxf} \& \ref{fig:0454_ppxf}.}
        \label{fig:0152_ppxf}
    \end{figure*}

\begin{table}[h!]
\section{Disk turbulence model}\label{ap:krumholz}

In this section we report the used parameter values for the model presented in \cite{krumholz_unified_2018}. Table \ref{table:params_theory} presents the rotation curve index $\beta$, the scaling factor for the turbulent dissipation rate $\eta$, the SF efficiency per free-fall time $\epsilon_{ff}$, the disk stability parameter $Q$, the fraction of the ISM in the SF cold molecular phase $f_{sf}$, the fraction of gas contributing to $Q$ $f_{g,Q}$, the fraction of gas self-gravity contributing to the mid-plane pressure $f_{g,P}$, the ratio of total pressure to turbulent pressure at mid-plane $\phi_{mp}$, the ratio of gas-to-stellar $Q$ plus one $\phi_Q$, the maximum SF timescale $t_{sf}$, and the mean momentum injected per unit mass of stars $\langle p_*/m_* \rangle$ (see also Table 1 in \citealt{krumholz_unified_2018}).
    \caption{Values used for the parameters of Eq. \ref{eq:sig_sfr} following the model description of \citealt{krumholz_unified_2018} for local spirals.}                 
    \label{table:params_theory}   
    \centering                      
    \begin{tabular}{c | c c c c c c c c c c }      
    \hline\hline              
    Parameter     &  $\beta $ & $\eta$  & $\epsilon_{ff}$ & $Q$ & $f_{sf}$ & $f_{g,P},f_{g,Q}$ & $\phi_{mp}$ & $\phi_Q$ & $t_{sf}$ & $\langle p_*/m_* \rangle$ \\
                  &           &         &                 &     &          &                   &             &          & Gyr      & Myr                       \\
    \hline 
    Value         &  0.0      & 1.5     & 0.015            & 1.0 & 0.5      & 0.5               & 1.4         & 2.0      & 2       & 3000                      \\
    \hline                                 
    \end{tabular}
    \end{table}

\twocolumn
\end{appendix}
\end{document}